\documentclass[lettersize,journal]{IEEEtran}
\usepackage{amsmath,amsfonts}
\usepackage{diagbox} % 加载绘制对角线表头所需的包
\usepackage{array}
 \usepackage{subfigure}
\usepackage{textcomp}
\usepackage{stfloats}
\usepackage{url}
\usepackage{verbatim}
\usepackage{graphicx}
\usepackage{cite}
\usepackage{algorithm}
 \usepackage{algpseudocode}
 \usepackage{amssymb}
 \usepackage{booktabs}
 \usepackage{multirow}
 \usepackage{epstopdf}
\renewcommand{\Require}{\item[\textbf{Input:}]}
\renewcommand{\Ensure}{\item[\textbf{Output:}]}
  \newcounter{mytempeqncnt}
\begin{document}

\title{Blockchain-Enabled Dynamic Spectrum Sharing for Satellite and Terrestrial
	Communication Networks}
 
	\author{Zixin Wang, Mingrui Cao, Hao Jiang, Bin Cao\IEEEauthorrefmark{1}~\IEEEmembership{Senior Member,~IEEE}, Shuo Wang, \\Chen Sun\IEEEauthorrefmark{1}, and Mugen Peng ~\IEEEmembership{Fellow,~IEEE}
	% <-this % stops a space
\thanks{This work was supported in part by the National Key Research and
Development Program of China under Grant 2021YFB1714100, in part by the
National Natural Science Foundation of China under Grant No. U22B2006,  and in part by Research and Development Center, Sony China. (*\textit{Corresponding author: Bin Cao, Chen Sun})}% <-this % stops a space
\thanks{Z. Wang, M. Cao, H. Jiang, B. Cao, and M. Peng are with the State Key Laboratory of Networking and Switching Technology,  Beijing University of Posts and Telecommunications, Beijing, 100876, China (Email: \{wangzx, caomingrui, jdsxjh, caobin,   pmg\}@bupt.edu.cn)}

\thanks{S. Wang and C. Sun are with Research \& Development Center Sony (China) Ltd., Beijing, 100028, China (Email: \{Shuo.Wang, Chen.Sun\}@sony.com)}

}
\markboth{Journal of \LaTeX\ Class Files,~Vol.~14, No.~8, August~2021}%
{Shell \MakeLowercase{\textit{et al.}}: A Sample Article Using IEEEtran.cls for IEEE Journals}


\maketitle

\begin{abstract}
	Dynamic spectrum sharing (DSS) between satellite and terrestrial networks has increasingly engaged the academic and industrial sectors. Nevertheless, facilitating secure, efficient and scalable sharing continues to pose a pivotal challenge. Emerging as a promising technology to bridge the trust gap among multiple participants, blockchain has been envisioned to enable  DSS  in a decentralized manner. However, satellites with limited resources may struggle to support the frequent interactions required by blockchain networks. Additionally, % due to the extensive coverage area of satellites, the differentiated spectrum sharing needs in various regions  make traditional blockchain approaches inadequate. 
	given the extensive coverage of satellites, spectrum sharing needs vary by regions, challenging traditional blockchain approaches to accommodate  differences. In this work, a partitioned, self-governed, and customized dynamic spectrum sharing approach (PSC-DSS) is proposed for spectrum sharing between satellite access networks and terrestrial access networks. This approach establishes a sharded and tiered architecture which allows various regions to manage spectrum autonomously while jointly maintaining a single blockchain ledger. Moreover,  a spectrum-consensus integrated mechanism, which decouples  DSS process and couples  it with   blockchain consensus protocol, is designed to enable regions to  conduct DSS transactions  in parallel and dynamically innovate spectrum sharing schemes without affecting others.  Furthermore, a theoretical framework  is derived to justify the stability performance of  PSC-DSS. Finally, simulations and experiments  are conducted to validate the advantageous performance  of  PSC-DSS in terms of low-overhead, high efficiency,  and  robust  stability.
\end{abstract}


\section{Introduction}
%3GPP, “Solutions for NR to support non-terrestrial networks (NTN): Non-terrestrial networks (NTN) related RF and co-existence aspects,” Technical Report 38.863, Version 0.1.0, 2021.

%ESAARTES project, ASCENT [4], has conducted field trials of a system sharing satellite-terrestrial spectrum.   H¨oyhty¨a, M., et al. “Licensed shared access field trial and a testbed for satellite-terrestrial communication including research directions for 5G and beyond.” 2020. International Journal of Satellite Communications and Networking.

%频谱资源是移动通信发展的基础
%\IEEEPARstart{S}{p}ectrum is the single most importance nature resource for wireless communication in cellular networks. 
\IEEEPARstart{S}{a}tellite access networks (SANs) are envisioned to be broadly  supplement to terrestrial  access networks (TANs) to realize global seamless coverage, operating in frequency bands below 6 GHz for providing mobile satellite services \cite{TR38811, tr38863}.
%With the increasing of wireless devices and  traffic, the demand on spectrum increase significantly.
%According to the vision of Groupe Speciale Mobile Association (GSMA), the communication spectrum demand will reach 2 GHz by 2030 \cite{SpctrumGSMA}. 
With the increasing of wireless devices and the expansion of  satellite constellations, there is a marked escalation in the demand for spectrum resources.
%Despite high-band spectrum resources continuing to be exploited as mobile technology advances, the below 6 GHz band still plays a significant role in mobile communications due to its wide coverage property. 
However, the spectrum resources at  below 6 GHz band have been almost exhaustively licensed while the Federal Communications Commission (FCC) reports that less than 85\% of this band is actually used \cite{9913217}. High demand and low utilization of spectrum motivate the development of spectrum sharing technologies that reallocate temporarily idle resources between SANs and TANs. 

The typical spectrum sharing solutions  include static spectrum sharing and dynamic spectrum sharing (DSS)  \cite{7500126}. Static spectrum sharing experiences low utilization efficiency due to its fixed and exclusive manner. Thus, DSS  is progressively becoming the mainstream  to further exploit the potential of  limited spectrum resources supply \cite{9751742}. Several DSS solutions have been developed, such as the well-known spectrum access system (SAS) for citizens broadband radio service (CBRS) system \cite{8737533}. %However, existing DSS solutions faces the following major challenges: 
 %However, existing DSS solutions faces major challenges of mandatory trust, constrained  flexibility, and limited scalability.
%However, existing DSS solutions faces major challenges in terms of security,    flexibility, and   scalability.
However, existing DSS solutions faces major challenges in terms of  security and scalability.

For security, all spectrum users must place absolute trust in the spectrum administrators, such as SAS administrators for CBRS \cite{10109160}, which are  presumed to be trustworthy to perform reasonable spectrum allocation decisions using a centralized database-based system. This  mandatory trust, however,  inevitably leads the risk of single point failure and  raises security concerns  over the malicious exploitation of critical nodes,  especially in  the evolving threat landscape of SANs.
%extensive and complex  SANs. %Second, centralized DSS models require all  participants adhere to a unified spectrum sharing scheme. It results in constrained  flexibility in the evolving SANs where  an expanding diversity of  participants introduces a spectrum of increasingly dynamic and varied requirements. 
For scalability,  centralized DSS models  impose excessive regulatory pressure on regulators as an  increasing number of heterogeneous participants from regions or countries become involved in SANs, consequently leading to limited scalability.
Therefore, a new DSS paradigm that is secure and scalable is  in high demand.

%As an emerging technology, blockchain has attracted attention from both academia and industry for DSS \cite{ZTE,5GZORRO,FCCblockchain}, due to its ability to enable trusted transaction processing and  immutable ledger keeping among mutually distrustful participants,  even if a certain portion of them behave  maliciously  \cite{10109160,8972381}.  
%As an emerging technology, blockchain has been envisioned as a viable technology to enhance the security of DSS due to its ability to enable trusted transaction processing and  immutable ledger keeping among mutually distrustful participants,  even if a certain portion of them behave  maliciously  \cite{10109160,8972381}.  
As an emerging technology, blockchain has shown the potential to improve the security of DSS due to its ability to enable trusted transaction processing and  immutable ledger keeping among mutually distrustful participants,  even if a certain portion of them behave  maliciously  \cite{10109160,8972381}.   The support for the self-executing smart contracts also empowers blockchain to improve the scalability of DSS by distributing responsibility and workload among various participants in a decentralized manner.
%The decentralization, collective maintenance, and support for the self-executing smart contracts also empower blockchain to improve the  scalability of DSS by distributing responsibility and pressure among  various participants.
%have made blockchain as a viable technology for  facilitating decentralized and scalable spectrum sharing \cite{8703084}. 
Many government agencies and organizations have voiced consider blockchain as a possible paradigm to enable  DSS in the future, such as FCC \cite{FCCblockchain}, China Communications Standards Association \cite{CCSAblockchain}, and l'Agence Nationale des Fr\'equences \cite{FaGuoblockchain}. 
 Meanwhile, several  concrete solutions  \cite{ZTE,5GZORRO} and  innovative studies have been proposed  \cite{8737533, 9679805,10109160}. However, employing blockchain for DSS in SANs  still faces the following important challenges:
 % in terms of high overhead, constrained  flexibility, limited throughput, and inadequate support for both forward and backward compatibility.
\begin{itemize}
	\item \textit{{High overhead}}:  Conventional blockchain-based DSS architecture requires that each transaction be validated by all blockchain nodes. Moreover,  additional cross-chain infrastructure is needed to enable interoperability and communication between different chains. This architecture is impractical for DSS in SANs, given the limited %communication, storage, and computational
	 resources available on satellites.% Although two-tier multi-zone architecture has been developed for DSS, participants are required to preserve both the global and local chain to enable interoperability and communication between difference local chains. %Moreover,  additional cross chain infrastructure is needed to enable interoperability and communication between difference local chains.
		\item \textit{Limited efficiency}: % The prevailing  blockchain-based DSS process currently entails sequentially distributing each step to distinct blocks, that is, a round consensus only perform one step of DSS.
		The prevailing blockchain-based DSS process currently involves distributing each step sequentially to distinct blocks, meaning that a round of consensus only performs one step of the DSS.
	However,  this process is time-consuming and power-intensive, which falls significantly short in meeting the crucial needs for both efficient and large-scale spectrum sharing.
	%However, considering the timeliness requirements and the scale of services demanded by SANs, this process falls significantly short in meeting the crucial needs for both efficient and large-scale spectrum sharing.
	%A complete spectrum sharing process involves many steps, such as status change,  requirement statement,  and specific algorithm execution. However, the most common practice currently is to sequentially distribute each step across different blocks.
	%\item \textit{Inadequate support for both forward and backward compatibility}: 	For practical purposes, 
	
	\item \textit{Constrained  flexibility}: Existing  blockchain-based DSS solutions  require all  participants adhere to a unified spectrum sharing scheme,  constraining flexibility in the evolving SANs that face a growing diversity of participants with dynamic and varied requirements. For practical purposes,  an optimal solution should support for both forward and backward compatibility, especially for SANs undergoing rapid evolution.
	%It would lead to constrained  flexibility in the evolving SANs where  an expanding diversity of  participants introduces a spectrum of increasingly dynamic and varied requirements. For practical purposes,  a optimal solution should be 
	%Although two-tier multi-zone architecture has been developed for 

\end{itemize}

Moreover, the characterization of the stability of blockchain systems in SANs is of importance because of ultra-expensive costs for deploying satellites and installing blockchain in satellites. An accurate theoretical   framework is   essential to thoroughly understand how such systems operate,  which kinds of system factors can affect their performance,  what the principles that these system factors influence the performance, and further obtain insights on network design guidance \cite{9841465}.
Different from wired networks, the features of SANs, including unstable channel, severe interference, untrusted physical entities, etc., pose many extra difficulties in both theoretical analysis and practical implementations.
Therefore, faced with such a complex environment, it is crucial but challenging to consider these features in analyzing  the stability of blockchain systems in SANs. However, the study that considers these features simultaneously when applying blockchain to DSS in SANs, is yet inadequate.

%Aiming at tackling the above challenges, we focus on developing a  low-overhead, highly flexible, and efficient blockchain-based DSS paradigm that leverages the principle of blockchain to bridge the gap among various parties  and establish healthy relationships among diverse spectrum sharing participants in SANs.
The above observations inspire us to develop  a partitioned, self-governed and customized
 DSS approach dubbed PSC-DSS, aiming to provide a low-overhead, highly  efficient and  flexible  DSS solution for SANs. This approach  leverages the principle of blockchain to bridge the gap among various parties  and establish healthy relationships among diverse spectrum sharing participants in SANs.
  In order to advance the understanding of the proposed PSC-DSS, we address three fundamental questions in this paper, as follows.
%we establish a novel two-tier multi-zone blockchain-based DSS architecture  with a single-chain structure, which is better suited to meet the low overhead, high flexibility, and  efficiency requirements of SANs. Following this architecture, we propose a new consensus mechanism that integrated the  
 \begin{itemize}
 	\item \textit{How to construct the PSC-DSS for SANs:}
 	To tackle this question, we establish a two-tier multi-region  blockchain-based DSS architecture  with a single-chain structure, where the  spectrum autonomy within each region  and global information synchronization achieved through upper-level interaction. Importantly,  this architecture allows different region s to adopt various spectrum sharing schemes and enables interaction across region s without the need for any additional cross-chain infrastructure.
 	\item \textit{How to perform the  PSC-DSS in SANs:}
 	To address this question, we design a spectrum-consensus integrated mechanism, which  couples blockchain consensus protocol  with spectrum sharing scheme.
	 This mechanism redesigns the consensus protocol and restructures the DSS procedure, enabling regions to parallelly process DSS transactions and dynamically innovate spectrum sharing schemes without affecting others.
		 Furthermore, the generalized workflow and  main functions are introduced to advance the understanding of the proposed mechanism.
 	
 	\item \textit{How to analyze the performance of the  PSC-DSS within SANs:} 
 	Faced with this question, we first build a theoretical framework to study the probability of system stability. Based on the derived closed-form expression, we further explore the influence of the unstable wireless environment of satellite-terrestrial communication on system stability using stochastic geometry. %These analyses can help us understand how the new paradigm works and provide insightful guidelines for further implementations and extensions. 
 	%Moreover, to the best of our knowledge, this work is the first to analyze blockchain system stability in conjunction with an unstable satellite-terrestrial environment. 
 	
 \end{itemize}

Furthermore, simulation and  experiment results demonstrate the performance of this work in terms of low-overhead, high efficiency,  and  robust  stability under various network parameters. %We further provide some pivotal insights and design guidelines into the design of this work.
Pivotal insights and design guidelines are  provided for further implementations and extensions.

The rest of this work is organized as
follows.  Section II  introduces the overview of PSC-DSS including architecture, entities, and workflow. Section III  presents the proposed spectrum-consensus integrated mechanism, detailing the main functions and procedures. Section IV analyses  the stability
performance. Simulations and  experiments are conducted in Section V.  Section VI reviews the existing related works. Finally, Section VII concludes this work.

\section{PSC-DSS Overview}
In this section,  the architecture of PSC-DSS is  first introduced. Then,  participants and main tasks are defined. Finally, the work flow of PSC-DSS is described.

\subsection{Two-tier Multi-region Architecture}

\begin{figure*}
	\centering
	\includegraphics[width=\linewidth]{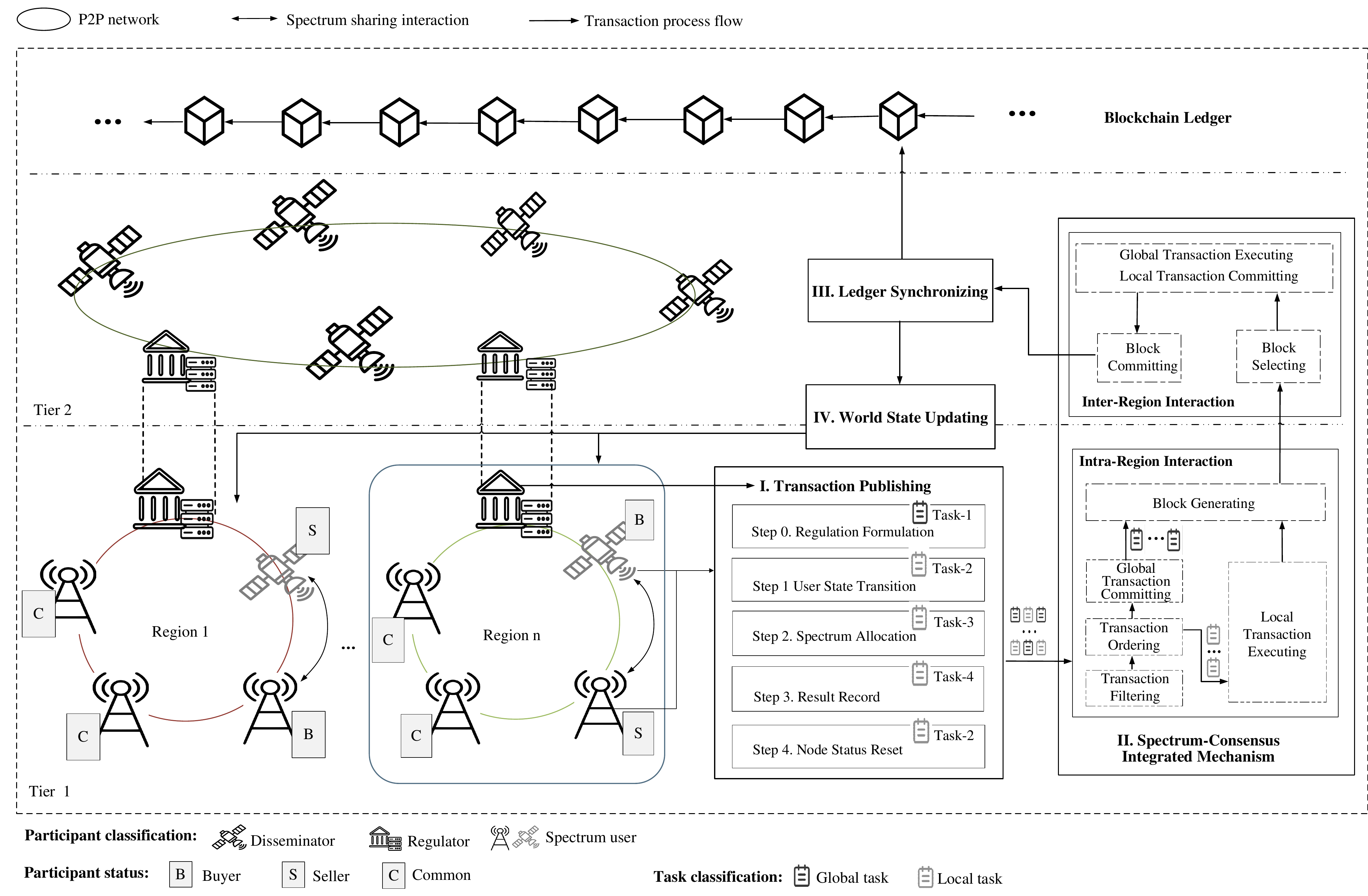}
	\caption{PSC-DSS architecture}
	\label{fig:architecture}
\end{figure*}

As illustrated in Fig. \ref{fig:architecture}, the proposed PSC-DSS is composed of two tiers and multiple regions. It is worth noting that all participants in PSC-DSS only maintain a 
single-chain, which is highly different with existing sharding-based DSS solutions. 

Tier 1 consists of multiple specific  regions, each including a regulator node,   base stations, and satellites. Each region autonomously manages its spectrum, encompassing the selection and dynamic evolution of suitable spectrum sharing schemes. Transactions related to a specific region are  packaged into a block and submitted to tier 2 after the intra-region interaction of the  proposed spectrum-consensus integrated mechanism is performed.   
Accordingly,  spectrum sharing can be undertaken separately and parallelly, and thus enabling  efficient and large-scale spectrum sharing in SANs. Furthermore, following this sharding-based design, spectrum management rights are devolved to regions, and promoting   more activity and flexible spectral business.

Tier 2, curated by all regulator nodes and many satellites,  is responsible for   receiving and disseminating the blocks  submitted by tier 1, then performing the inter-region interaction of the proposed spectrum-consensus integrated mechanism. After that, all  regulator nodes  update their world state and blockchain ledger, which are then synchronized with base stations and satellites in tier 1. %Therefore, all spectrum sharing participants record only one ledger, which reduce the excessive consumption resource caused by maintaining a multi-region blockchain networks. Since any DSS transaction in any region can be traced back in blockchain, thus enabling simple but trustworthy cross-region supervision and increases recognition of on-chain DSS transactions.
Therefore, all participants record only one ledger, which reduces resource consumption caused by maintaining multi-region blockchain networks, especially for the resource-constrained satellites. 
This single-chain design enables any DSS transaction in any region can be traced back through the blockchain, thus realizing simple but trustworthy cross-region supervision and enhancing the recognition of on-chain DSS transactions.
%Since any DSS transaction in any region can be traced back through the blockchain, this single-chain structure  enables simple but trustworthy cross-region supervision and increases the recognition of on-chain DSS transactions

Moreover, PSC-DSS demonstrates robust backward and forward compatibility, allowing for integration with existing distributed spectrum sharing systems. An example of PSC-DSS instantiation is seen in CBRS, where the concept of a ``region" aligns with a CBRS ``zone" and the ``tier" concept corresponds to the relationship between the FCC and CBDS. 
Furthermore, PSC-DSS supports the dynamic evolution of spectrum sharing schemes for each region by updating a key function in the proposed spectrum-consensus integrated mechanism (as detailed in Section \ref{Spectrum-Consensus Integrated Mechanism}), without impacting others.

\iffalse
Consider a cellular network consisting of multiple gNBs and services, with mobile users located within the coverage area. It is worth to note that the gNBs are identified by spectrum provider (SPs) and spectrum demander (SDs) rather than being distinguished as primary and secondary users, thus our model can be applied in difference scenarios, is not limited to vertical sharing (between licensed users and unlicensed users) and horizontal sharing (among licensed users).

\begin{figure}
	\centering
	\includegraphics[width=1\linewidth]{FIG/DSS.pdf}
	\caption{System model}
	\label{fig:dss}
\end{figure}

 \begin{align}
 	 {I_{i,j}} = {\left( {\frac{\lambda }{{4\pi {d_{i,j}}}}} \right)^\alpha }p_i^{trams}{G_{Tx}}{G_{Rx}} 
 \end{align}
\fi
\subsection{Participants}
PSC-DSS involves 3 types of participants: regulators, spectrum users, and disseminators. These participants each play a critical role in spectrum sharing system  and blockchain system, respectively.  

\textbf{Regulators} are spectrum management entities that publish regulations
on spectrum sharing in their jurisdiction, and provide regulation-compliant spectrum allocation service to spectrum users in their regions. For example, regulators can be distributed FCC entities to make regulations for spectrum sharing in U.S., or they can be SAS servers that implement specific spectrum sharing schemes in certain regions. In blockchain networks, regulators are responsible for starting a round of intra-region interaction, and submitting the generated block to satellites in tier 2. Besides, regulators are also in charge of  performing the inter-region interaction in tier 2.
%publishing transactions that include a series of DSS tasks, packaging these transactions into a block, and then submitting the block to satellites in tier 2. Besides, regulators are also in charge of block executing and block committing in tier 2.

\textbf{Spectrum users} include base stations and satellites in tier 1 for spectrum  access assignment. Additionally, this category can be expanded to encompass other entities such as access points, road side units, and campus hotspots. A spectrum user are  categorized into 3 status: buyer,  and seller, common.  Buyer status indicates that a spectrum user needs additional spectrum to meet specific demands, while seller status represents one who has surplus spectrum available for sublease. Common status is assigned to the spectrum users with stable demand who neither need to buy nor sell spectrum.
In blockchain network, spectrum users are responsible for performing intra-region interactions as directed by the regulator in their region.  

\textbf{Disseminators} are consisted of   satellites that have no spectrum  spectrum  access assignment needs in their regions, considering potential security risks from interest entanglements. They are in charge of  receiving blocks submitted by  regulators and propagating these blocks into all  participants in tier 2. A bootstrapper is opened in a selected disseminator  to order the blocks and bootstrap  regulators for a new round of inter-region interaction.

Disseminators do not participate in spectrum sharing and consensus in blockchain. Therefore, among all participants in the PSC-DSS, only regulators and spectrum users are defined as \textbf{blockchain participants}. They engage in transaction verification, task execution, and ledger updates.

\subsection{Main Tasks} 
In PSC-DSS, spectrum sharing-related tasks are typically classified into 4 types, where Task-1 is a global task that needs coordinated implementation in tier 2, while Tasks 2 to 4 are local tasks that can be executed within a specific region.
Each type of task can be derived into many independent blockchain transactions, and a blockchain transaction can contain multiple types of tasks.

\textbf{Task-1 Regulation formulation:} PSC-DSS allows regulators to issue rules on various tasks, such as identifying spectrum regions, specifying interference models, and updating participant's information. It is the action guideline for spectrum sharing schemes, and all subsequent spectrum sharing-related tasks are performed on this basis. %Considering the dynamic feature of SANs,  information updates such as registration and revocation of participants are also performed by publishing this task.

\textbf{Task-2 User status transition:} Spectrum users specify their spectrum needs, including frequency band, current status and target status for this frequency band,   price, duration of use,  and other relevant details. Regulators publish this task to reset the status of spectrum  users after completing a round of  spectrum allocation process.

\textbf{Task-3 Spectrum allocation:} PSC-DSS performs spectrum sharing based on user requests,  including parameters for predefined spectrum sharing scheme, user information contained in Task-2, and other required details. Various customize spectrum sharing schemes are allowed to lunched in difference regions to suit their actual situation.

\textbf{Task-4 Result record:} Irreversible spectrum  requests, allocation results, and  current spectrum status are recorded. This task provides a complete track of the spectrum sharing process for all frequency band. Thus, efficient supervisions and audits can be implemented based on this task.

To reduce the overhead of base stations and satellites, PSC-DSS stipulates that only Task-2 is issued by spectrum users during their spectrum application stage, while all other tasks are issued by regulators.

\subsection{Main Workflow}

Since the PSC-DSS architecture is envisioned to be broadly inter-operative and to support multiple advanced wireless services and standards, this work focuses on the most basic spectrum sharing approach for which the procedure is shown in Fig. \ref{fig:architecture}.
 
 In step 0, the preparation for spectrum sharing, regulators should issue transactions for Task-1  to publish regulations on spectrum usage and management. 
This step may be performed periodically or irregularly, depending on updates to spectrum sharing regulations, changes in current spectrum sharing requirements, and  evaluations for spectrum sharing implementation plans.

 In step 1,   spectrum users  initiate status requests to regulators  to publish transactions for Task-2, according to their own needs. For example, if a user has idle spectrum available for sublease, its status in the corresponding frequency band is changed from common to seller.
 Details about the current information of users and parameters for subsequent spectrum allocation in step 3 are submitted to the system.

In step 2, transactions for Task-3 are published by  regulators to trigger the specific spectrum sharing schemes.  An explicit spectrum allocation solution is obtained for all frequency bands on sale, encompassing spectrum access decisions,  operational parameters, business settings, etc.

In step 3, regulators publish transactions for Task-4 to record the whole spectrum sharing process. For a specific frequency band,   transfer details such as  involved users and price, and operational parameters like transmission power, will be comprehensively recorded. For a spectrum user, specifics in request, transitions in status,  changes in assets are captured.
%Regulators can use this step to synchronize the status of spectrum management, and service providers can use this step to assess their own benefits. At the same time, based on this step, data and case support can be provided for any subsequent compensation scheme.

In step 4, the status of users  are reset as common through the transactions pertain to Task-2 published by  regulators.  This step  marks the end of this round of spectrum sharing. If a user  has a persistent requirement to maintain a certain status, it needs to re-initiate Task-2 as a new transaction in step1, considering the  rapid changes in terms of   network topology, channel usage, and participants  in SANs.
%If there are no specific requests from both parties, the status of both parties is reset to common. However, if there are special requirements from spectrum users during this round of allocation, such as the seller/buyer status continuing for a specific round, the system will reset their status to the ideal status at this stage in accordance with the requirements of these users. This approach offers the advantage of skipping step 1 in the next round of spectrum sharing, thereby reducing overhead and latency.

Each of the above steps is accomplished by executing a blockchain transaction. Therefore, from the perspective of blockchain, the process that each step undergoes is described as follows. 
Step I is for transaction publishing.  All transactions for tasks are published by regulators with the aim of reducing the operational overhead of base stations and satellites. Although Task-2 is issued by users to actively transition their status, users merely send task requests to regulators. After a fixed period of time or once a certain number of requests have been collected, regulators generate transactions for these requests.
Step II is to perform the proposed spectrum-consensus integrated mechanism. % which consists of intra-region interaction  and inter-region interaction. 
 During this step,  transactions are packaged into blocks and  executed by blockchain participants. Specifically, transactions related to global tasks are executed by regulators, while transactions for local tasks are  executed by blockchain participants within corresponding regions, with only the results of these executions being recorded in blocks.
In step III, each blcokchain participant updates its local ledger copy to match the latest blockchain state on the network. %This process ensures that all blcokchain participants possess the same and up-to-date block in ledger. It achieves consistency and completeness in the spectrum sharing process across the network, guaranteeing that all transactions and blocks are correctly recorded and confirmed.
Accordingly, in step IV, the world state is first updated in all regulators and then propagated to base stations and satellites within each region. Details pertaining to all tasks are recorded and updated to ensure the accuracy and currency of spectrum management in the PSC-DSS.

\section{Spectrum-Consensus Integrated Mechanism}
\label{Spectrum-Consensus Integrated Mechanism}
In this section,  the proposed  spectrum-consensus integrated mechanism based on the PSC-DSS architecture are detailed. First, the main  components for this mechanism are introduced. Then, key functions and corresponding procedures are described.

\begin{figure*}
	\centering
	\includegraphics[width=1\linewidth]{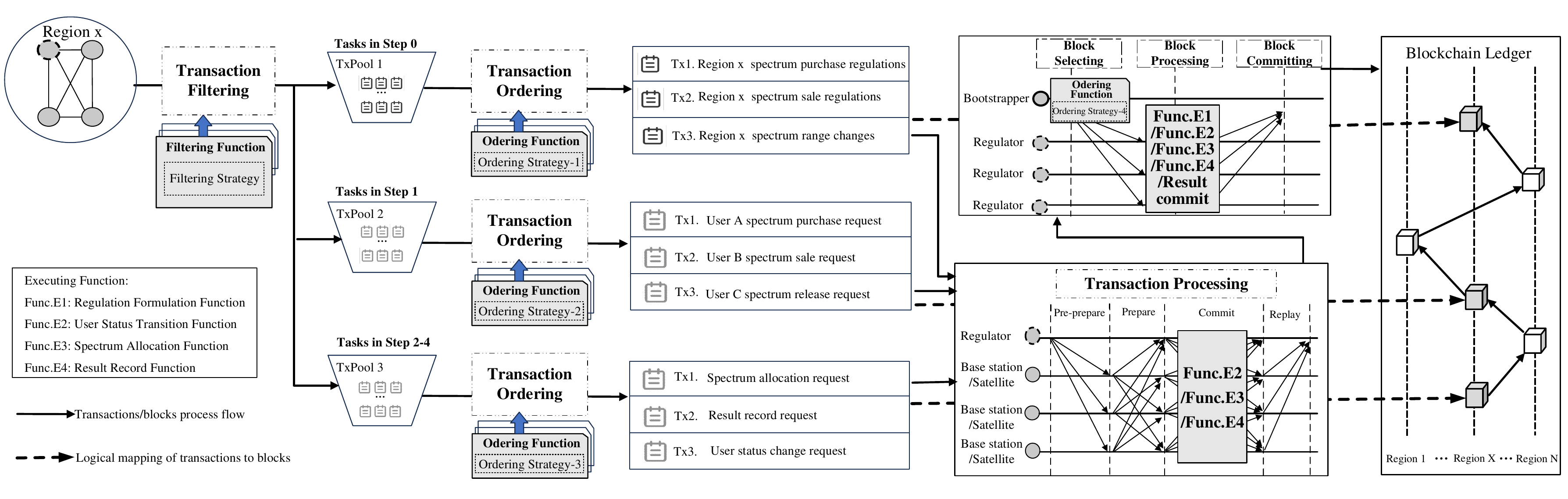}
	\caption{The process of spectrum-consensus integrated mechanism}
	\label{fig:mechanism}
\end{figure*}

\subsection{Main Components }
%The proposed  mechanism  deeply  couples the spectrum sharing scheme with the blockchain consensus protocol. Meanwhile, this mechanism support  the dynamic evolution of spectrum sharing schemes by updating 

The proposed mechanism shown in Fig. \ref{fig:mechanism}, comprises intra-region and inter-region interactions, which couple   spectrum sharing scheme with  blockchain consensus protocol. %while decoupling transaction execution from the entire blockchain network.
Details of each component are described as follows.
 
\subsubsection{\textbf{Intra-region Interaction}}
\
\newline
\indent The process of intra-region interaction consists of 3 steps: transaction filtering, transaction ordering, and  transaction processing. Through these steps, transactions for local tasks are executed and packaged into blocks.

\textit{Transaction filtering:} %As a regulator issued   transactions for tasks, it first filters the transactions into 
As the execution modes and participants of tasks vary across the steps 1-4,  regulators employ  \textit{Filtering Function} to filter transactions corresponding to tasks into 3 distinct transaction pools (TxPools). 
TxPool1 stores transactions for Task-1 in step 0,   TxPool2 manages transactions for Task-2 in step 1, and TxPool 3 is responsible for transactions related to Tasks 2 to 4 across steps 2 to 4. Furthermore, transactions from Tasks 2 to 4  which aims to apply for a cross-region spectrum sharing, are also put into TxPool1.
This filtering facilitates asynchronous transaction processing and parallel resource allocation.
%transactions corresponding to tasks are filtered through the filtering strategy at this stage. transactions are filtered to three TxPools. There are three execution modes.

\textit{Transaction Ordering:} According to Section III-D, the sequence of executing spectrum sharing tasks critically influences the sharing process.    Additionally, the priority of task execution varies and develops across different regions. Consequently, regulators  utilize \textit{Ordering Functions} to sequence the transactions taken from TxPools,  determining which transactions are packaged into a block and their order.
It is worth noting that the transactions in TxPool3 and the cross-region spectrum sharing  transactions in TxPool1 involved three types tasks (i.e., Tasks 2-4), which are processed sequentially.  Therefore, theses transactions     should be ordered following a fixed rule and be packaged into one block. 
 Other transactions can correspond to a different ordering strategies, allowing for adaptive  and logical management at each step of the spectrum sharing process. 

\textit{Transaction Processing:} %The PBFT based consensus mechanism will validate and execute the transactions at this stage. 
Each blockchain participant invokes \textit{Executing Functions} to execute transactions sequentially. These functions are specifically identified as Func.E2, Func.E3, and Func.E4, detailed in Section \ref{Key Functions}. Notably, transactions in TxPool1 involve global tasks and requires the participation from all regulators, so there is no corresponding  executing function  Func.E1 in this step. Instead, transactions in TxPool1 are packaged directly into blocks after consensus verification and submitted to tier 2 for execution. 
To be specific, a regulator packages the ordered transactions into a block and broadcasts it to blockchain participants within its region. For the block  containing transactions from TxPool1,  the practical byzantine fault tolerance (PBFT) consensus protocol is initiated. For the block  with transactions from TxPools2 or TxPools3, during the \textit{commit} stage of PBFT consensus protocol, blockchain participants trigger the corresponding executing functions (e.g., Func.E2 for TxPool2, Func.E3, Func.E4, Func.E2 sequentially for TxPool3). Then, the processed results  replace the original  transactions in the block  to form a new block. 
Subsequently,  commit message   for this new block is generated and broadcast to all blockchain participants.
Once the regulator receives commit messages  from more than $\frac{2}{3}$ blockchain participants in its region, it confirms the validity of the new block. 
Upon validation, the block is then broadcast to disseminators, marking the completion of intra-region interactions and triggering the start of the next round.

\begin{algorithm}[t]
	\caption{Ordering Function}
	\label{OrderingFunction}
	\begin{algorithmic}[1]
		\Require  {TxPool1}, {TxPool2}, {TxPool3}, {BlPool}
		
		\Ensure Transaction Lists $TxLists$, candidate block $CadiBl$
		%	\State  $TxLists \leftarrow \{\}$ 
		\If{$TxPool1$ is not empty}
		\State Order $Tx$ following FCFS model
		\State Append them into  $TxLists(1)$ 
		\EndIf

		\If{$TxPool2$  is not empty}
		\State Order $Tx$  following FCFS model
		\State Append them into  $TxLists(2)$ 
		\EndIf
		
		\If{$TxPool3$ is not empty}
		\State   Order   \textit{SpecAllo\_Intra} type $Tx$   into  $AlloTxlist $;
		\For {each $Tx$  in  $AlloTxlist $}
		\State Order   the corresponding   \textit{ResRecord }  and  \textit{StatusRest } type sequentially after the $Tx$   
		\State Append  $AlloTxlist $ into $TxLists(3)$
		\EndFor
		\EndIf
		
		\If{$BlPool$  is not empty}
		\State Order blocks following FCFS model
		\State Append them into  $CandiBlList$
		
		\EndIf
		
		\State  \textbf{Return} $TxLists $,  $ CandiBl \leftarrow CandiBlList(1)$
	\end{algorithmic}
\end{algorithm}
 
\subsubsection{\textbf{Intra-region Interaction}}
\
\newline
\indent After receiving blocks from regulators, disseminators  transmit these blocks to all disseminators and regulators through inter-satellite links (ISL). Then, the following steps are performed sequentially.

\textit{Block Selecting:} The bootstrapper activated in a selected disseminator triggers the \textit{Ordering Function} to propose a candidate block, and transmit it to all regulators.

\textit{Block Processing:} If the candidate block contains transactions from TxPool1, regulators invoke \textit{Executing Function} (i.e., Func. E1-Func. E4). Then, the execution results replace the regional transactions in the candidate block. Conversely, if the candidate block contains execution results from transactions in TxPool2 or TxPools3, regulators verify these results.
 
 \textit{Block Committing:}  After receiving feedback from $\frac{1}{2}$ regulators, the bootstrapper   sends a confirmation message to indicate that the  candidate block has been authenticated by the whole network. Then, the blockchain ledger and  the world state are updated, marking the completion of inter-region interactions and triggering the start of the next round.

 \begin{algorithm}[t]
	\caption{Func.E1 for processing Task-1}
	\label{Func.E1}
	\begin{algorithmic}[1]
		\Require Candidate block $CandiBl$
		\Ensure  Processed candidate block $  ProCandiBl $
		
		\For{each transaction $Tx$ with $ Global$ type in $CandiBl$}
		
		\State Trigger  $SC.GloFunc$ to process the  $Tx$ 
		% \Else
		%\State  Trigger  $SC(AlloFunc)$, $SC(RecFunc)$, and $SC(StaFunc)$ to process the  $Tx$.
		
		\State The process result  replaces the original $Tx$ in  $CandiBl$;
		\EndFor
		\State  \textbf{return}  $ ProCandiBl \leftarrow CandiBl$ 
	\end{algorithmic}
\end{algorithm}

\subsection{Key Functions} 
\label{Key Functions}
As shown in Fig. \ref{fig:mechanism}, the proposed mechanism involves 3 types of key functions,  which  enable specialized %and parallelized
 transaction execution. Importantly, these functions can vary across different regions and can be dynamically updated without impacting other regions. In this work,  algorithms corresponding to these key functions are presented  as use cases to advance the understanding of PSC-DSS.

\subsubsection{\textbf{Filtering Function}}
\
\newline
\indent This function filters transactions to corresponding TxPools according to their types. Based on Section II-C,   \textit{Global} tasks for global operations are placed into TxPool1.   \textit{StatusTrans} tasks for user status transitions are allocated to TxPool2.   \textit{SpecAllo}, \textit{ResRecord}, and \textit{StatusReset} tasks, which are for spectrum allocation, result recording, and status resetting respectively, are assigned to TxPool3.
%which can be presented as Alg. \ref{Filtering Function}.

\subsubsection{\textbf{Ordering Function}}
\
\newline
\indent This function sequences transactions taken from TxPools, and decides how the boostrapper selects the next candidate block from multiple blocks transmitted by regulators.  As presented in Alg. \ref{OrderingFunction},  
this work adopts  first-come-first-serve (FCFS) model to order transactions in TxPool1 and TxPool2, appending them to $TxLists(1)$ and $TxLists(2)$. Similarly, a candidate block $CandiBl$ is selected from the bootstrapper's block pool $BlPool$.
%These transactions are appended into the transaction lists $TxLists(1)$ and  $TxLists(2)$, respectively.
Especially, for transactions in TxPool3, transactions  with  \textit{SpecAllo} type are first ordered following the FCFS model. %Then, all the ordered \textit{SpecAllo} type transactions are collected to form a new transaction $AlloTx$. After this, all \textit{ResRecord}  type transactions and \textit{StatusReset}  type transactions associated with \textit{SpecAllo} type transactions in $AlloTx$ are compiled into the new transaction $RecTx$ and $ResTx$, respectively, maintaining the order of  $AlloTx$.
 Then, for each \textit{SpecAllo} transaction, the corresponding \textit{ResRecord} and \textit{StatusReset} transactions follow sequentially and are appended into $TxLists(3)$.
 %Then, for each of  \textit{SpecAllo} type transactions,    corresponding transactions of type \textit{ResRecord} and \textit{StatusReset} follow sequentially.  These transactions are appended into $TxLists(3)$. 
 %Similarly, transactions in TxPool1 with type \textit{SpecAllo-Intra}, \textit{ResRecord-Intra},   and \textit{StatusReset-Intra} are ordered and appended into $TxLists(4)$.
Finally, a transaction list $TxLists$ including ordered transactions for TxPools is generated. Each element of $TxLists$ (e.g. $TxLists(1)$) will be packaged into
one block to be processed. 
 %Here, $TxLists(1)$ and $TxLists(4)$ are allowed being  packaged into  one block.
 
 \begin{algorithm}[t]
	\caption{Func.E2 for processing Task-2}
	\label{Func.E2}
	\begin{algorithmic}[1]
		\Require   $TxLists$,   $RecResultTx$,  $AlloResultTx$  
		\Ensure  Region block $RegBl$,  shareable spectrum list $SharedSpecList$, terminal list $AwaitTerList$
		
		%\State Initialize the region block  $RegBl$;
		\If  {$TxLists(2)$ is not empty}
		%\State Initialize  $ SharedSpecList$,   $AwaitTerList$
		%	\State Initialize $AwaitTerList$;
		
		\For{each $Tx$}
		\If {$Tx$ is for $seller$ status}
		\State  Add seller's spectrum  to $ SharedSpecList$;
		
		\ElsIf {$Tx$ is for $buyer$ status}
		\State Add    buyer's  terminal   to $AwaitTerList$;
		
		\EndIf
		\State Invoke  $ SC.StaFunc$, obtain processed result ${Tx}^{'}$
		\State Add  ${Tx}^{'}$ into  $RegBl$ 
		\EndFor
		\State \textbf{return} $RegBl$,  $SharedSpecList$, $AwaitTerList$   
		\EndIf
		
		\If  {$TxLists(3)$  is not empty}
	%	\State Initialize a list  $ResetTxList  $;
		\State Collect all $Tx$  with  $StatusReset $  type
		\State invoke $SC.StaFunc$  to get ${Tx}^{'}$ 
		\State Add   $AlloResultTx$, $RecResultTx$, ${Tx}^{'}$  into $RegBl$ 
		\State \textbf{return}  $RegBl$ 
		\EndIf 
	 
	\end{algorithmic}
\end{algorithm}

\begin{algorithm}[t]
	\caption{Func.E3 for processing Task-3}
	\label{Func.E3}
	\begin{algorithmic}[1]
		\Require   $TxLists$, $ SharedSpecList$, $AwaitTerList$ 
		\Ensure $AlloResultTx$,  $SpecRecList$
		\If{$TxLists(3)$  is not empty}
	%	\State Initialize a list  $AlloTxList$
		\State Add $Tx$  with $SpecAllo$ type into  $AlloTxList$
		\State Perform \textit{SpecSche($AlloTxList$, $ SharedSpecList$, $AwaitTerList$)}    
		\State Obtain $SpecAlloSolution$ and $SpecRecList$
		\State  Invoke $SC.AlloFunc $ to get the  result $AlloResultTx$

		\EndIf       
		\State  \textbf{return}  $AlloResultTx$,  $SpecRecList$
		
	\end{algorithmic}
\end{algorithm}

\begin{algorithm}[t]
	\caption{Func.E4 for processing Task-4}
	\label{Func.E4}
	\begin{algorithmic}[1]
		\Require   $TxLists$, $SpecRecList$
		\Ensure $RecResultTx$
		\If{$TxLists(3)$   is not empty}
	%	\State Initialize a list  $ResRecordTxList  $;
		\State Add $ResRecord$ type $Tx$   into  $ResRecordTxList $ 
		\State Add  $SpecRecList$ into $ResRecordTxList $  and  match each $Tx$  
		\State Invoke $ SC.RecFunc$  to process $ResRecordTxList $, obtain result $RecResultTx$
		\EndIf    
		\State  \textbf{return}  $RecResultTx$		
	\end{algorithmic}
\end{algorithm}

\subsubsection{\textbf{Executing Function}}
\
\newline
\indent There are 4 functions for transaction execution.   Func. E1-Func. E4  are used to execute the Tasks 1 to 4, respectively.
%Func.E1 is for regulators  to publish spectrum regulations by executing transactions derived from Task-1.  Func.E2  (detailed in Alg.\ref{Func.E2}) addresses transactions from Task-2, allowing  spectrum users to publish their demands while enabling regulators to reset the status of spectrum users.  Func.E3  (detailed in Alg.\ref{Func.E3})  is dedicated to processing transactions from Task-3, implementing a specific algorithm to achieve an optimal spectrum allocation scheme. Func.E4   processes transactions derived from Task-4, thoroughly recording the entire spectrum sharing process from the initial requests for spectrum up to the current status. %ensuring a complete record of spectrum utilization and management
Here, a smart contract which  includes 4 sub-function (i.e., $GloFunc$, $StaFunc$, $AlloFunc$, $RecFunc$) is used to perform the information change for Tasks 1 to 4.  

In Func.E1 (shown in Alg.\ref{Func.E1}), the sub-function $SC.GloFunc$ in  smart contract  is triggered  to process transactions $Tx$ with $Global$ type in $CandiBl$. Then the process results will be added to $CandiBl$ to replace the original transactions to form a new processed candidate block.

In Func.E2 (shown in Alg.\ref{Func.E2}), transactions in $TxList(2)$ are processed by the sub-function in the smart contract ($SC.StaFunc$) to change the status of users included in each $Tx$. The processed results, ${Tx}^{'}$, are then added to the region block $RegBl$.
Additional information, including the details of idle spectrum available for sharing and the information about terminals that require spectrum, are recorded by seller nodes in $SharedSpecList$ and by buyer nodes in $AwaitTerList$, respectively. This information is used to facilitate the execution of the spectrum sharing scheme. 
For transactions in $TxLists(3)$, all  $StatusReset$ type transactions are  collected and  processed by $SC.StaFunc$ to get a result ${Tx}^{'}$. Then, the  transaction $AlloResultTx$   returned by Func.E3,  $RecResultTx$    returned by Func.E4, and the  ${Tx}^{'}$ are  sequentially added to $RegBl$. Here, $AlloResultTx$  contains the spectrum allocation results, and $RecResultTx$ details the entire spectrum sharing process.  %which indicates the transaction processing has been done in intra-region interaction. 
% Besides, for transactions with $StatusReset\_Inter$ type in $CandiBl$, the address $SC.StaFunc$ contained in $Tx$  will trigger smart contract to reset the status of users, and then the results replace the original $Tx$ to form a proceeded candidate block $ProCandiBl$.

In Func.E3 (shown in Alg. \ref{Func.E3}), all $SpecAllo$ type transactions in $Txlists(3)$ are first added to the list $AlloTxList$. Then, a   spectrum sharing scheme $SpecSche$ is performed based on   $AlloTxList$, $SharedSpecList$, and $AwaitTerList$, obtaining an optimal spectrum sharing solution $SpecAlloSolution$. According to $SpecAlloSolution$, the sub-function $SC.AlloFunc$ in the smart contract is invoked to update the spectrum information to get the result $AlloResultTx$. Besides, details of the entire spectrum sharing process, including requests, parameters for $SpecSche$, and the current spectrum situation, are recorded in $SpecRecList$.

In Func. E4 (shown in Alg.\ref{Func.E4}),, all $ResRecord $ type transactions in  $Txlists(3)$ are collected into a list $ResRecordTxList $. Then, $SpecRecList$ is appended to $ResRecordTxList$ by matching its elements. Then, $ SC.RecFunc$ is used to process $ResRecordTxList $ and get a result $RecResultTx$.

\section{Performance Analysis}
In this section, the stability performance of PSC-DSS is analyzed in the perspective of the success rate of reaching consensus, $P_S$. Network model and   the  definition of the success rate of reaching consensus are first introduced. Then, the closed form of $P_S$  is derived.

\subsection{Network Model}
\begin{figure}[!t]
	\centering
	\subfigure[Downlink satellite-terrestrial networks]{
		\includegraphics[width=0.47\linewidth]{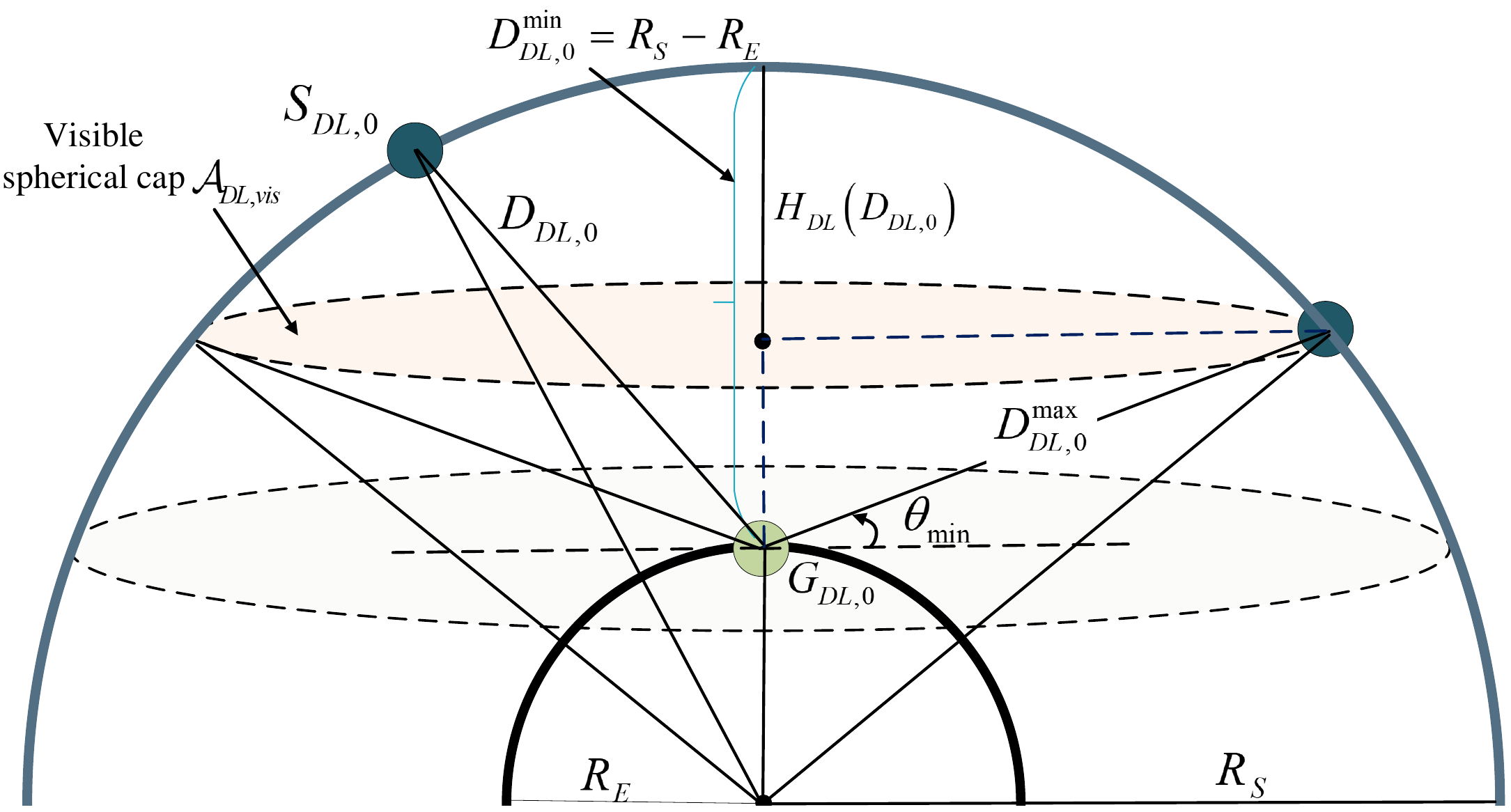}\label{fig:dlfig}}	
	\subfigure[Uplink satellite-terrestrial networks]{
		\includegraphics[width=0.47\linewidth]{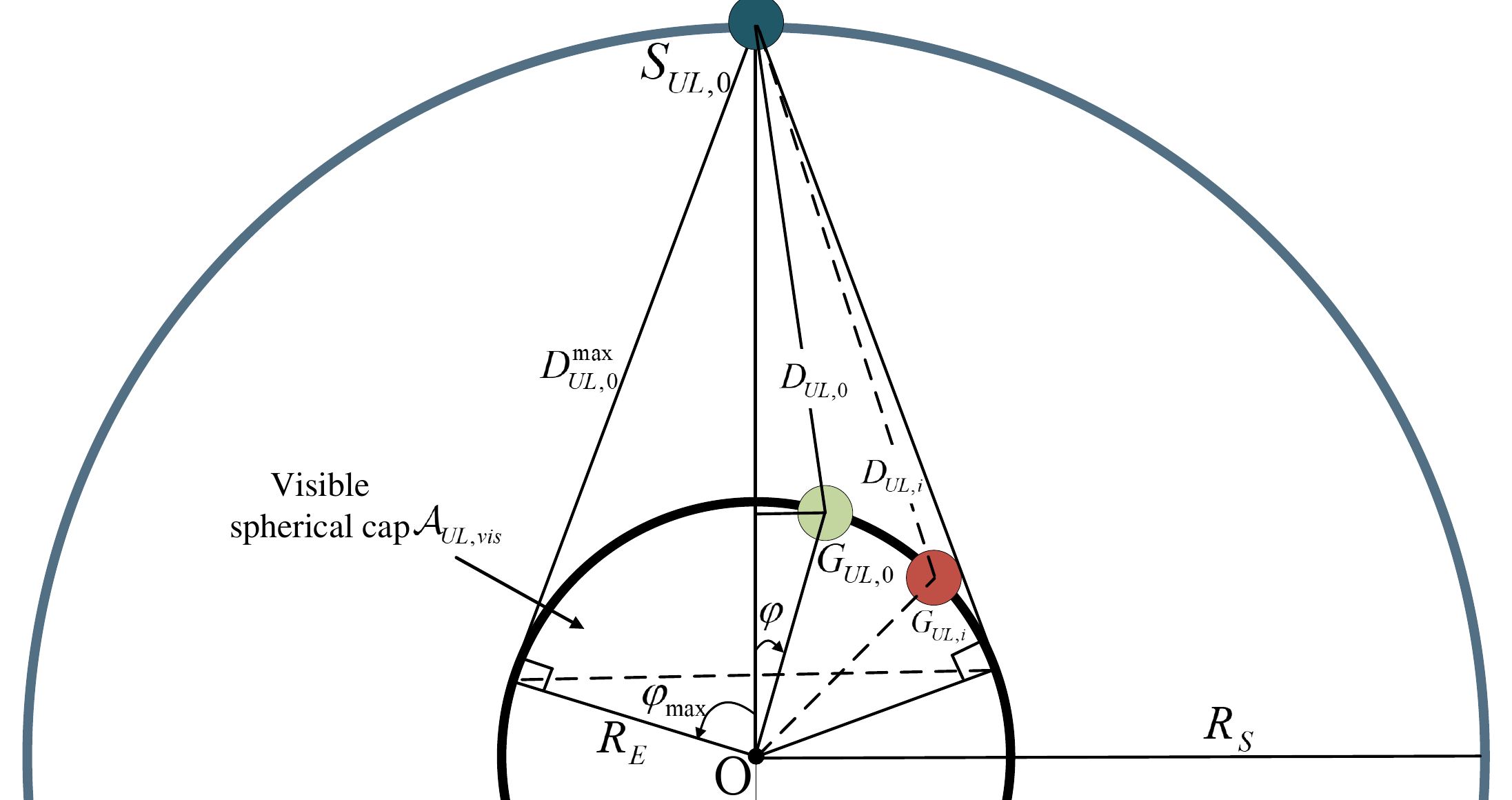}\label{fig:ulfig}}	
	\caption{Illustration of the geometry of  downlink and  uplink satellite-terrestrial networks}
\end{figure}

Consider a satellite-terrestrial communication system with $M$ regions, each one contains $N_s$ satellites and $N_g$ ground nodes (including a regulator's entity and base stations). We assume that the satellites  are located on the surface of the sphere with radius $R_S$  according to a homogeneous spherical Poisson point process (SPPP) $\Phi _S$ with density $\lambda _s$.  The locations of ground nodes are further assumed to be distributed on the surface of the Earth with radius $R_E$ according to a homogeneous SPPP $\Phi _G$ with density $\lambda _g$. For the downlink communication shown in Fig. \ref{fig:dlfig}, the visible spherical cap of a ground node $G_{DL,0}$ with minimum elevation angle $\theta _{\min}$  is expressed as  ${\cal A}_{DL,vis}$. The distance between $G_{DL,0}$  and its nearest satellite $S_{DL,0}$ is denoted as $D_{DL,0}$. %Since $G_{DL,0}$  communicates with $S_{DL,0}$, all other gNBs would be considered as interference satellites.  
The distance between $G_{DL,0}$  and the  $i$th  interference satellite (i.e., ${S_{DL,i}} \in {\Phi _S} \cap {{\cal A}_{DL,vis}}\backslash {S_{DL,0}}$) is denoted as  $D_{DL,i}$.
For the uplink communication shown in Fig. \ref{fig:ulfig}, the visible  spherical cap of  a satellite $S_{UL,0}$ with maximum Earth-centered zenith angle $ \varphi _{\max }$ is expressed as  ${\cal A}_{UL,vis}$. The distance between  $S_{UL,0}$ and its nearest ground node $G_{UL,0}$ is denoted as $D_{UL,0}$, and the distance between  $S_{UL,0}$  and the  $j$th  interference ground node (i.e., ${S_{UL,j}} \in {\Phi _S} \cap {{\cal A}_{UL,vis}}\backslash {S_{UL,0}}$) is denoted as  $D_{UL,j}$.

For the channel model, the Shadowed-Rician fading model is used for the channels between satellites and ground nodes \cite{9813714}. %which is widely adopted in various frequency bands, such as L-band, S-band, and Ka-band \cite{9813714}. 
 Following with \cite{9918046,7056548,8068989}, a gamma random variable is used to   approximate the PDF of channel gain ${{{\left| {{h}} \right|}^2}}$, which is expressed as 
\begin{align}
	{f_{{{\left| {{h }} \right|}^2}}}\left( x \right) = \frac{1}{{{\beta ^\alpha }\Gamma \left( \alpha  \right)}}{x^{\alpha  - 1}}\exp \left( { - \frac{x}{\beta }} \right),
\end{align}
where $\Gamma(.)$ is the gamma function, $\alpha  = \frac{{m{{\left( {2{b_0}\Omega } \right)}^2}}}{{4mb_0^2 + 4mb_0^2\Omega  + {\Omega ^2}}}$ and $\beta  = \frac{{2{b_0} + \Omega }}{\alpha }$ is  the shape and scale parameters, respectively. $m$, $2b_0$, and $\Omega$ denote  the Nakagami fading coefficient, the average power of scattered component, and the average power of line-of sight component, respectively.

Hence,  the revived  signal-to-interference-plus-noise ratio (SINR) at $G_{DL,0}$ for the dowlink, can be expressed as 
\begin{align}\label{gammaDL}
	\gamma_{DL,0}= %\frac{{{{\left| {{h_{UL,0}}} \right|}^2}D_{UL,0}^{ - 2}}}{{\bar \sigma _S^2 +  {{\bar I}_S}}} < \gamma 
 	 {\frac{{P_t^S g_{DL,0}{{\left| {{h_{DL,0}}} \right|}^2}D_{DL,0}^{ - 2}}}{{\sigma _G^2 + \sum\limits_{i \in {\Phi _S^{'}} } {P_t^S{{\left| {{h_{DL,i}}} \right|}^2}g_{DL,i}D_{DL,i}^{ - 2}} }}}, 
\end{align}
 where  $\Phi _S^{'}={\Phi _S} \cap {{\cal A}_{DL,vis}}\backslash {\cal A}_{DL,vis}^{'}\left( {{D_{DL,0}}} \right) $ and ${\cal A}{^{'}_{DL,vis}}\left( D_{DL,0} \right)$ denotes the visible spherical cap with the maximization distance $D_{DL,0}$ between  $G_{DL,0}$  and  satellites, $P_t^S$ is the transmit power at satellites, $\sigma _G^2$ is the variance of the AWGN at   $G_{DL,0}$,  ${{\left| {{h_{DL,x}}} \right|}^2}$ denotes the channel gain between $G_{DL,0}$ and $S_{DL,x}$  $\left( x \in \left\{0,i\right\}\right)$, and $g_{DL,x}$ is the effective antenna gain \cite{9861782} for the signal path from   $S_{DL,x}$ to $G_{DL,0}$. Here, $g_{UL,0}=g_t^S g_r^G\left(\frac{c}{4\pi f_c} \right)^2$ and  $g_{UL,i}=\bar{g}_{DL} g_{DL,0}$, where $g_t^S$ is the transmit antenna gain of satellites, $g_r^G$ is the receive antenna gain of ground nodes, $c$ is the speed of light, and $f_c$ is the carrier frequency,  $\bar{g}_{DL} \in \left[0,1\right]$ is the interference mitigation factor for downlink \cite{9313025}.

 Similarly, the SINR at  $S_{UL,0}$  for the uplink, can be expressed as  
 \begin{align} \label{gammaUL}
 	\gamma_{UL,0}={\frac{{P_t^G g_{UL,0}{{\left| {{h_{UL,0}}} \right|}^2}D_{UL,0}^{ - 2}}}{{\sigma _S^2 + \sum\limits_{j \in{\Phi _G^{'} }} {P_t^G{{\left| {{h_{UL,j}}} \right|}^2}g_{UL,j}D_{UL,i}^{ - 2}} }}}, 
 \end{align}
 where $\Phi _G^{'} ={\Phi _G} \cap {{\cal A}_{UL,vis}}\backslash {\cal A}_{UL,vis}^{'}\left( {{D_{UL,0}}} \right) $ and ${\cal A}{^{'}_{UL,vis}}\left( D_{UL,0} \right)$ denotes the visible spherical cap with the maximization distance $D_{UL,0}$ between $S_{DL,0}$  and  ground nodes,  $P_t^G$ is the transmit power at ground nodes, $\sigma _G^2$ is the variance of the AWGN at   $S_{UL,0}$,  ${{\left| {{h_{UL,y}}} \right|}^2}$ denotes the channel gain between $S_{UL,0}$ and $G_{UL,y}$ $\left( y \in \left\{0,j\right\}\right)$,  $g_{UL,y}$ is the effective antenna gain \cite{9861782} for the signal path from   $G_{UL,y}$ to $S_{UL,0}$. Besides,  $g_{UL,0}=g_t^G g_r^S\left(\frac{c}{4\pi f_c} \right)^2$ and  $g_{UL,j}=\bar{g}_{UL} g_{UL,0}$, where $g_t^G$ is the transmit antenna gain of ground nodes, $g_r^S$ is the receive antenna gain of satellites, and $\bar{g}_{UL} \in \left[0,1\right]$ is the interference mitigation factor for uplink \cite{9313025}.

%Accordingly, the transmission outage probability 

\subsection{Performance Metric}
To quantitatively measure the stability performance of PSC-DSS in SANs, the introduction of a specific performance metric is indispensable. Considering the complex environment of SANs, such as unstable channel and sever interference, satellites and ground nodes are inevitably faced with   faulty probabilities,  denoted as $P_f^S$ and $P_f^G$, respectively, which significantly affect the consensus reaching process in  intra-region interaction and the inter-region interaction.

%Hence, the success rate of reaching  consensus in both intra-region interaction and the inter-region interaction, $P_S$, is defined as  the stability probability for 
Hence, the success rate of reaching  consensus in both intra-region interaction and the inter-region interaction, $P_S$, is defined as  the stability performance metric for PSC-DSS in SANs. %$P_S$ reflects the stability performance of SPC-DSS in such a unstable environment. 
%Hence, the stability probability, $P_S$, is defined as  theperformance metric for SPC-DSS in SANs. $P_S$ presents the success rate of reaching    consensus in both intra-region interaction and the inter-region interaction.
In PSC-DSS, %a transaction published in a region need experience the intra-region interaction and the inter-regin interaction to be recorded into blockchain. 
the intra-region interaction tolerates no more than $\left\lfloor {\frac{N_s+N_g}{3}} \right\rfloor$ faulty nodes (including satellites and ground nodes) based on PBFT consensus protocol, and the inter-regin interaction requires no more than $\left\lfloor {\frac{M}{2}} \right\rfloor$ faulty regulators to commit  a block. 
In such case,   $P_S$ is given as 
 \begin{align} \label{PS}
 		P_{S} &=\sum\limits_{i = 0}^{\left\lfloor {\frac{{{N_g} + {N_s}}}{3}} \right\rfloor } {\left[ {\left( {\sum\limits_{n = 0}^{\min (i,{N_g} - {\rm{1}})} {\left( {1 - P_f^G} \right)   {P_{n0}}} } \right)     \left( {\sum\limits_{j = 0}^{\left\lfloor {\frac{M}{2}} \right\rfloor } {{P_j}} } \right)} \right]} \notag\\
 	&	+  \sum\limits_{i = 1}^{\left\lfloor {\frac{{{N_g} + {N_s}}}{3}} \right\rfloor } {\left[ {\left( {\sum\limits_{n = 1}^{\min (i,{N_g})} {P_f^G  {P_{n1}}} } \right)     \left( {\sum\limits_{j = 0}^{\left\lfloor {\frac{M}{2}} \right\rfloor  - 1} {{P_j}} } \right)} \right]} ,
 \end{align}
with 
 \begin{align}
  	{P_{n0}} &= C_{{N_g} - {\rm{1}}}^n{\left( {P_f^G} \right)^n}{\left( {1 - P_f^G} \right)^{{N_g} - 1 - n}} \times {\textbf{1}_{\left\{ {i - n < {N_s}} \right\}}} \notag\\
  & \;\;\;\;	\times C_{{N_s}}^{i - n}{\left( {P_f^S} \right)^{i - n}}{\left( {1 - P_f^S} \right)^{{N_s} - \left( {i - n} \right)}},
 \end{align}

 \begin{align}
	{P_{n{\rm{1}}}}&=C_{{N_g} - {\rm{1}}}^{n - {\rm{1}}}{\left( {P_f^G} \right)^{n - {\rm{1}}}}{\left( {1 - P_f^G} \right)^{{N_g} - n}} \times {\textbf{1}_{\left\{ {i - n < {N_s}} \right\}}} \notag\\
	& \;\;\;\; \times C_{{N_s}}^{i - n}{\left( {P_f^S} \right)^{i - n}}{\left( {1 - P_f^S} \right)^{{N_s} - \left( {i - n} \right)}},
\end{align}
and
 \begin{align}
	{P_j}= {C_{M - 1}^j{{\left( {P_f^G} \right)}^j}{{\left( {1 - P_f^G} \right)}^{M - 1 - j}}},
\end{align}
where $ {\textbf{1}_{\left\{ . \right\}}}$ is the indicator function, and $C_a^b$ is the binomial coefficient. Details are given in Appendix A.
Fig. \ref{fig:anasimupabfixedpgpstest1} shows that the analytical results for $P_S$ closely match the simulation results,  verifying the correctness of the analysis in (\ref{PS}).
 We can see that the success rate $P_S$ decreases with $P_f^G$ and $P_f^S$. Given the conditions $M=60$, $N_g=40$, and $N_s=20$, PSC-DSS can a faulty probability of over 0.7 for satellites when $P_f^G=0.03$ and  over 0.3 for satellites when $P_f^G=0.2$.
 
 \begin{figure}
	\centering
	\includegraphics[width=0.7\linewidth]{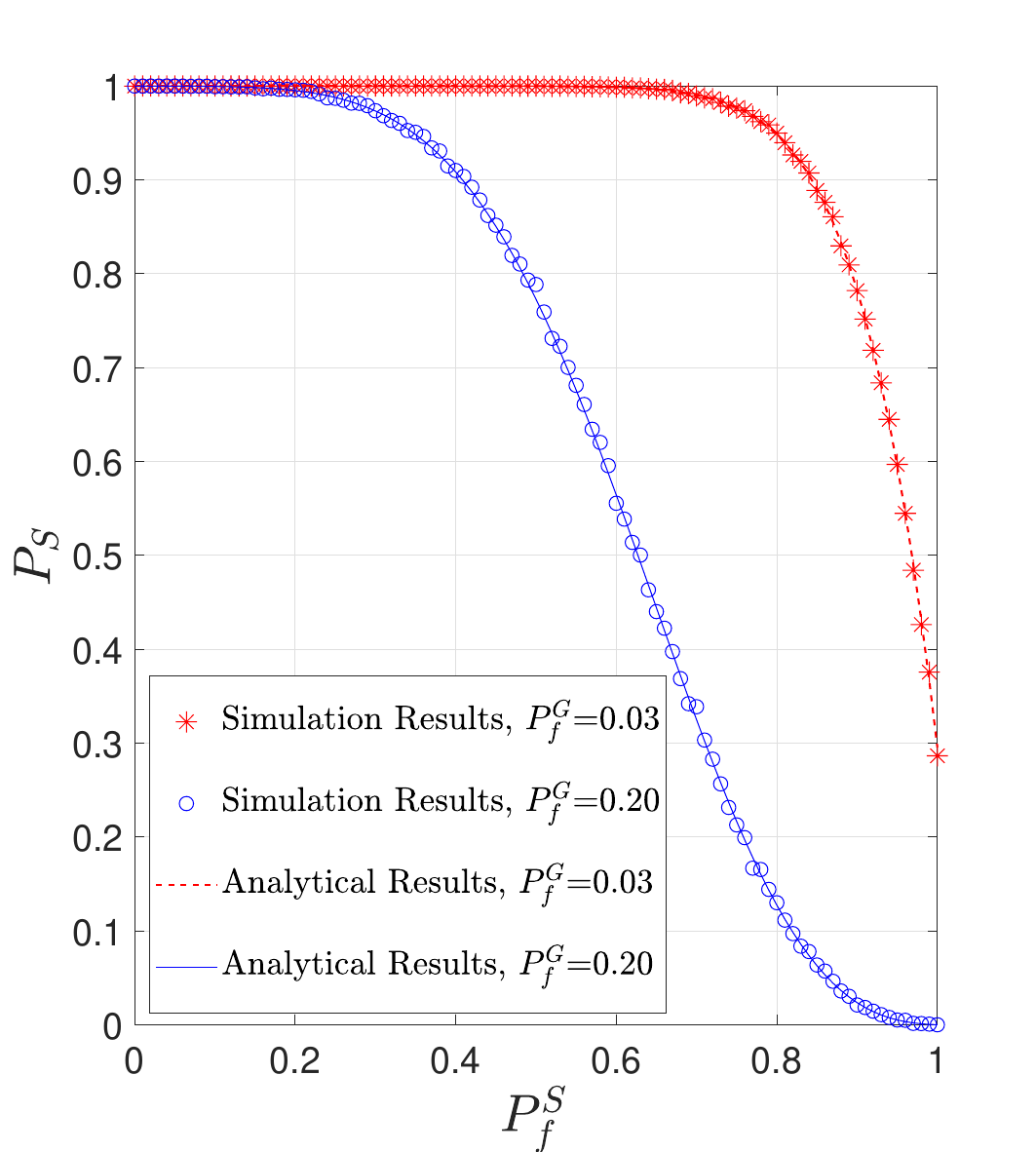}
	\caption{Success rate $P_S$ versus faulty probability of satellites $P_f^G$}
	\label{fig:anasimupabfixedpgpstest1}
\end{figure}

However, $P_f^G$ and $P_f^S$ are influenced by various factors, such as security outage and transmission outage probabilities.  
Satellite and base station as communication infrastructure, its security performance is highly valued by countries and regions. Therefore, this work mainly measures the faulty probability of nodes from the perspective of transmission outage\footnote{It is worth noting that $P_f^G$ and $P_f^S$ can be characterized by both security outage and transmission outage probabilities. For example, $P_f^G$ is expressed as $P_f^G = P_{so}^G + P_{out}^G - P_{so}^G \times P_{out}^G$, where $P_{so}^G$ is the security outage probability.}. For satellites, a fault occurs when they cannot receive signals from ground nodes through ISLs  and satellite-terrestrial communication links. For ground nodes, a fault occurs when they fail to receive signals through wired  and satellite-terrestrial communication links. Therefore, $P_f^G$ and $P_f^S$ can be expressed as $	{P_f^G}= P_{out}^{DL} \times  P_{out}^{WL} $ and ${P_f^S}= P_{out}^{UL} \times  P_{out}^{ISL}$, respectively.
 \iffalse
 \begin{align}
	{P_f^G}= P_{out}^{DL} \times  P_{out}^{WL},
\end{align}
and 
 \begin{align}
		{P_f^S}= P_{out}^{UL} \times  P_{out}^{ISL},
\end{align}
respectively. 
\fi
Here, $P_{out}^{DL}$ and $P_{out}^{WL}$ are the transmission outage probabilities at ground nodes through the downlink satellite-terrestrial and wired communication links, while $P_{out}^{UL}$ and $P_{out}^{ISL}$ are the outage probabilities at satellites via the uplink satellite-terrestrial link and ISLs. 
Next,  $P_{out}^{DL}$ and $P_{out}^{UL}$ are analyzed in SANs.

\subsection{Outage Analysis}
%In this section, the expressions of  $P_{out}^{DL}$ and $P_{out}^{UL}$ are first described. Then, the corresponding derivations are detailed, in terms of the conditional nearest distance distribution, and the conditional Laplace transform of the aggregated interferences.

The outage probability is characterized when there is at least one transmitter in visible spherical cap, i.e., ${\Phi _S({{\cal A}_{DL,vis}}) >0} $ and ${\Phi _G({{\cal A}_{UL,vis}})>0} $.  Thus,    $P_{out}^{DL}$ and $P_{out}^{UL}$, are expressed as 
\begin{align} 
	\label{POUTDL}
	P_{out}^{DL}\left( \gamma  \right) &= \Pr \left( {\gamma _{DL,0}^{} < \gamma \left| {{\Phi _S}\left( {{{\cal A}_{DL,vis}}} \right) > 0} \right.} \right) \notag\\
 	&{\rm{  }} \times \Pr \left( {{\Phi _S}\left( {{{\cal A}_{DL,vis}}} \right) > 0} \right).  %\notag\\
%	&=\Pr \left( {\frac{{{{\left| {{h_{DL,0}}} \right|}^2}D_{DL,0}^{ - 2}}}{{\bar \sigma _G^2 +  {{\bar I}_G}}} < \gamma \left| {{\Phi _S}\left( {{{\cal A}_{DL,vis}}} \right) > 0} \right.} \right) \notag\\
%& \times \Pr \left( {{\Phi _S}\left( {{{\cal A}_{DL,vis}}} \right) > 0} \right),
	\end{align}
and	
\begin{align}
	\label{POUTUL}
P_{out}^{UL}\left( \gamma  \right) &= \Pr \left( {\gamma _{UL,0}^{} < \gamma \left| {{\Phi _G}\left( {{{\cal A}_{UL,vis}}} \right) > 0} \right.} \right) \notag\\
&{\rm{  }} \times \Pr \left( {{\Phi _G}\left( {{{\cal A}_{UL,vis}}} \right) > 0} \right),  %\notag\\
%&=\Pr \left( {\frac{{{{\left| {{h_{UL,0}}} \right|}^2}D_{UL,0}^{ - 2}}}{{\bar \sigma _S^2 +  {{\bar I}_S}}} < \gamma \left| {{\Phi _G}\left( {{{\cal A}_{UL,vis}}} \right) > 0} \right.} \right)  \notag \\
%& \times \Pr \left( {{\Phi _G}\left( {{{\cal A}_{UL,vis}}} \right) > 0} \right),
\end{align}
\iffalse respectively, with  $  {\bar \sigma _G^2{\rm{ = }}\frac{{\sigma _G^2}}{{P_t^Sg_{GL,0}}}}$ and  $  {\bar \sigma _S^2{\rm{ = }}\frac{{\sigma _S^2}}{{P_t^Gg_{UL,0}}}}$. 
Here, ${{\bar I}_G}$ and ${{\bar I}_S}$ represent the conditional aggregated interference at $G_{DL,0}$ and $S_{UL,0}$, respectively. These are given as 
\begin{align}
	{{\bar I}_G} = \sum\limits_{i \in{\Phi _S^{'}}  } {{{\bar g}_{DL}}{{\left| {{h_{DL,i}}} \right|}^2}D_{DL,i}^{ - 2}} ,
   \end{align}
   and 
\begin{align}
	{{\bar I}_S} = \sum\limits_{j \in {\Phi _G^{'}}} {{{\bar g}_{UL}}{{\left| {{h_{UL,j}}} \right|}^2}D_{UL,j}^{ - 2}} .
\end{align}
%where $ {\bar g_{DL}{ { = }}\frac{{g_{DL,i}}}{{g_{DL,0}}}}$, and $ {\bar g_{UL}{ { = }}\frac{{g_{UL,j}}}{{g_{UL,0}}}}$.
\fi

Following  (\ref{POUTDL}) and (\ref{POUTUL}), and based on (\ref{gammaDL}) and (\ref{gammaUL}), there are three fundamental distributions (i.e., the probability of transmitter-visibility,  the conditional nearest transmitter distance distribution, and the conditional Laplace transform of the aggregated interfere) for  providing the general expressions for   $P_{out}^{DL}$ and $P_{out}^{UL}$. Thus, three lemmas are introduced for these fundamental distributions as follows.

\textit{Lemma 1 (The Probability of Transmitter-Visibility)}: The probability that there is at least one satellite in visible spherical cap of a ground node, and the  probability that there is at least one ground node in visible spherical cap of a satellite, are respectively given as 
\begin{align}
&\Pr \left( {\Phi_S \left( {{{\cal A}_{DL,vis}}} \right) > 0} \right)= 1 - \exp \left(-\lambda_s \left| {{{\cal A}_{DL,vis}} } \right| \right) \notag\\
 &=1 - \exp \left( { - \pi{\lambda _s} {R_S}\frac{{{{\left( {D_{DL,0}^{\max }} \right)}^{\rm{2}}} + 2{R_E}{R_S} - R_E^2 - R_S^2}}{{{R_E}}}} \right),
\end{align}
and 
\begin{align}
&\Pr \left( {\Phi_G \left( {{{\cal A}_{UL,vis}}} \right) > 0} \right)= 1 - \exp \left(-\lambda_g \left| {{{\cal A}_{UL,vis}}} \right| \right) \notag\\
&= {1 - \exp \left( { - 2\pi {\lambda _g}R_E^2\left( {1 - \cos {\psi _{\max }}} \right)} \right)},
\end{align}
where  $ \left| {{{\cal A} } } \right|$ denotes the area of a visible spherical  $   {{{\cal A} } } $, and ${D_{DL,0 }^{\max}} =  \sqrt {{{\left( {R_E^{}\sin {\theta _{\min }}} \right)}^2} + R_S^2 - R_E^2}  - R_E^{}\sin {\theta _{\min }}$ is the maximum distance between $G_{DL,0}$ and satellites in the downlink communicates.  %and $ \left| {{{\cal A}_{UL,vis}} } \right|$ is the area of $   {{{\cal A}_{UL,vis}}} $ in the uplink communicates.

\textit{Proof:} See Appendix \ref{appendixB}.
	
	\begin{figure}[!t]
		\centering
		\subfigure[$P_{out}^{DL}$ versus $\gamma_{out}^{DL}$]{
			\includegraphics[width=0.48\linewidth]{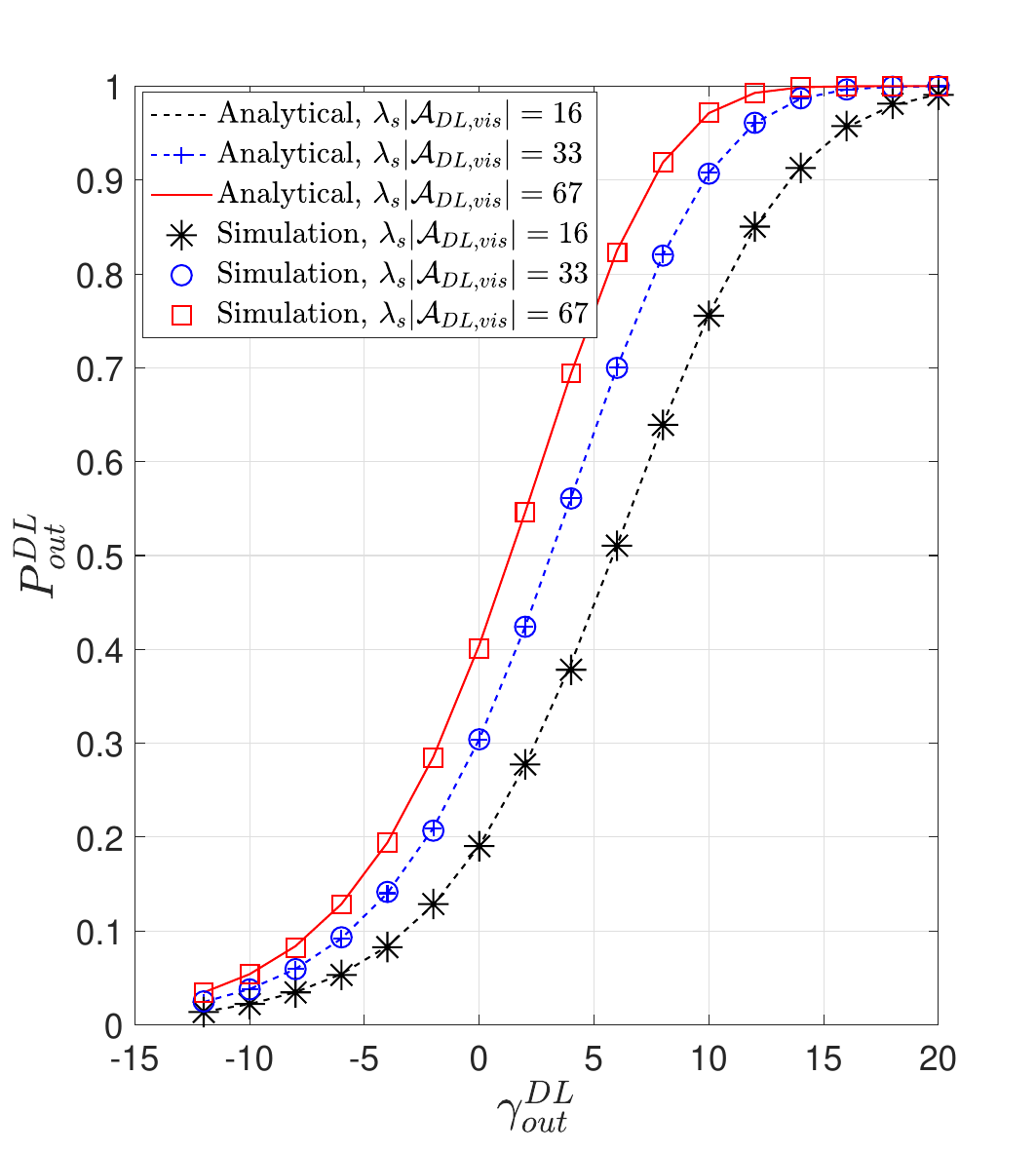}}
		\subfigure[$P_{out}^{UL}$ versus $\gamma_{out}^{UL}$]{
			\includegraphics[width=0.48\linewidth]{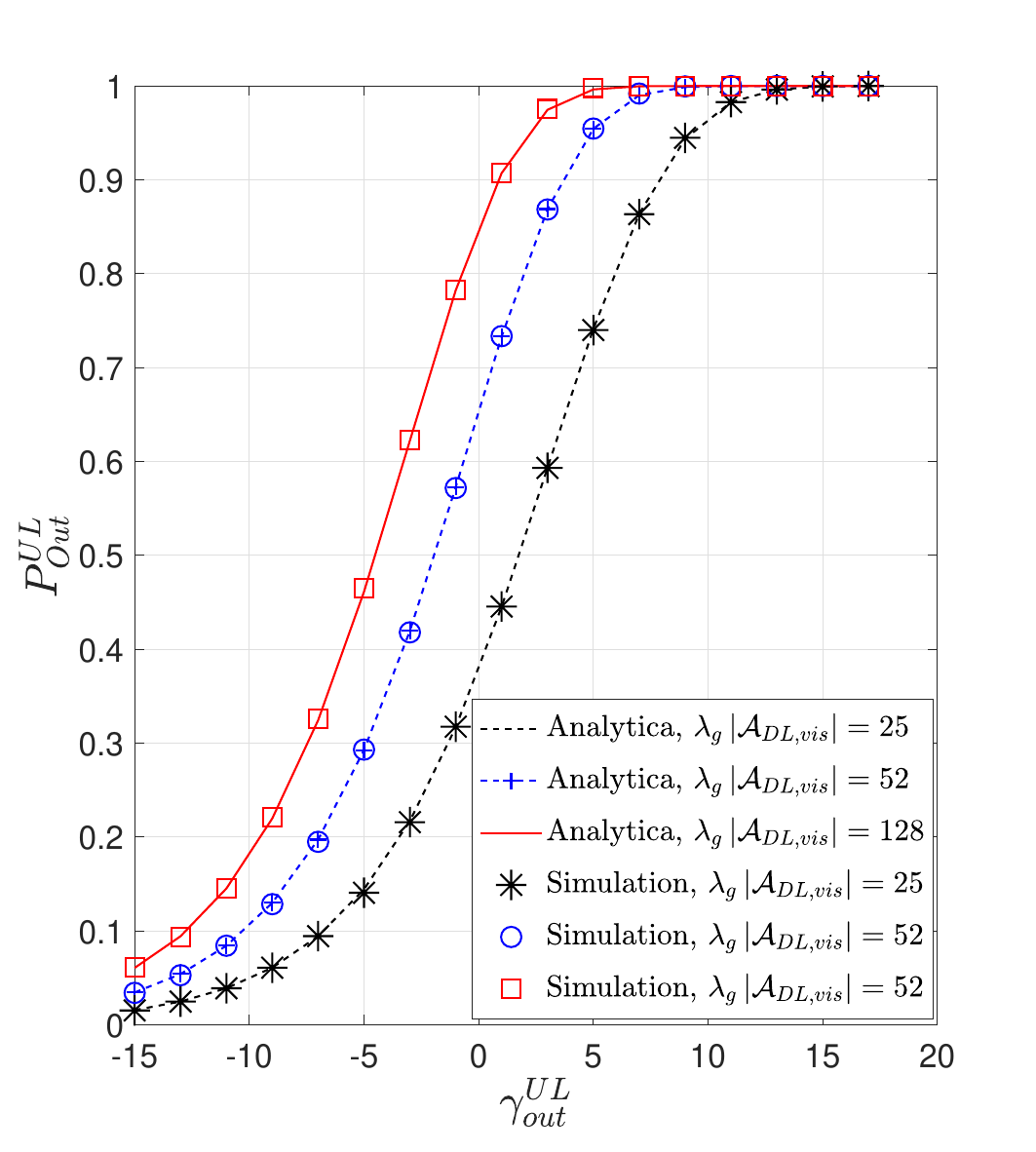}}
		\caption{Outage analysis for fixed $R_S-R_E=500$ Km,  and other parameters are same with Section V-A.}
		\label{OPverification}
	\end{figure}

		\begin{figure*}[b]
					\setcounter{mytempeqncnt}{\value{equation}}
			% Set the equation number to one less than the one
			% desired for the first equation here.
			% The value here will have to changed if equations
			% are added or removed prior to the place these
			% equations are referenced in the main text.
			\setcounter{equation}{15}
			\hrulefill
			\vspace*{8pt}
					\begin{align}\label{laplace}
				  {\cal G}(s, \lambda, R_1, R_2, d_{max}, d, \bar{g})&=\exp \left( { - \pi {\lambda  }\frac{{{R_1}}}{{{R_2}}}\left( {{{  d_{\max }^2   }} - d ^2} \right)}
			   + {\pi {\lambda  }\frac{{{R_1}}}{{{R_2}}}\frac{{ {{  d_{\max }^{2\left( {\alpha  + 1} \right)}  }}}}{{{\left( {s{{\bar g} }\beta } \right)}^{ \alpha }(\alpha  + 1)}}{ \times  {{   _2}{F_1}\left( {\alpha,\alpha  + 1;2 + \alpha ; - \frac{{{d_{\max} ^2}}}{{s  \bar{g} \beta  }}} \right)}} }   \right)  \\  \notag
			 &	\div \exp \left( {\pi {\lambda}\frac{{{R_1}}}{{{R_2}}}\frac{{d ^{2\left( {\alpha  + 1} \right)}}}{{{\left( {s{{\bar g} }\beta } \right)}^{  \alpha }(\alpha  + 1)}}{ \times  { {   _2}{F_1}\left( {\alpha,\alpha  + 1;2 + \alpha ; - \frac{{{d ^2}}}{{s  \bar{g}  \beta   }}} \right)}} } \right)   
				\end{align}
      %    \vspace*{2pt}
			\setcounter{equation}{\value{mytempeqncnt}}
			
		\setcounter{mytempeqncnt}{\value{equation}}
		% Set the equation number to one less than the one
		% desired for the first equation here.
		% The value here will have to changed if equations
		% are added or removed prior to the place these
		% equations are referenced in the main text.
		\setcounter{equation}{18}
		\hrulefill
	%	\vspace*{6pt}
		\begin{align}\label{OP1}
			\eta \left( {\gamma ,\delta ,{d_{\max }},{d_{\min }},g,\kappa ,\lambda ,} \right) = \int_{{d_{\min }}}^{{d_{\max }}} {\exp\left( { - {\beta ^{ - 1}}\gamma \delta d_{}^2} \right) \times \upsilon \left( {\gamma ,d,d_{\max},\delta, g } \right)}  \times  \zeta \left( {d,{d_{\max }},\kappa ,\lambda } \right) {\rm{d}}d
		\end{align}
		%	\hrulefill
	%	\vspace*{2pt}
	 
	%	\hrulefill
%		\vspace*{6pt}
		\begin{align}\label{OP2}
			\zeta \left( {d,{d_{\max }},\kappa ,\lambda } \right) = \frac{{2d\pi \lambda \kappa  \times \exp \left( { - \pi \lambda \kappa \left( {{d^2} + 2{R_E}{R_S} - R_E^2 - R_S^2} \right)} \right)}}{{1 - \exp \left( { - \pi \lambda \kappa \left( {d_{\max }^2 + 2{R_E}{R_S} - R_E^2 - R_S^2} \right)} \right)}}
		\end{align}
		%	\hrulefill
	%	\vspace*{2pt}
	 
	%	\hrulefill
	%	\vspace*{6pt}
		\begin{align}\label{OP3}
			\upsilon \left( {\gamma ,d,d_{max}, \delta, g} \right) = \sum\limits_{m = 0}^{\alpha  - 1} {\left[ {\frac{1}{{m!}} \times \sum\limits_{k = 0}^m {C_m^k\delta _{}^{2\left( {m - k} \right)}{{\left( {{\beta ^{ - 1}}\gamma d_{}^2} \right)}^m} \times {{\left( { - 1} \right)}^k}\frac{{{{\rm{d}}^k}\chi \left( {\kappa ,\lambda,d, {d_{\max }},g} \right)}}{{{\rm{d}}{{\left( {{\beta ^{ - 1}}\gamma d_{}^2} \right)}^k}}}} } \right]} 
		\end{align}
		%	\hrulefill
	%	\vspace*{2pt}
	 
	%	\hrulefill
	%	\vspace*{6pt}
		\begin{align}\label{OP4}
			\begin{array}{c}
				\chi \left( {\kappa ,\lambda ,d,{d_{\max }},g} \right) = \exp \left( {\pi \lambda \kappa \left( {d_{\max }^2 - d_{}^2 + \frac{{{{\left( {d_{}^2   g\gamma } \right)}^{ - \alpha }}}}{{\alpha  + 1}}d_{\max }^{2\left( {\alpha  + 1} \right)}{ \times _2}{F_1}\left( {\alpha ,\alpha  + 1;2 + \alpha ; - \frac{{d_{\max }^2}}{{d_{}^2g\gamma }}} \right)} \right)} \right)\\
				\times \exp {\left( {\pi {\lambda _{}}\kappa \frac{{{{\left( {g\gamma } \right)}^{ - \alpha }}}}{{\alpha  + 1}}d_{}^2{ \times _2}{F_1}\left( {\alpha ,\alpha  + 1;2 + \alpha ; - {{\left( {g\gamma } \right)}^{ - 1}}} \right)} \right)^{ - 1}}
			\end{array}
		\end{align}
		%	\hrulefill
	 
		\setcounter{equation}{\value{mytempeqncnt}}
	\end{figure*}

\textit{Lemma 2 (The Conditional Nearest Transmitter Distance Distribution)}: 
 Let $D_{DL,0}$ denotes the distance from $G_{DL,0}$ and its nearest transmission satellite $S_{DL,0}$ in  $  {{{\cal A}_{DL,vis}} } $, and  $D_{UL,0}$ denotes the distance from $S_{UL,0}$ and its nearest transmission ground node $G_{UL,0}$ in  $  {{{\cal A}_{UL,vis}} } $. 
Then, the probability density functions (PDF) of $D_{DL,0}$ and $D_{UL,0}$,  are expressed as 
 \begin{align}
	{f_{{D_{DL,0}}}}\left( {{d_{DL,0}}} \right) = {\xi _{DL}}{d_{DL,0}}\exp \left( { - \pi {\lambda _s}\frac{{{R_S}}}{{{R_E}}}d_{DL,0}^2} \right),
\end{align}
and
\begin{align}
	{f_{{D_{UL,0}}}}\left( {{d_{UL,0}}} \right) 	=  {{\xi _{UL}}{d_{UL,0}}\exp \left( { - \pi {\lambda _g}\frac{{{R_E}}}{{{R_S}}}d_{UL,0}^2} \right)},
\end{align}
respectively, where $ d_{DL,0} \in \left[ {{D_{DL,0}^{min}},{D_{DL,0}^{max}}} \right]$,  $ d_{UL,0} \in \left[ {{D_{UL,0}^{min}},{D_{UL,0}^{max}}} \right]$,  with ${D_{DL,0}^{min}}  = {D_{UL,0}^{min}}={R_S} - {R_E}$, %${D_{DL,0}^{max}}=  \sqrt {{{\left( {R_E^{}\sin {\theta _{\min }}} \right)}^2} + R_S^2 - R_E^2}  - R_E^{}\sin {\theta _{\min }}$,  
${D_{UL,0}^{max}}  =  \sqrt {R_E^2 + R_S^2 - 2{R_S}{R_E}\cos {\varphi_{\max }}} $, 
$	{\xi _{DL}} = \frac{{\frac{{2\pi {\lambda _s}{R_s}}}{{{R_E}}}\exp \left( { - \pi{\lambda _s} {R_S}\frac{{2{R_E}{R_S} - R_E^2 - R_S^2}}{{{R_E}}}} \right)}}{{1 - \exp \left( { - \pi {\lambda _s}{R_S}\frac{{{{\left( {D_{DL,0}^{\max }} \right)}^{\rm{2}}} + 2{R_E}{R_S} - R_E^2 - R_S^2}}{{{R_E}}}} \right)}}$, and ${\xi _{UL}} = \frac{\frac{{2\pi {\lambda _g}{R_E}}}{{{R_S}}}{\exp \left( { - \pi {\lambda _g}{R_E}\frac{{2{R_E}{R_S} - R_E^2 - R_S^2}}{{{R_S}}}} \right)}}{{1 - \exp \left( { - \pi {\lambda _g}{R_E}\frac{{{{\left( {D_{UL,0}^{\max }} \right)}^{ {2}}} + 2{R_E}{R_S} - R_E^2 - R_S^2}}{{{R_S}}}} \right)}}$.

\textit{Proof:} See Appendix \ref{appendixC}.

 \textit{Lemma 3 (The Laplace Transform of the Conditional  Aggregated Interference)}: Given the conditional aggregated interference for downlink and uplink as  ${{\bar I}_G} = \sum\limits_{i \in{\Phi _S^{'}}  } {{{\bar g}_{DL}}{{\left| {{h_{DL,i}}} \right|}^2}D_{DL,i}^{ - 2}}$ and ${{\bar I}_S} = \sum\limits_{j \in {\Phi _G^{'}}} {{{\bar g}_{UL}}{{\left| {{h_{UL,j}}} \right|}^2}D_{UL,j}^{ - 2}}$,  respectively. The Laplace transforms are given as  
  \begin{align}
  		&{{\cal L}_{{{\bar I}_S}}}\left( s \right) ={\cal G}(s, \lambda_s, R_S, R_E, D_{DL,0}^{max}, d_{DL,0}, \bar{g}_{DL}),
  \end{align}
and 
  \begin{align}
	&{{\cal L}_{{{\bar I}_G}}}\left( s \right) ={\cal G}(s, \lambda_g, R_E, R_S, D_{UL,0}^{max}, d_{UL,0}, \bar{g}_{UL}),
\end{align}
\iffalse
  \begin{align}
 	&{{\cal L}_{{{\bar I}_S}}}\left( s \right) =\exp \left( { - \pi {\lambda _s}\frac{{{R_S}}}{{{R_E}}}\left( {{{\left( {D_{DL,0}^{\max }} \right)}^2} - d_{DL,0}^2} \right)} \right) \notag\\
 	&	\times \exp \left( {\pi {\lambda _s}\frac{{{R_S}}}{{{R_E}}}\frac{{ {{\left( {D_{DL,0}^{\max }} \right)}^{2\left( {\alpha  + 1} \right)}}}}{{{\left( {s{{\bar g}_{DL}}\beta } \right)}^{ \alpha }(\alpha  + 1)}}{ \times  {J\left( {s,{{\bar g}_{DL}},D_{DL,0}^{\max }} \right)}} } \right) \notag\\
 	&	\div \exp \left( {\pi {\lambda _s}\frac{{{R_S}}}{{{R_E}}}\frac{{d_{DL,0}^{2\left( {\alpha  + 1} \right)}}}{{{\left( {s{{\bar g}_{DL}}\beta } \right)}^{  \alpha }(\alpha  + 1)}}{ \times   {J\left( {s,{{\bar g}_{DL}},d_{DL,0} } \right)}} } \right),
 \end{align}
 and
  \begin{align}
 	&{{\cal L}_{{{\bar I}_G}}}\left( s \right) =\exp \left( { - \pi {\lambda _g}\frac{{{R_E}}}{{{R_S}}}\left( {{{\left( {D_{UL,0}^{\max }} \right)}^2} - d_{UL,0}^2} \right)} \right) \notag\\
 	&	\times \exp \left( {\pi {\lambda _g}\frac{{{R_E}}}{{{R_S}}}\frac{{ {{\left( {D_{UL,0}^{\max }} \right)}^{2\left( {\alpha  + 1} \right)}}}}{{{\left( {s{{\bar g}_{UL}}\beta } \right)}^{ \alpha }(\alpha  + 1)}}{ \times  {J\left( {s,{{\bar g}_{UL}},D_{UL,0}^{\max }} \right)}} } \right) \notag\\
 	&	\div \exp \left( {\pi {\lambda _g}\frac{{{R_E}}}{{{R_S}}}\frac{{d_{UL,0}^{2\left( {\alpha  + 1} \right)}}}{{{\left( {s{{\bar g}_{UL}}\beta } \right)}^{  \alpha }(\alpha  + 1)}}{ \times   {J\left( {s,{{\bar g}_{UL}},d_{UL,0} } \right)}} } \right),
 \end{align}
\fi
respectively, %with  $ J\left( {\varpi ,\tau ,\upsilon } \right){ = _2}{F_1}\left( {\alpha,\alpha  + 1;2 + \alpha ; - \frac{{{\upsilon ^2}}}{{\varpi \tau \beta }}} \right)$, 
where $\cal G(.)$ is given in (\ref{laplace}) and $ _2{F_1}\left( {.,. ;. ; .} \right)$  is the Gaussian hypergeometric function.
	
	\textit{Proof:} See Appendix \ref{appendixD}.
	
	Based on the lemmas provided above, the outage probabilities, $P_{out}^{DL}$ and $P_{out}^{UL}$ are obtained as  
  		\setcounter{mytempeqncnt}{\value{equation}}
 \setcounter{equation}{16}
	\begin{align} \label{OPDLFULL}
	&	P_{out}^{DL}\left( \gamma  \right) %&= \Pr \left( {\gamma _{DL,0}^{} < \gamma \left| {{\Phi _S}\left( {{{\cal A}_{DL,vis}}} \right) > 0} \right.} \right) \notag\\
		%&{\rm{  }} \times \Pr \left( {{\Phi _s}\left( {{{\cal A}_{DL,vis}}} \right) > 0} \right) \notag\\
		 = 1 - \eta \left( {\gamma ,\bar \sigma _G^2,D_{DL,0}^{\min },D_{DL,0}^{\max },{{\bar g}_{DL}},\frac{{{R_S}}}{{{R_E}}},{\lambda _s}} \right) \notag\\
		&\;\;\;\; \times \left( {1 - e^  { - \pi {R_S}{\lambda _s}\left( {\frac{{{{  \left({D_{DL,0}^{\max }} \right) }^{ {2}}} + 2{R_E}{R_S} - R_E^2 - R_S^2}}{{{R_E}}}} \right)} } \right),
	\end{align} 
	and
	\begin{align} \label{OPULFULL}
		P_{out}^{UL}\left( \gamma  \right) %&= \Pr \left( {\gamma _{UL,0}^{} < \gamma \left| {{\Phi _G}\left( {{{\cal A}_{UL,vis}}} \right) > 0} \right.} \right) \notag\\
	%	&{\rm{  }} \times \Pr \left( {{\Phi _G}\left( {{{\cal A}_{UL,vis}}} \right) > 0} \right) \notag\\
		&	= 1 - \eta \left( {\gamma ,\bar \sigma _S^2,D_{UL,0}^{\min },D_{UL,0}^{\max },{{\bar g}_{UL}},\frac{{{R_E}}}{{{R_S}}},{\lambda _g}} \right) \notag\\
		&	 \times \left( {1 - \exp \left( { - 2\pi {\lambda _g}R_E^2\left( {1 - \cos {\varphi _{\max }}} \right)} \right)} \right),
	\end{align} 
		respectively,  where $  {\bar \sigma _G^2{\rm{ = }}\frac{{\sigma _G^2}}{{P_t^Sg_{GL,0}}}}$ and  $  {\bar \sigma _S^2{\rm{ = }}\frac{{\sigma _S^2}}{{P_t^Gg_{UL,0}}}}$.  The expression of  $\eta \left( {\gamma ,\delta ,{d_{\max }},{d_{\min }},g,\kappa ,\lambda ,} \right)$ and the corresponding sub-functions are given in (\ref{OP1})-(\ref{OP4}).  The derivations are provides in Appendix \ref{AppendixE}.
	Fig. \ref{OPverification}  shows that the analytical expressions obtained
	in (\ref{OPDLFULL}) and (\ref{OPULFULL}) exactly match  with the simulation results, verifying  the correctness of the analyses.

\iffalse	
	\begin{figure}[t]
		\centering
		\begin{minipage}[t]{0.45\linewidth}
			\includegraphics[width=\linewidth]{FIG/OPDL_DensityE_SINRUL}
			\caption{Success rate $P_S$ versus faulty probability of satellites $P_f^G$}
			\label{fig:anasimupabfixedpgpstest1}
		\end{minipage}
		\hfill
		\begin{minipage}[t]{0.45\linewidth}
			\includegraphics[width=\linewidth]{FIG/OPUL_DensityS_SINRDL}
			\caption{Success rate $P_S$ versus the density of satellites $\lambda_s$}
			\label{fig:PlotPABDensitySandSINRDLSIMUANA1}
		\end{minipage}
	\end{figure}
\fi

\subsection{Success Rate of Reaching  Consensus}

%Based on the above analyses, the closed-form of success rate of reaching  consensus, $P_S$, which reflects the impacts of the complex SANs' environment in SPC-DSS. 

 Based on the aforementioned analysis, the closed-form expression for the consensus success rate, $P_S$, reflecting the impact of the complex SANs environment on the stability of the PSC-DSS system, can  been derived as 
 		\setcounter{mytempeqncnt}{\value{equation}}
 \setcounter{equation}{22}
\begin{align} \label{PSclosedform}
 &P_S  
 	= \sum\limits_{i = 0}^{\frac{{{N_s} + {N_g}}}{3}} {\left[ {\sum\limits_{n = 0}^{\min \left( {i,{N_g}} \right)} {C_{{N_g} - {\rm{1}}}^n{P_{n,G}}{P_{n,S}}} \sum\limits_{j = 0}^{\left\lfloor {\frac{M}{2}} \right\rfloor } {{P_{j,G}}} } \right]} \notag\\
 &	+ \sum\limits_{i = 1}^{\frac{{{N_s} + {N_g}}}{3}} {\left[ {\sum\limits_{n = 0}^{\min \left( {i,{N_g} - {\rm{1}}} \right)} {} C_{{N_g} - {{1}}}^{n - 1}{P_{n,G}}{P_{n,S}}\sum\limits_{j = 0}^{\left\lfloor {\frac{M}{2}} \right\rfloor  - 1} { {P_{j,G}}} } \right]},
\end{align}
with 
\begin{align}
 &{P_{n,G}} =
 	\Bigg[ {1 - \eta \left( {\gamma ,\bar \sigma _G^2,D_{DL,0}^{\min },D_{DL,0}^{\max },{{\bar g}_{DL}},\frac{{{R_S}}}{{{R_E}}},{\lambda _s}} \right)} \Bigg.\notag\\
 &	\times {\left. {\left( {1 - {e^{ - \pi {R_S}{\lambda _s}\frac{{{{\left( {D_{DL,0}^{\max }} \right)}^2} + 2{R_S}{R_S} - R_E^2 - R_S^2}}{{{R_E}}}}}} \right){P_{out}^{WL}}} \right]^n} \notag\\
 &	\times \Bigg[ {1 - \left( {1 - \eta \left( {\gamma ,\bar \sigma _G^2,D_{DL,0}^{\min },D_{DL,0}^{\max },{{\bar g}_{DL}},\frac{{{R_S}}}{{{R_E}}},{\lambda _s}} \right)} \right)} \Bigg.  \notag\\
 &	\times {\Bigg. {\left( {1 - {e^{ - \pi {R_S}{\lambda _s}\frac{{{{\left( {D_{DL,0}^{\max }} \right)}^2} + 2{R_S}{R_S} - R_E^2 - R_S^2}}{{{R_E}}}}}} \right)   P_{out}^{WL}} \Bigg]^{{N_g} - n}},
\end{align}

\begin{align}
		 &{P_{n,S}} =C_{{N_s}}^{i - n} \Bigg[ {1 - \eta \left( {\gamma ,\bar \sigma _S^2,D_{UL,0}^{\min },D_{UL,0}^{\max },{{\bar g}_{UL}},\frac{{{R_E}}}{{{R_S}}},{\lambda _g}} \right)} \Bigg.\notag\\
	 &	\times {\Bigg. {\left( {1 - \exp \left( { - 2\pi {\lambda _g}R_E^2\left( {1 - \cos {\varphi _{\max }}} \right)} \right)} \right) {P_{out}^{ISL}}}  \Bigg]^{i - n}} \notag\\
	 &	\times \Bigg[ {1 - \left( {1 - \left( {\gamma ,\bar \sigma _S^2,D_{UL,0}^{\min },D_{UL,0}^{\max },{{\bar g}_{UL}},\frac{{{R_E}}}{{{R_S}}},{\lambda _g}} \right)} \right)} \Bigg.\notag\\
	 &	\times {\Bigg. {\left( {1 - \exp \left( { - 2\pi {\lambda _g}R_E^2\left( {1 - \cos {\varphi _{\max }}} \right)} \right)} \right)P_{out}^{WL}} \Bigg]^{{N_s} -  {i +n} }}\notag \\
	 &\times  {{\textbf{1}_{\left\{ {i - n < {N_s}} \right\}}}},
\end{align}
and
\begin{align}
	&{P_{j,G}} =C_{M - 1}^j
	\Bigg[ {1 - \eta \left( {\gamma ,\bar \sigma _G^2,D_{DL,0}^{\min },D_{DL,0}^{\max },{{\bar g}_{DL}},\frac{{{R_S}}}{{{R_E}}},{\lambda _s}} \right)} \Bigg.\notag\\
	&	\times {\left. {\left( {1 - {e^{ - \pi {R_S}{\lambda _s}\frac{{{{\left( {D_{DL,0}^{\max }} \right)}^2} + 2{R_S}{R_S} - R_E^2 - R_S^2}}{{{R_E}}}}}} \right){P_{out}^{WL}}} \right]^j} \notag\\
	&	\times \Bigg[ {1 - \left( {1 - \eta \left( {\gamma ,\bar \sigma _G^2,D_{DL,0}^{\min },D_{DL,0}^{\max },{{\bar g}_{DL}},\frac{{{R_S}}}{{{R_E}}},{\lambda _s}} \right)} \right)} \Bigg.  \notag\\
	&	\times {\Bigg. {\left( {1 - {e^{ - \pi {R_S}{\lambda _s}\frac{{{{\left( {D_{DL,0}^{\max }} \right)}^2} + 2{R_S}{R_S} - R_E^2 - R_S^2}}{{{R_E}}}}}} \right)   P_{out}^{WL}} \Bigg]^{{M-1} -j}},
\end{align}

\section{Simulations and Experiments}

In this section,  simulations and experiments  are conducted to evaluate the performance of the proposed PSC-DSS in terms of overhead, efficiency,  and stability. The evaluation criteria are defined as follows: overhead is assessed through consensus latency, efficiency is measured by transactions per second,   and stability by the   success rate of reaching consensus.
%First, xx. Then, the stability performance is evaluated in terms of xx.

 \subsection{Setup}
 
 In this work, the common  parameter settings are listed as follows on the basis of \cite{9861782,9813714,9841465,9678973}.  
$b_0=0.851$, $m=2.91$, $\Omega=0.278$, $\theta_{\min}= {10^ \circ }$, $ {\varphi_{\max }} = \arccos \left( {\frac{{{R_E}}}{{{R_S}}}} \right)$, $f_c=2$ GHz, 
$\sigma_G^2=\sigma_S^2=-174$ dBm/Hz, $R_E=6371$ Km, $R_S=6871$ Km, $P_t^S=30$ dBw, $P_t^G=33$ dBm, $g_t^G=38.5$ dBi, $g_t^S=38$ dBi, $g_r^S=37.8$ dBi, $g_r^G=39.7$ dBi, $\bar{g}_{UL}=\bar{g}_{DL}=0.1$.

For simulations, to evaluate the consensus latency and TPS, we collect ground stations of American from the website: https://satellitemap.space, and use the two-line element set of Starlink \cite{TLE} to construct  SANs. Here, ground stations are treated as  regulators,  base stations are random distributed  with the ground station as the center and each region includes 15 ground nodes. To evaluate the stability of PSC-DSS for further large-scale deployment,  the distributions of ground nodes and satellites are
realized by the generation of SPPP and Monte-Carlo simulation
is used.  The transmission rate is $ 200$ Mbps \cite{STARLINKSPECIFICATIONS}, processing speed is $ 2.4$ GHz \cite{9918062}, the costs of generating and verifying the consensus messages are $ 4$M CPU cycles \cite{9918062}. The size of block header in intra-region interaction and inter-region processes are $B_{hIntra}=39$ bytes and $B_{hInter}=321$  bytes, respectively. The size of transactions for Tasks 1 to 4 are $ 700$,  $ 390$, $ 200$, $ 650$ bytes,  respectively.  The size of messages for pre-prepare, prepare, commit and vote (in tier 2) are $800, 200, 215$, and $800$ bytes, respectively.

For experiments,   considering the  heterogeneous characteristics of SANs, P3-Chain \cite{p3chain} with the advantages of   partitioning on transaction, parallelizing on block, and programming on consensus, is applied as the blokckchain platform for this work. We run the PSC-DSS   
on a cloud virtual machine with equipped with 128 cores and 256GB of RAM, 
%on a laptop with equipped with an Intel(R) Core(TM) i7-10510U CPU @1.80GHz and 16 GB RAM and configured to accommodate up to xx node instances. Here, PSC-DSS 
 and implement  two distinct spectrum sharing schemes\footnote{https://github.com/rebear077/DSS\_RelevantCode}: one to maximize spectrum revenue and the other to minimize aggregated interference, with respective time complexities of $O(nlk+n\log n)$  and $O(nlk+nl\log l)$, where $n$, $l$,  and  $k$ are the number of terminals that require spectrum,   idle spectrum, and  the existing terminals that using a specific spectrum, respectively.

Benchmarks are given as follows: \textit{CBRS} approach \cite{sas}   performs spectrum using a centralized SAS in each region,  \textit{Single-Chain} approaches \cite{9913217,9512423} are the approaches that use blockchain with PBFT consensus protocol to enable spectrum sharing,  \textit{Multi-Chain} approach  \cite{8737533} performs spectrum allocation in local chains while records allocation information in global chain, and \textit{Cross-Chain}  \cite{9785462} approach provides a two-phase-confirmation scheme for communication and synchronization between two local chains. To achieve a unified consensus protocol across different approaches, both the \textit{Multi-Chain} and \textit{Cross-Chain}  approaches employ PBFT.  The number of transactions in a block is the same as the number of nodes (i.e., $N_g+N_s$) in a region.

\subsection{Simulations}
%\subsubsection{Overheads}

%In this experiment, we collect ground stations of American from the website: https://satellitemap.space, and use the two-line element set of Starlink \cite{TLE} to construct  SANs. Here, base stations are random distributed  with the ground station as the center and each region includes 15 ground nodes. %and the minimum elevation angle of ground station is set as $\theta_{\min}= {20^ \circ }$ to avoid satellites  on the edge of the region.
\subsubsection{Consensus Latency}
\begin{figure}
 	\centering
 	\includegraphics[width=\linewidth]{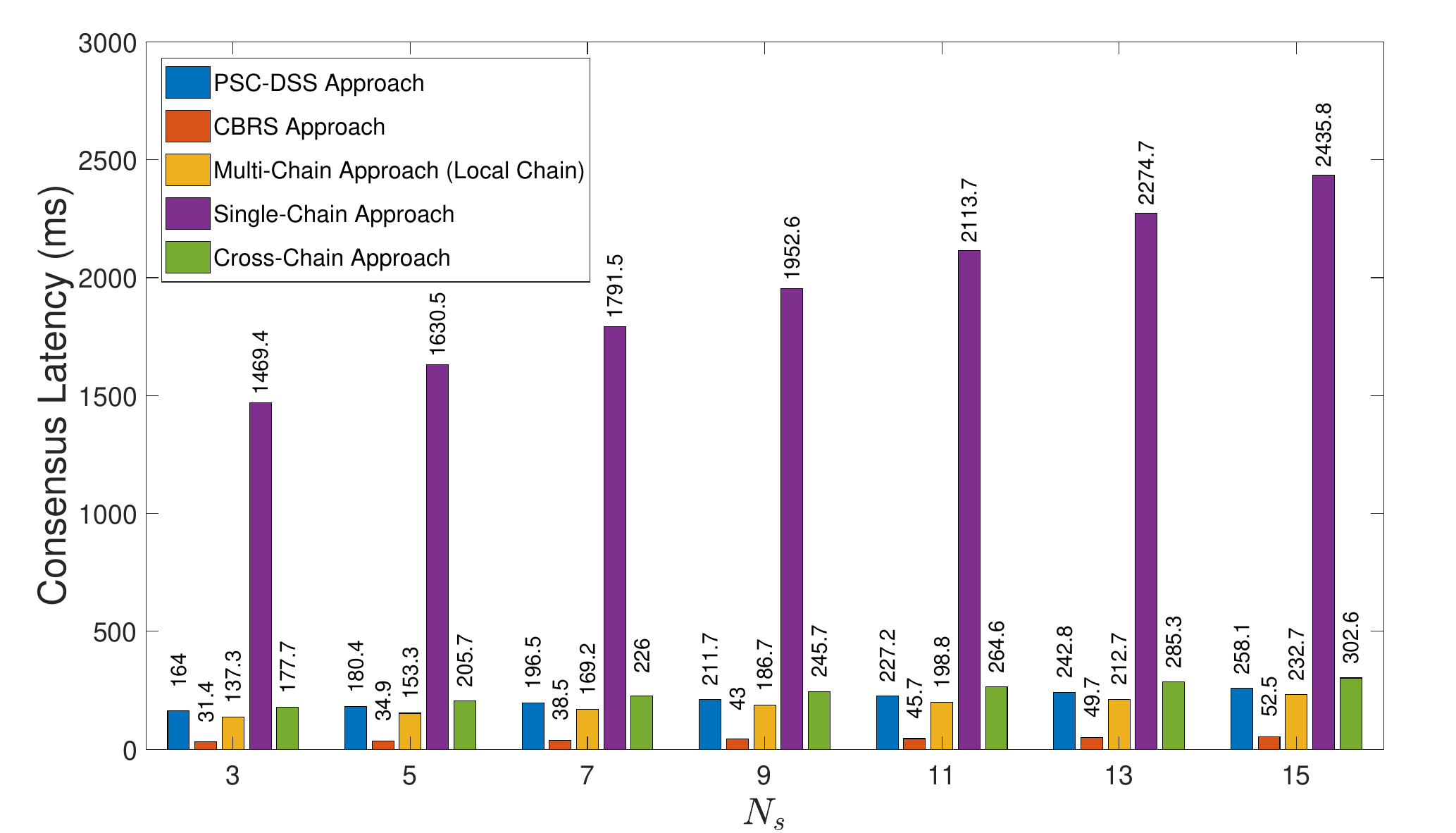}
 	\caption{Consensus latency versus $N_s$}
 	\label{fig:latency}
 \end{figure}

\
\newline
\indent
Fig. \ref{fig:latency} shows consensus latency increases with $N_s$ when $M=11$. This increase is due to the fact that a larger $N_s$ leads to greater consensus complexity, which in turn raises the time consumed in propagation,  transmission and computing. Compared with CBRS approach, PSC-DSS requires more time  to reach consensus due to the necessary communications among blockchain participants. However, the increase in latency is relatively minor, and  PSC-DSS not only enables CBRS sharing within the region but also facilitates inter-region information sharing. Similarly, since PSC-DSS requires interactions among all regulatory  to achieve global information synchronization, the latency is slightly increased compared to the local sharing in multi-chain approach that lack global synchronization. Besides, this figure shows a significant gap between PSC-DSS and single-chain approach, because PSC-DSS reduces the need for extensive communications among all blockchain participants by implementing sharding and tiering. Furthermore, compared to the cross-chain approach, the latency in PSC-DSS is slightly reduced because PSC-DSS does not require consensus within two separate regions as needed in the cross-chain approach, but only among cross-region regulators.

 \begin{figure}
 	\centering
 	\includegraphics[width=1\linewidth]{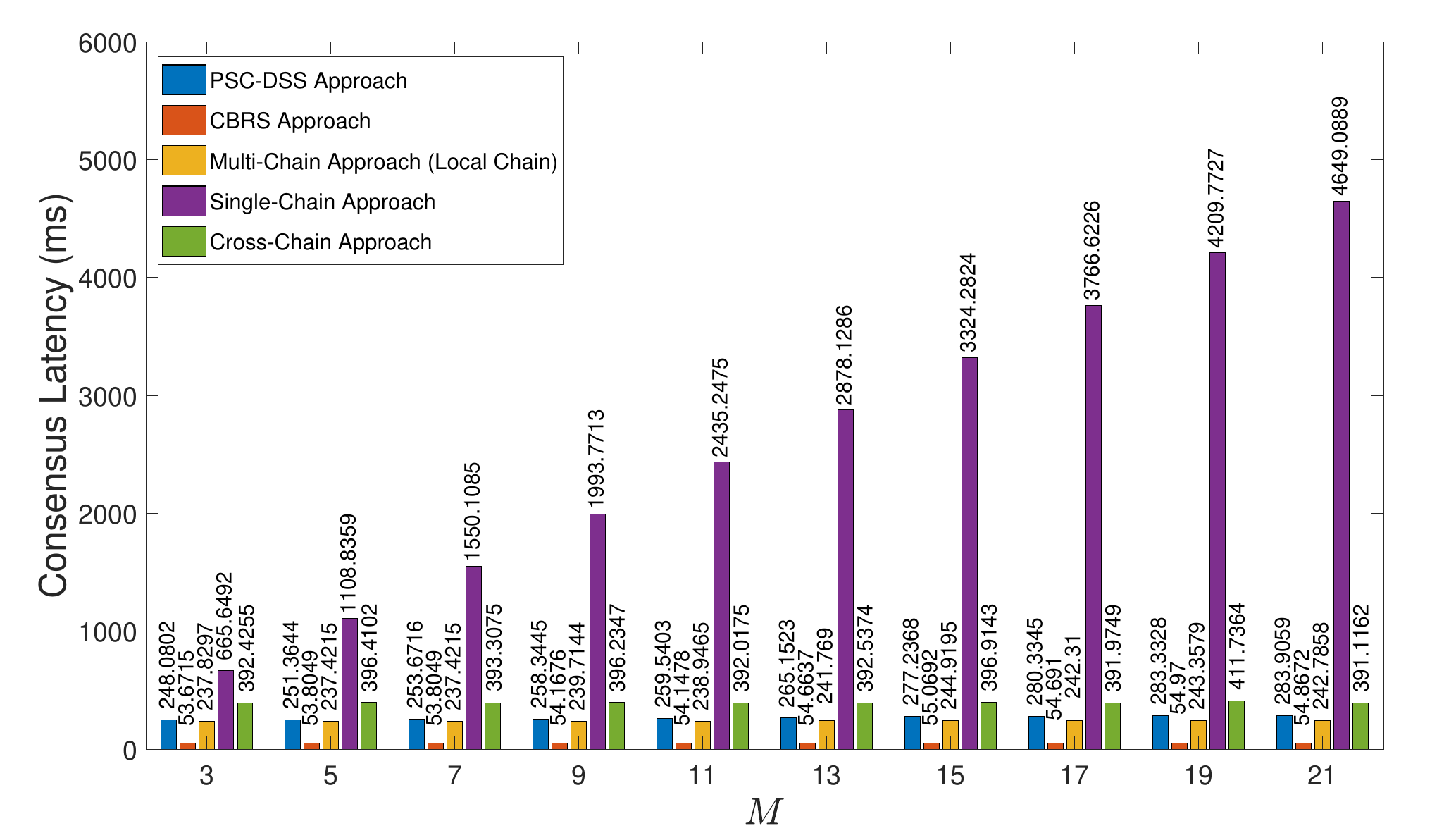}
 	\caption{Consensus latency versus $M$}
 	\label{fig:latencymonly}
 \end{figure}

Fig. \ref{fig:latencymonly} shows consensus latency versus $M$. We can find that the CBRS and multi-chain approaches are almost unaffected by the change of $M$, because both approaches  carry out information dissemination or consensus within their region. However, the slight fluctuations are due to randomly generated base station locations, with a small extent of approximately 2-3 ms. 
A similar situation occurs in  the Cross-Chain approach because this approach perform consensus protocol only in the relevant two regions and transmits cross-region information through  satellites.
In the single-chain approach, consensus latency increases rapidly with the growth of $M$, as the number of participating nodes also expands significantly. However, although the consensus latency of PSC-DSS increases with $M$ due to the growing number of nodes participating in  tier 2 consensus protocol, it still  achieves slow latency growth even as $M$ increases, which demonstrates the excellent scalability of PSC-DSS.

\subsubsection{Efficiency}
\begin{figure}
	\centering
	\includegraphics[width= 0.7\linewidth]{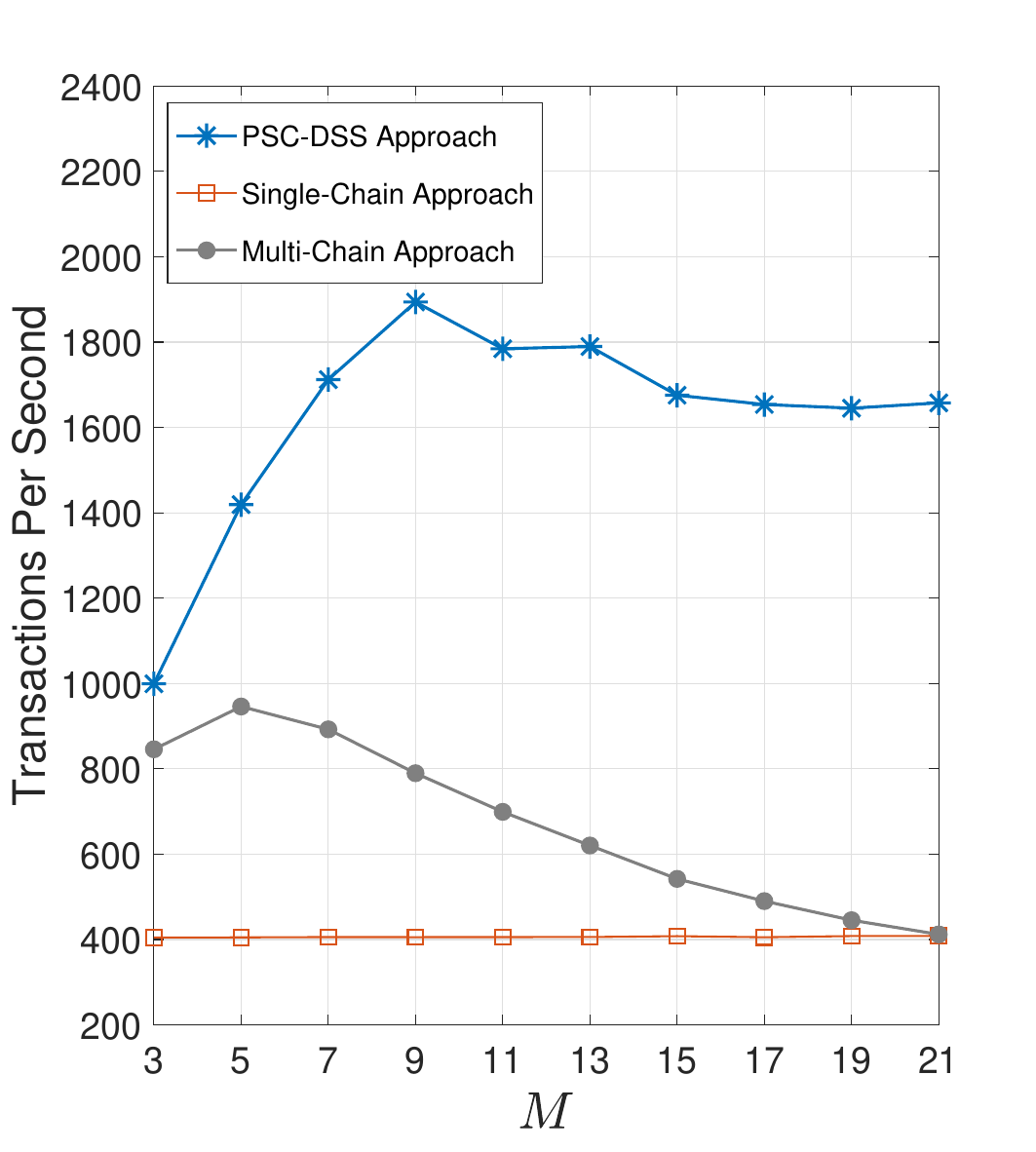}
	\caption{Transactions per second versus $M$}
	\label{fig:tpsmonly}
\end{figure}

\
\newline
\indent
Fig. \ref{fig:tpsmonly} shows the transactions per second (TPS) versus the number of regions $M$.   For the single-chain approach, as the number of regions $M$ increases, not only does the number of blockchain participants and transactions rise, but the consensus latency also escalates rapidly due to the transmission, propagation, and processing of information across multiple nodes. This results in a difficulty in increasing TPS even in this experiment omitting the network congestion.  For the multi-chain approach, TPS decreases with $M$, because allocation record transactions need wait for spectrum allocation transactions to complete a consensus within the region before proceeding to inter-region consensus.  Furthermore, as $M$ increases, more nodes need to participate in the global chain's consensus process, leading to longer consensus latency and consequently reducing TPS.  
Similarly, the TPS of PSC-DSS slightly decreases with $M$, because   the waiting time for candidate blocks submitted by regions to the bootstrapper, as well as the consensus latency among all regulators in tier 2, also increases with  $M$. 
However, PSC-DSS can achieve a relatively stable TPS even as $M$ increases. This is because PSC-DSS consolidates transactions such as spectrum allocation and recording into a single block for consensus, and requires consensus only among   nodes within a region and regulators across regions. This approach avoids the need for all nodes to participate in the consensus, and eliminates the waiting times associated with sequential block generation for different transactions.

\subsubsection{Stability}
 
\
\newline
\indent
In this experiment, we evaluated the success rate of reaching consensus in dynamic and fixed network topologies with different network parameters, which revealed how many satellite nodes share downlink channels and how many ground nodes share uplink channels for a stable consensus process can be tolerated by the PSC-DSS from the perspective of resource conservation. 
For dynamic   network topology, the number of blockchain participants are various with the density of satellites, where all satellites in  ${\cal A}_{DL,vis}$ are treated as blockchain participants and  share a same downlink channel. For  fixed network topology, the number of blockchain participants are fixed, where the density of satellites is only related with the number of satellites share a same downlink channel. Since this experiment is mainly to explore the effect of satellite-ground wireless network environment on $P_S$,   we set  $P_{out}^{ISL}=P_{out}^{WL}=1$.  Besides, the number of regions is set as $M=10$, and the ground nodes as the blockchain participants is set as $N_g=15$.%We first explore the effect of satellite-ground wireless network environment on $P_S$, thus we set   $P_{out}^{ISL}=P_{out}^{WL}=1$. 

	\begin{figure}[!t]
	\centering
	\subfigure[$\lambda_s$ versus $\gamma_{out}^{DL}$]{
		\includegraphics[width=0.48\linewidth]{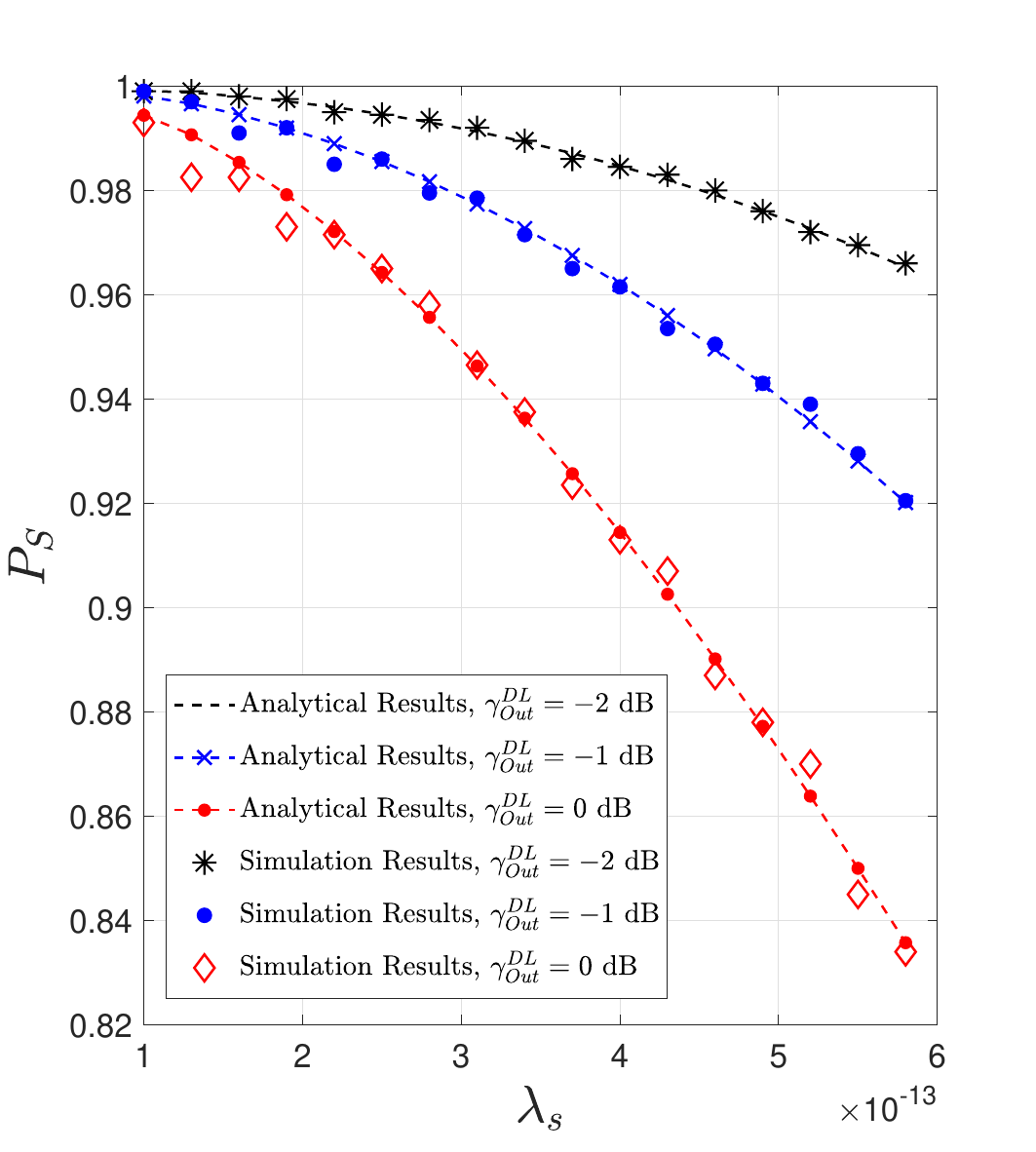}\label{PlotPABDensitySandSINRDLFig}}
	\subfigure[$\lambda_s$ versus $h$]{
		\includegraphics[width=0.48\linewidth]{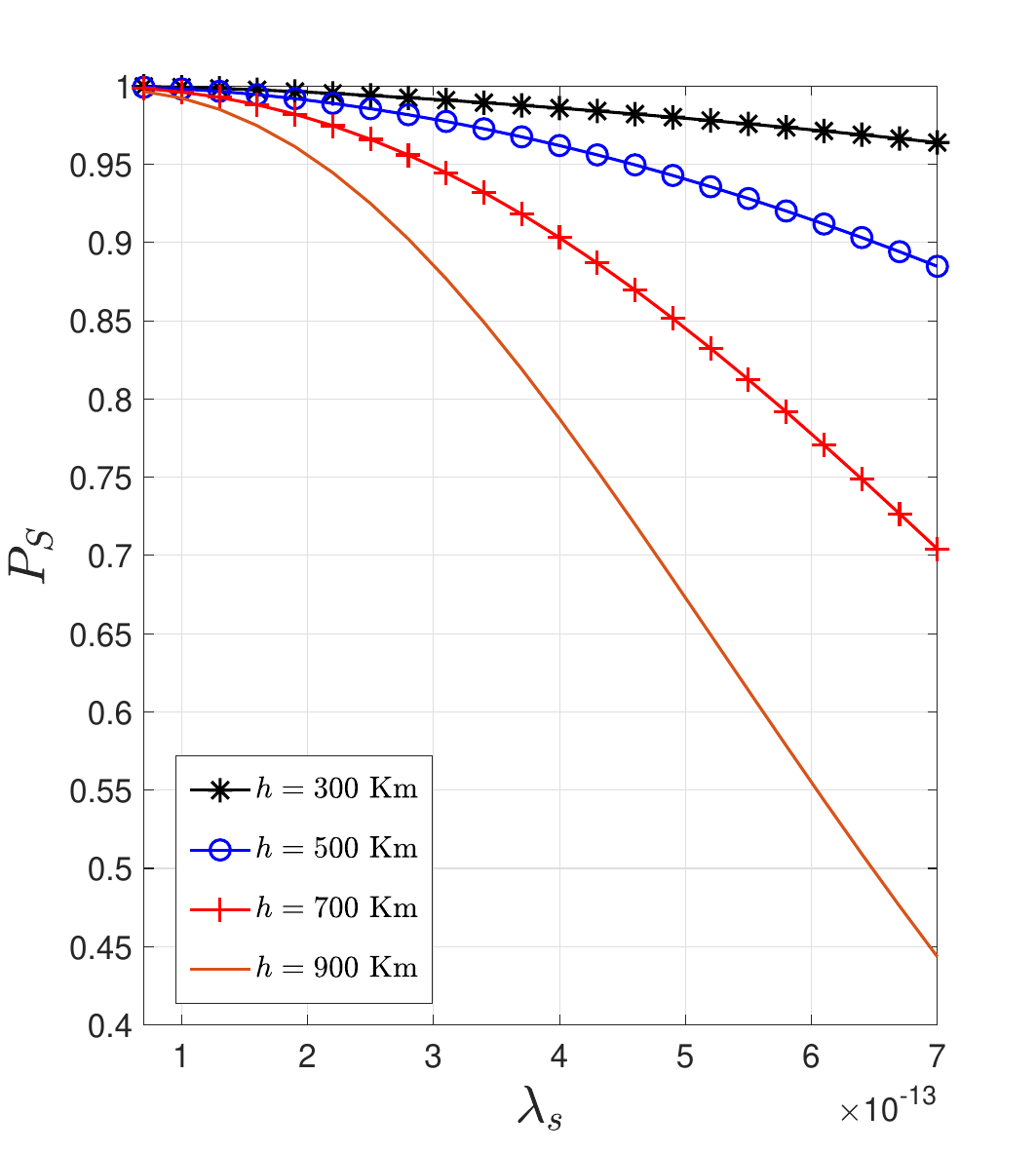}\label{Plot_PAB_DensityS_and_HeightDL_ANA}} 
	\subfigure[$\lambda_s$ versus $M$]{
	\includegraphics[width=0.48\linewidth]{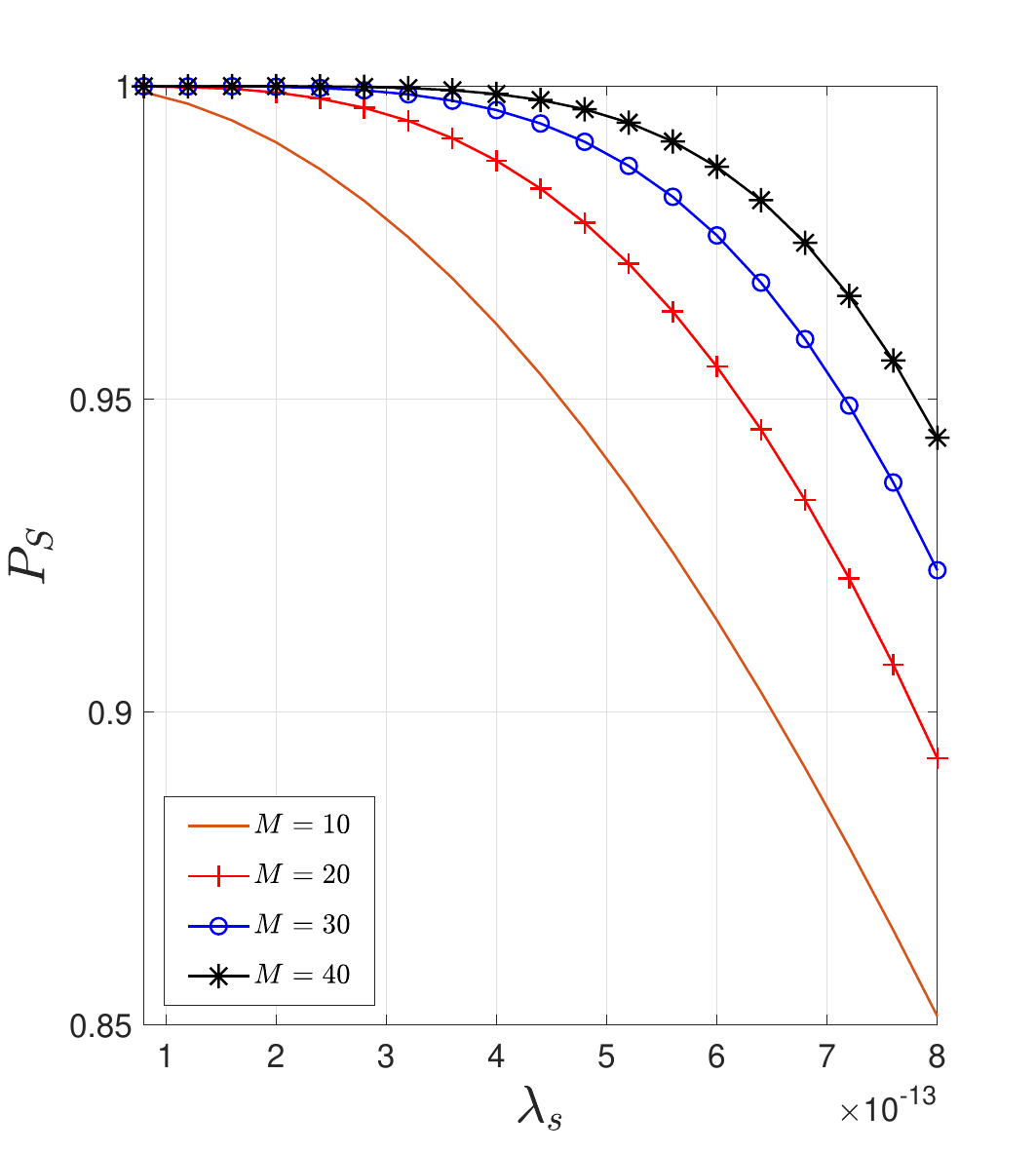}\label{Plot_PAB_DensityS_and_NumZonesDL_ANA0604}}	
		\subfigure[$\lambda_s$ versus $ {\lambda_g}$]{
	\includegraphics[width=0.48\linewidth]{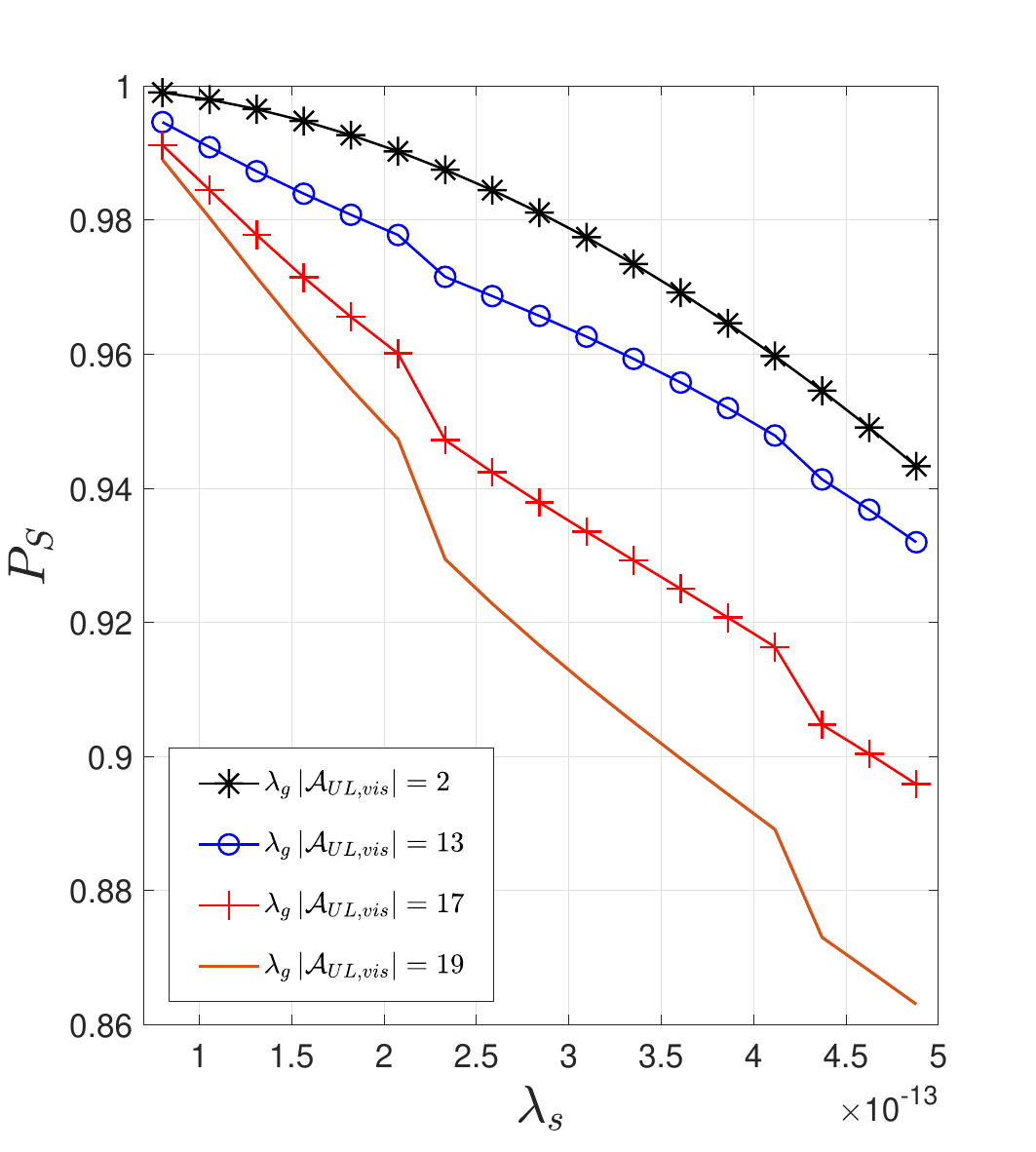}\label{Plot_PAB_DensityS_and_DensityGDL_ANA}}
	\caption{Success rate of reaching consensus versus satellite density at different network parameters in the dynamic network topology.}
	\label{PSdynamic}
\end{figure}

	\begin{figure}[!t]
	\centering
	\subfigure[$\lambda_g$ versus $\gamma_{out}^{UL}$]{
		\includegraphics[width=0.48 \linewidth]{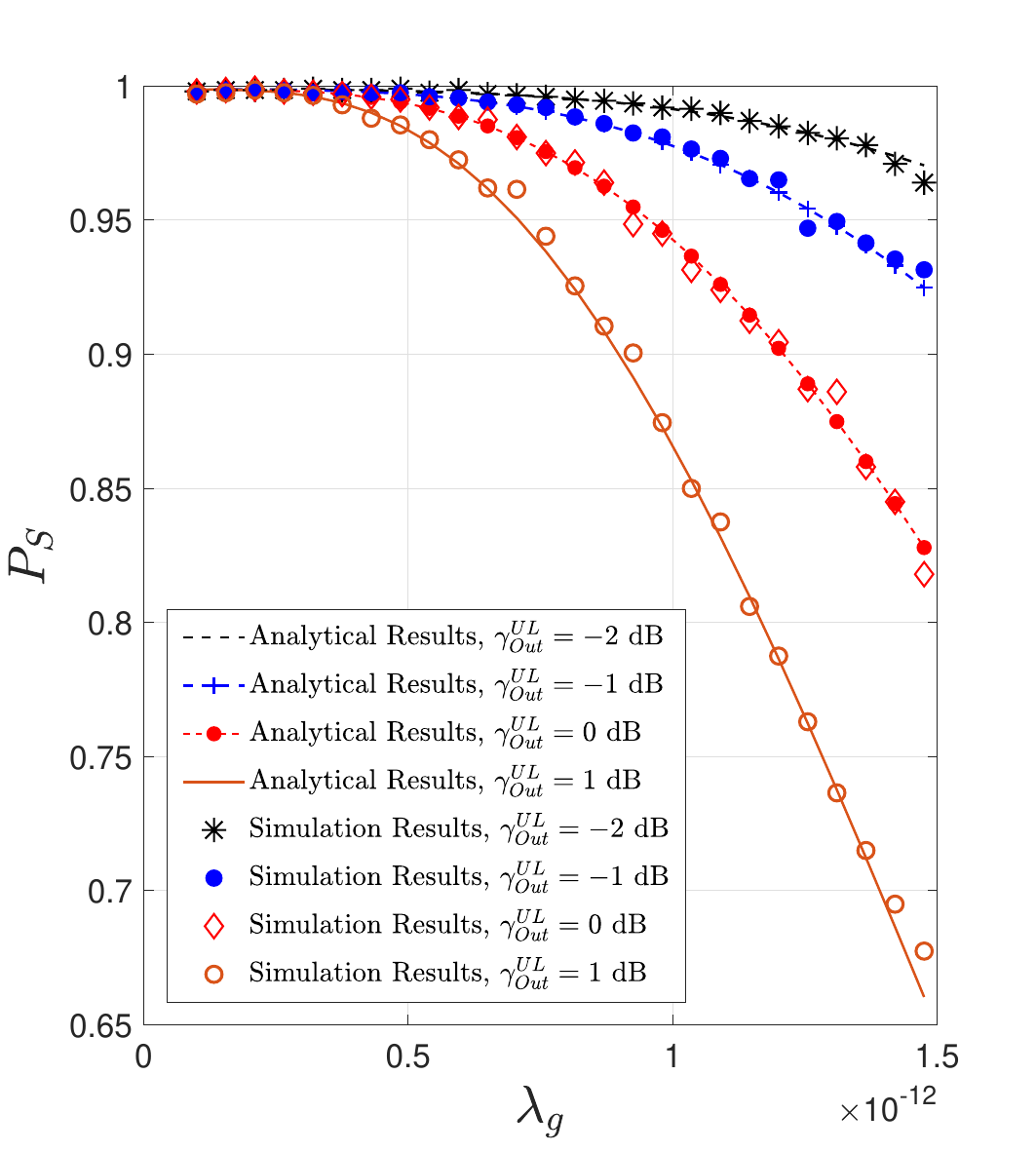}\label{Plot_PAB_DensityS_and_SINRUL}}
	\subfigure[$\lambda_g$ versus $M$]{
		\includegraphics[width=0.48 \linewidth]{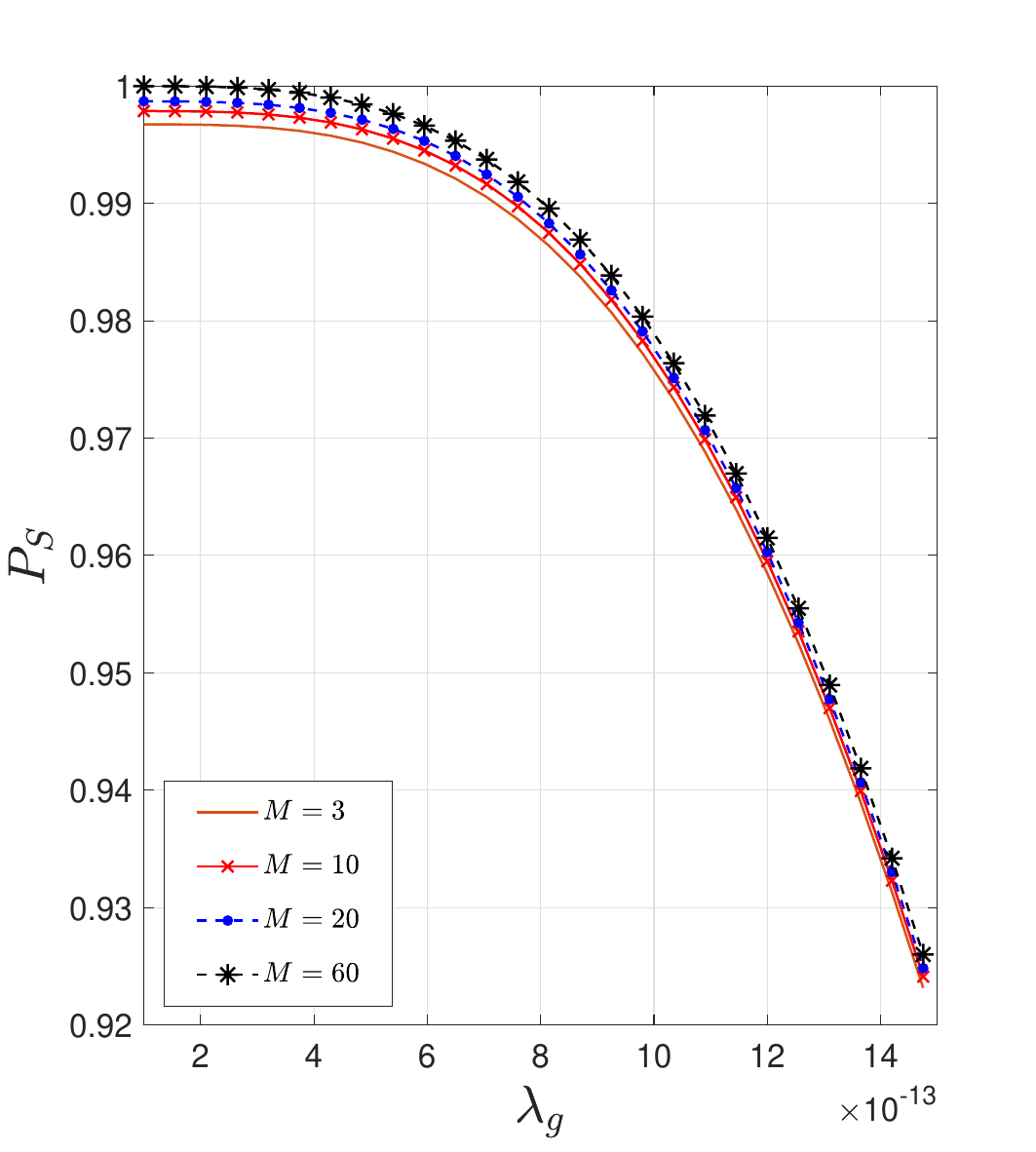}\label{Plot_PAB_DensityS_and_NumZonesUL_ANA}}
	\caption{Success rate of reaching consensus versus satellite density at different network parameters in the fixed network topology.}
	\label{PSFIXED}
\end{figure}

\begin{table*}[!t] 
	\centering
	\caption{The average CPU usage (\% core)}
	\renewcommand{\arraystretch}{1.5} % 增加行高
	\begin{tabular}{|c|c|c|c|c|c|c|c|c|}
		\hline
		\diagbox{Participants}{$M\times N_r$}   & { \; 2 $\times$ 4 \; } &  { \; 4 $\times$ 4 \; } &  { \; 4 $\times$ 8 \; } &  { \; 8 $\times$ 4 \; } &  { \; 8 $\times$ 8 \; } &  { \; 8 $\times$ 16 \; } &{ \; 16 $\times$ 8 \; } &{ \; 16 $\times$ 16 \; } \\
		\hline
		Bootstrapper        & 1.5 & 3.4 & 5.5   &6.85 & 6.7 & 43.85 &38.3& 53.65 \\
		\hline
		Regulator           & 2.1 & 4       &5.7  & 5.2  & 8.13 & 17.35 &13.5& 54.25\\
		\hline
		Spectrum user   & 2.5 & 2.85 & 4.95&5       & 5.3 & 19.8  &15.45 & 49 \\
		\hline
	\end{tabular}
	\label{CPU}
\end{table*}

\begin{table*}[!t]
	\centering
	\caption{The average memory usage (MB)}
	\renewcommand{\arraystretch}{1.5} % 增加行高
	\begin{tabular}{|c|c|c|c|c|c|c|c|c|}
		\hline
		\diagbox{Participants}{$M\times N_r$}   & { \; 2 $\times$ 4 \; } &  { \; 4 $\times$ 4 \; } &  { \; 4 $\times$ 8 \; } &  { \; 8 $\times$ 4 \; } &  { \; 8 $\times$ 8 \; } &  { \; 8 $\times$ 16 \; } &{ \; 16 $\times$ 8 \; } &{ \; 16 $\times$ 16 \; } \\
		\hline
		Bootstrapper   &54.3&60.95 & 56.3 &	59.9  & 58.25 &	67.9&	67.35&	100.05 \\
		\hline
		Regulator        &73.7   &77.05& 76.4 &	70.6  &80.15 & 84.85&	81.4&	100.45 \\
		\hline
		Spectrum user  &66.4& 71.95& 70&	76.95&	74.9&84.8&	79.55&	98.75 \\
		\hline
	\end{tabular}
	\label{memory}
\end{table*}

%\subsubsection{Dynamic Network Topology:}  
 
\textit{Dynamic Network Topology:} Here, every three ground nodes share a uplink channel with uplink SINR threshold set as $\gamma_{out}^{UL}=-1$ dB. From Fig. \ref{PSdynamic}, we can see that $P_S$ decreases with $\lambda_s$. The higher $\lambda_s$, the more satellites share the downlink channel, thus leading to a poorer $P_f^G$.  %Other finding are discuss as follows.

Fig. \ref{PlotPABDensitySandSINRDLFig} first demonstrates the accuracy of the success rate of reaching consensus, where the analytical results are computed from (\ref{PSclosedform}), and the  
simulation results are obtained by independent Monte Carlo trails. 
Although $P_S$ decreases with the  downlink  SINR threshold $\gamma_{Out}^{DL}$, the black curve shows that  PSC-DSS is capable of accommodating 72 satellites (i.e., $\lambda_s=6\times 10^{-13} per \; m^2$) sharing a downlink channel to reach consensus for spectrum sharing. For the following experiments, we set $\gamma_{Out}^{DL}=-1$ dB.  

Fig. \ref{Plot_PAB_DensityS_and_HeightDL_ANA} shows $P_S$ decreases with the altitude of satellites, $h=R_S-R_E$. As the number of satellites in ${\cal A}_{DL,vis}$ increases with altitude $h$,   $P_{out}^{DL}$ for ground nodes also increases, thus leading to a higher  $P_f^S$. However, this figure shows that the PSC-DSS  can support over 68, 49, 37, and 27 satellites sharing a downlink channel at respective altitudes of 300 Km, 500 Km, 700 Km, and 900 Km.

 Fig. \ref{Plot_PAB_DensityS_and_NumZonesDL_ANA0604} shows $P_S$ increases with the number of regions $M$. When $P_{out}^{DL}$ is large  but less than 0.5   (such as  $\lambda_{s}=6\times 10^{-13} per \; m^2$ for $P_{out}^{DL}=0.3416 $),  
  the stability of the system can be enhanced by increasing the number of regions. This is because the increase in the number of regions reduces the probability that more than half of the tier 2 ground nodes will fail, thereby enhancing the success rate of consensus in inter-domain interactions. %Therefore, to 

Fig. \ref{Plot_PAB_DensityS_and_DensityGDL_ANA} shows $P_S$   decreases with the number of  ground nodes which share a same uplink channel, represented by $\lambda_{g} \left|{\cal A}_{UL, vis} \right|$. From this figure, we can observe that a slight change in $\lambda_{g} \left|{\cal A}_{UL, vis} \right|$ leads to significant changes in $P_S$. This is because that  the   greater  $\lambda_{g} \left|{\cal A}_{UL, vis} \right|$, the higher $P_{out}^{UL}$. However, with the increase in $\lambda_{g} $,   more satellites experiencing a high $P_{out}^{UL}$ will participate in the consensus process, thus resulting in a rapid decrease in $P_S$.

%\subsubsection{Fixed Network Topology:}  
 
\textit{Dynamic Network Topology:} In this experiment, every 10 satellites share a downlink with the downlink SINR threshold  $  \gamma_{Out}^{DL}=-1 $ dB,  $N_g=15$ and $N_s=20$. Similar with above results,  Fig. \ref{PSFIXED} shows $P_S$ decreases with $\lambda_g$ due to the growth in number of ground nodes significantly affect $P_{out}^{UL}$ for satellites. %Since the number of satellites as blockchain participants are fixed, 

Fig. \ref{Plot_PAB_DensityS_and_SINRUL} demonstrates   the analytical results  exactly matches with the simulation results  for all the cases of $ \gamma_{Out}^{UL}$. In such case, the block curve shows the PSC-DSS can support over 27 ground nodes share a uplink channel while maintaining system stability. 
Fig. \ref{Plot_PAB_DensityS_and_NumZonesUL_ANA} shows that the increasing the number of regions $M$ in this case does not have a significant effect on enhancing system stability $P_S$. %which is different with Fig. \ref{Plot_PAB_DensityS_and_NumZonesDL_ANA0604}. 
This is because that, PSC-DSS can maintain a stable consensus process due to the major ground nodes in tier 2 is no faulty when $P_{out}^{DL}$ is small (about 0.1483).
% Fig. \ref{Plot_PAB_DensityG_and_DensitySUL_ANA.eps} shows the more satellites share a dowlink channel 

\subsection{Experiments}
In this section, we evaluate the performance of PSC-DSS in terms of  computing consumption and memory consumption.
\subsubsection{Computing Consumption}
 \
\newline
\indent
Table \ref{CPU} shows the average CPU usage of the different participants under various number of regions (i.e., $M$) and the number of nodes in a region (i.e., $N_r=N_g+N_s$). We can find that the average CPU usage increases with $M$ and $N_r$, attributable to the heightened number of transactions to be executed and the augmentation in communication information to be processed. PSC-DSS starts with CPU usage below 7\% for up to 64 nodes, which then rises gradually as it scales to 256 nodes and  demonstrates a modest upward trend. Furthermore, we can find that  bootstrappers are most affected by the increase in the number of nodes, followed by regulators, and finally spectrum users. This is because that spectrum users only process transactions and performs consensus within a region, while regulators need to execute the both, and bootstrappers need to organize network and coordinate the consensus process in tier 2.

\subsubsection{Memory Consumption}
\
\newline
\indent
%Table \ref{memory} shows the average memory usage of different  participants under various number of regions and  the number of nodes in a region. From this table, we can find that the average memory usage increases with the number of nodes. The memory usage of regulators  and spectrum users is similar because they both participate in the consensus and keep the blockchain ledger. Beside, the memory usage of the bootstrapper is generally less than that of regulators and spectrum users. This is because the bootstrapper does not need to store  blockchain ledger, but only the blocks submitted by each region. However, as the number of regions  increases, the number of submitted blocks also increases, leading to higher memory usage for the bootstrapper.
Table \ref{memory} illustrates the average memory usage for various participants under different configurations of regions and nodes within those regions.  From this table, we can find that average memory usage escalates with the increase in the number of nodes. Both regulators and spectrum users exhibit similar patterns in memory consumption. This similarity arises because both types of participants are involved in executing transaction   performing consensus process and maintaining a copy of the blockchain ledger.
In contrast, the memory usage of the bootstrapper is  lower than that of the regulators and spectrum users. The primary reason for this difference is that bootstrappers are not required to execute transaction  and  maintain a copy of the blockchain ledger. Instead, they only need to store the blocks that have been submitted by each region and boost the consensus process in tier 2. However, as the number of regions increases, so does the number of blocks each bootstrapper needs to manage. This escalation in the number of blocks directly contributes to an increase in memory usage for bootstrappers.

\section{Related Works}

%Applying blockchain in DSS has attracted attention from both industry and academia.

Recently, efforts have  highlighted blockchain's potential in DSS, focusing on developing smart contracts, blockchain architectures, and consensus mechanisms.
%discuss in detail those research activities and 

Using smart contract to  perform spectrum sharing schemes is a common approach. Jiang \textit{et al.} \cite{9334436} propose a smart contract-enabled permissioned blockchain-based dynamic spectrum acquisition scheme.
Boateng \textit{et al.}  \cite{9833469} design  smart contracts to perform decentralized spectrum trading. % among spectrum providers and spectrum requesters. 
Ayepah-Mensah  \textit{et al.} \cite{10201899}    employ  smart contracts to execute resource allocation and trading process among RANs. 
Xu \textit{et al.} \cite{10466714} record the spectrum auction results   into a smart contract. 
Although these efforts enable  DSS  efficiently, they  require all participants to follow a conventional architecture  and a unified spectrum sharing scheme. As participant numbers grow, these efforts inevitably face significant bottlenecks due to limited scalability and constrained flexibility in meeting the dynamic and varied needs of participants.

Aiming to improve the scalability of  blockchain-based DSS systems,    an emerging trend  is to  establish new architectures incorporating tiered and sharded features.
Hu \textit{et al.} \cite{9382024}  introduce a two-tier hierarchical blockchain architecture   for  DSS,  consisting  of a global chain in tier 1 and multiple local chains in tier 2. Local chains are updated to the global chain at a fixed or a dynamic frequency.  
Xiao \textit{et al.} \cite{9679805,10109160} propose a blockchain-based decentralized
SAS architecture, with a global chain used for spectrum regulatory tasks and several local chains  for automating spectrum access assignment.  Also, the state of local chains  is updated  on global chain at a fixed frequency.
However,  these approaches % is similar to the   CBRS, which 
cannot realize the global real-time synchronization of spectrum information. 
Grissa \textit{et al.} \cite{8737533} present a trustworthy framework for SAS. In each cluster, secondary users maintain a local blockchain and validators maintain a global blockchain. Members within clusters  hold a light copy of the global blockchain containing the latest status of the system. 
%Cheng \textit{et al.} \cite{10013766} propose a multiple-blockchain framework for CBRS, where multiple coexistence blockchains are formed by citizens broadband radio service devices (CBDSs) and a decision blockchain is formed by SASs or CBSDs with the best trading reputation.  However, the cross-chain transactions are only validated by decision blockchain, which cannot  guarantee the 
However, these efforts are not optimal for SANs because they depend on a multi-chain structure and require additional operations for cross-chain transactions, demanding significant resources from participants. 
  %Moreover, there is a lack of concrete solutions to support different spectrum sharing schemes within a blockchain system.

In order to reduce the overhead of blockchain in DSS, many efforts have focused on developing innovative consensus mechanisms.
Zhu \textit{et al.}  \cite{9751742} integrate the computation of the deep reinforcement learning-based method  for solving the winner determination problem with the proof-of-work consensus mechanism. %aiming to  prevent the abuse of  computing capacity in blockchain networks. 
Fernando  \textit{et al.} \cite{9762480} propose a proof-of-sense consensus mechanism, where %spectrum sensors continuously  perform spectrum sensing, and 
the spectrum first sensor to successfully recovers a secret key  transmitted  randomly in a frequency band is rewarded. 
Ye \textit{et al.} \cite{9680721} introduce  a proof-of-trust consensus mechanism that links participants' trust values with mining difficulty to reduce overhead in DSS. 
%Zhang \textit{et al.}  \cite{9528845} design a proof-of-strategy consensus mechanism to incentive  miners to propose strategies. Only the block which contains the strategy with optimal global return can be added into blockchain. 
Notwithstanding the progress, these  efforts still face  challenges in terms of constrained flexibility and weak compatibility, and thus failing to be  effectively applied in the involving SANs. 

To analyze the performance of blockchain in wireless networks, theoretical models are required. %However, there are a few works consider the theoretical analysis of blockchain in  wireless networks. 
Sun \textit{et al.} \cite{9841465} establish an analytical
model for the blockchain-enabled wireless system. %and analyze blockchain transaction successful rate and overall throughput based on the derivation of the distribution of signal-to-interference-plus-noise ratio. 
 Xu \textit{et al.} \cite{8982036} investigate the security performance of wireless blockchain networks in the presence of malicious jamming for RAFT consensus mechanism. Wang \textit{et al.} \cite{9785462} analyze the probability of forking events in the intra-shard-transactions by considering the fading channel.  However, the existing studies is  for terrestrial networks and can not be directly applied in SANs due to the differences between   terrestrial  networks and SANs in terms of channel model, spatial distribution, and coverage condition.

To face the aforementioned challenges, we construct  a  blockchain-based architecture with a two-tier and multi-region design, however, it requires maintaining only one blockchain. Then,  a spectrum-consensus integrated mechanism is proposed to enhance DSS efficiency and enable regions to dynamically innovate spectrum sharing schemes. Moreover, a theoretical framework which considers the unsteady and complex features of SANs is built to analyze the stability of  PSC-DSS.

\section{Conclusion}
 This work proposes a partitioned, self-governed, and customized dynamic spectrum sharing approach for spectrum sharing between satellite access networks and terrestrial access networks. First, a sharded and tiered architecture is established to allows various regions to manage spectrum autonomously while jointly maintaining a single blockchain ledger. Then, 
a spectrum-consensus integrated mechanism is designed to enable regions to parallelly conduct DSS transactions and dynamically innovate spectrum sharing schemes without affecting others. Finally,  a theoretical framework using stochastic geometry is derived to justify the stability performance of  the proposed approach. Simulations and experiments  are conducted to validate the advantageous performance  of  PSC-DSS in terms of low-overhead, high efficiency,  and  robust  stability.
 %In this paper, a partitioned, self-governed, and customized dynamic spectrum sharing approach (PSC-DSS) is proposed for spectrum sharing between satellite access networks and terrestrial access networks. This approach establishes a sharded and tiered architecture which allows various regions to manage spectrum autonomously while jointly maintaining a single blockchain ledger. Moreover,  a spectrum-consensus integrated mechanism is designed to enable regions to parallelly conduct DSS transactions and dynamically innovate spectrum sharing schemes without affecting others.  Furthermore, a theoretical framework using stochastic geometry is derived to justify the stability performance of  PSC-DSS. Finally, simulations and experiments  are conducted to validate the advantageous performance  of  PSC-DSS in terms of low-overhead, high efficiency,  and  robust  stability.

 \appendices
 \section{}
 In PSC-DSS, %a transaction published in a region need experience the intra-region interaction and the inter-regin interaction to be recorded into blockchain. 
the intra-region interaction tolerates no more than $\left\lfloor {\frac{N_r}{3}} \right\rfloor$ (i.e., $N_r=N_s+N_g$)faulty nodes (including satellites and ground nodes) based on PBFT consensus protocol (denoted as EVENT A), and the inter-regin interaction requires no more than $\left\lfloor {\frac{M}{2}} \right\rfloor$ faulty regulators to commit  a block (denoted as EVENT B).  Furthermore, EVENT A includes two case: the regulator is not faulty (denoted as EVENT A0), and the  regulator is  faulty (denoted as EVENT A1).
However, EVENT A and EVENT B are not independent. If a regulator in 
intra-region interaction is faulty, it will affect the the consensus reaching process in the inter-region interaction. %Therefore, we have $ P_S= P(A) \times  P\left( {A  \left| {B} \right.} \right)   $ .
In such case, %we assume there are $i$ faulty nodes in  intra-region interaction.
we have 
\begin{align}
	&	 P_S= \sum\limits_{i = 0}^{\left\lfloor {\frac{{{N_r}}}{3}} \right\rfloor } {\Pr\left( {A0} \right) \times \Pr\left( {A = i\left| {A0} \right.} \right) \times } \Pr\left( {B\left| {A = i,A0} \right.} \right) \notag\\
		&+ \sum\limits_{i = 1}^{\left\lfloor {\frac{{{N_r}}}{3}} \right\rfloor } {\Pr\left( {A1} \right)   \times \Pr\left( {A = i\left| {A1} \right.} \right)  } \Pr\left( {B\left| {A = i,A1} \right.} \right),
\end{align}
Here, $\Pr(A0)=1-P_f^G$ and $\Pr(A1)=P_f^G$, $\Pr \left( {A = i\left| {A1} \right.} \right) $  and $\Pr \left( {A = i\left| {A0} \right.} \right) $  presents  the  probabilities that there are $i$ faulty nodes in  intra-region interaction under the conditions that the regulator is faulty and not faulty, respectively. These probabilities are given as 
\begin{align}
	\Pr\left( {A = i\left| {A0} \right.} \right) = \sum\limits_{n = 0}^{\min (i,{N_g} - {\rm{1}})} {  P_{n0} },
 \end{align}
 and 
 \begin{align}
	\Pr\left( {A = i\left| {A1} \right.} \right) = \sum\limits_{n = 1}^{\min (i,{N_g})} { P_{n1}}, 
\end{align}
with
\begin{align}
		{P_{n0}} &= C_{{N_g} - 1}^n{\left( {P_f^G} \right)^n}{\left( {1 - P_f^G} \right)^{{N_g} - 1 - n}} \times {\textbf{1}_{\left\{ {i - n < {N_s}} \right\}}} \notag\\
	&\;\;\;\;	\times C_{{N_s}}^{i - n}{\left( {P_f^S} \right)^{i - n}}{\left( {1 - P_f^S} \right)^{{N_s} - \left( {i - n} \right)}},
   \end{align}

   \begin{align}
	   {P_{n{\rm{1}}}}&=C_{{N_g} - {\rm{1}}}^{n - {\rm{1}}}{\left( {P_f^G} \right)^{n - 1}}{\left( {1 - P_f^G} \right)^{{N_g} - n}} \times {\textbf{1}_{\left\{ {i - n < {N_s}} \right\}}} \notag\\
	& \;\;\;\;	\times C_{{N_s}}^{i - n}{\left( {P_f^S} \right)^{i - n}}{\left( {1 - P_f^S} \right)^{{N_s} - \left( {i - n} \right)}},
   \end{align}
where $ {\textbf{1}_{\left\{ . \right\}}}$ is the indicator function.

 $ \Pr\left( {B\left| {A = i,A0} \right.} \right)$ and $ \Pr\left( {B\left| {A = i,A1} \right.} \right)$  indicates   the  probabilities of EVENT B occurring under the conditions that there are already $i$ faulty nodes in  intra-region interaction, with a faulty regulator and without a faulty regulator, respectively. The expressions for these probabilities are given as   
\begin{align}
	\Pr\left( {B\left| {A = i,A0} \right.} \right)={\sum\limits_{j = 0}^{\left\lfloor {\frac{M}{2}} \right\rfloor } {P_j} },
\end{align}
and
\begin{align}
	\Pr\left( {B\left| {A = i,A1} \right.} \right)={\sum\limits_{j = 0}^{\left\lfloor {\frac{M}{2}} \right\rfloor  - 1} {P_j} },
\end{align}
with
\begin{align}
 P_j={C_{M - 1}^j{{\left( {P_f^G} \right)}^j}{{\left( {1 - P_f^G} \right)}^{M - 1 - j}}}.
\end{align}

Thus, the expression of $P_S$ can be given as (\ref{PS}).

  \section{Proof of Lemma 1}
  \label{appendixB}
  Since the locations of satellites are distributed according to   homogeneous SPPP with density, the number of satellites in  ${{\cal A}_{DL,vis}}$ follows Poisson random variable with mean $\lambda_s   \left| {{{\cal A}_{DL,vis}} } \right|$.
  Therefore, the probability that there is at least one satellite in $  {{{\cal A}_{DL,vis}} } $ can be computed by 
  \begin{align}
  	  	\label{P0DL}
  	\Pr \left( {\Phi \left( {{{\cal A}_{DL,vis}}} \right) > 0} \right) & =1- 	\Pr \left( {\Phi_S \left( {{{\cal A}_{DL,vis}}} \right) = 0} \right)  \notag \\
  	&	= 1 - \exp \left(-\lambda_s \left| {{{\cal A}_{DL,vis}}} \right| \right),
  \end{align}
  Similarly, the probability that there is at least one ground node in $  {{{\cal A}_{UL,vis}} } $ can be computed by 
  \begin{align}
  	  	\label{P0UL}
  	\Pr \left( {\Phi \left( {{{\cal A}_{UL,vis}}} \right) > 0} \right)&=1- 	\Pr \left( {\Phi_G \left( {{{\cal A}_{UL,vis}}} \right) = 0} \right)  \notag \\
  	&= 1 - \exp \left(-\lambda_g \left| {{{\cal A}_{UL,vis}}} \right| \right).
  \end{align}
  
  Based on Archimedes’ Hat-Box Theorem  \cite{9861782},  $ \left| {{{\cal A}_{DL,vis}} } \right|$, and  $ \left| {{{\cal A}_{UL,vis}} } \right|$, respectively, are given as 
    \begin{align}	\label{areaofDL}
       \left| {{{\cal A}_{DL,vis}} } \right|=2\pi R_S H_{DL}(D_{DL,0}^{\max}),
    \end{align}
and 
\begin{align}	\label{areaofUL}
	\left| {{{\cal A}_{UL,vis}}} \right| =2\pi R_E H_{UL}(\varphi _{\max }),
\end{align}
where ${D_{DL,0 }^{\max}} =  \sqrt {{{\left( {R_E^{}\sin {\theta _{\min }}} \right)}^2} + R_S^2 - R_E^2}  - R_E^{}\sin {\theta _{\min }}$ is the maximum distance between $G_{DL,0}$ and satellites in downlink communications. Here, $H_{DL}(D_{DL,0})$ is the height of  spherical crown  shown in Fig. \ref{fig:dlfig}, and $H_{UL}(\varphi )$  is the height of  spherical crown  shown in Fig. \ref{fig:ulfig}, which are given by $ H_{DL}(D_{DL,0})= \frac{{D_{DL,0}^2 + 2{R_E}{R_S} - R_E^2 - R_S^2}}{{2{R_E}}}$ and $H_{UL}(\varphi )=R_E   \left( {1 - \cos {\varphi }} \right)$.
 
%$H_{DL}(D_{DL,0})= \frac{{D_{UL,0}^2 + 2{R_E}{R_S} - R_E^2 - R_S^2}}{{2{R_E}}}$ is the height of  spherical crown  shown in Fig. \ref{fig:dlfig}, ${D_{DL,0 }^{\max}} =  \sqrt {{{\left( {R_E^{}\sin {\theta _{\min }}} \right)}^2} + R_S^2 - R_E^2}  - R_E^{}\sin {\theta _{\min }}$ is the maximum distance between $G_{DL,0}$ and satellites in downlink communications, $H_{UL}(\psi)=R_E   \left( {1 - \cos {\psi }} \right)$  is the height of  spherical crown  shown in Fig. \ref{fig:ulfig}.

By  substituting (\ref{areaofDL}) and  (\ref{areaofUL}) into (\ref{P0DL}) and (\ref{P0UL}), respectively, the proof is completed.

 \section{Proof of Lemma 2}
  \label{appendixC}
  The  conditional cumulative density function (CDF) of $ D_{DL,0} $, $F_{D_{DL,0}} (d_{DL,0})$, is characterized as   $\Pr  \left( {D_{DL,0} \le d_{DL,0}\left|{\Phi \left( {{{\cal A}_{DL,vis}}} \right) > 0} \right.} \right)$, indicating that the distance between $G_{DL,0}$ and $S_{DL,0}$ is less than $ d_{DL,0}$ conditioned that at least
  more than one satellite exists in ${\cal A}_{DL,vis}$. 
Following  derivations in \cite{9861782}, the  $F_{D_{DL,0}} (d_{DL,0})$ is given by 
  \begin{align}
  &	{F_{{D_{DL,0}}}}\left( {{d_{DL,0}}} \right) = \Pr \left( {{D_{DL,0}} \le {d_{DL,0}}\left| {{\Phi _S}\left( {{{\cal A}_{DL,vis}}} \right) > 0} \right.} \right) \notag\\
  &	=1 - \Pr \left( {{D_{DL,0}} > {d_{DL,0}}\left| {{\Phi _S}\left( {{{\cal A}_{DL,vis}}} \right) > 0} \right.} \right)\notag\\
  	&= 1 - {\left[ {1 - \exp \left( { - {\lambda _s}\left| {{{\cal A}_{DL,vis}}} \right|} \right)} \right]^{ - 1}} \notag\\
 &  \;\;\;\;\;\; \;\times \exp \left( { - {\lambda _s}\left| {{{{\cal A}'}_{DL,vis}}\left( {{d_{DL,0}}} \right)} \right|} \right) \notag\\
  &	 \;\;\;\;\;\; \; \times \left[ {1 - \exp \left( { - {\lambda _s}\left( {\left| {{{\cal A}_{DL,vis}}} \right| - \left| {{{{\cal A}'}_{DL,vis}}\left( {{d_{DL,0}}} \right)} \right|} \right)} \right)} \right] \notag\\
  &	\mathop  = \limits^{\left( a \right)}  \frac{{1 - \exp \left( { - \pi {\lambda _s}{R_S}\frac{{d_{DL,0}^2 + 2{R_E}{R_S} - R_E^2 - R_S^2}}{{{R_E}}}} \right)}}{{1 - \exp \left( { - \pi {\lambda _s}{R_S} {\frac{{{{\left( {D_{DL,0}^{\max }} \right)}^{\rm{2}}} + 2{R_E}{R_S} - R_E^2 - R_S^2}}{{{R_E}}}} } \right)}},
  \end{align}
where (a) follows from (\ref{areaofDL}) and $ \left| {{{{\cal A}'}_{DL,vis}}\left( {{d_{DL,0}}} \right)} \right|=2 \pi R_S H_{DL}(d_{DL,0})$.
  
  Similarly, the conditional CDF of $ D_{UL,0} $, $F_{D_{UL,0}} (d_{UL,0})$,
   is characterized as   $\Pr  \left( {D_{UL,0} \le d_{UDL,0}\left|{\Phi \left( {{{\cal A}_{UL,vis}}} \right) > 0} \right.} \right)$, indicating that the distance between $S_{UL,0}$ and $G_{UL,0}$ is less than $ d_{UL,0}$ conditioned that at least
   more than one ground node  exists in ${\cal A}_{UL,vis}$. 
   The  $F_{D_{UL,0}} (d_{UL,0})$ is given by 
  \begin{align}
  	&	{F_{D_{UL,0}^{}}}\left( {{d_{UL,0}}} \right) = \Pr \left( {{D_{DL,0}} \le {d_{DL,0}}\left| {{\Phi _G}\left( {{{\cal A}_{UL,vis}}} \right) > 0} \right.} \right)\notag\\
  		&=1 - \Pr \left( {{D_{UL,0}} > {d_{UL,0}}\left| {{\Phi _G}\left( {{{\cal A}_{UL,vis}}} \right) > 0} \right.} \right)\notag\\
  %	&	= 1 - {\left[ {1 - \exp \left( { - {\lambda _{\rm{g}}}\left| {{{\cal A}_{UL,vis}}} \right|} \right)} \right]^{ - 1}}\notag\\
  	% &  \;\;\;\;\;\; \; \times \exp \left( { - {\lambda _g}\left| {{{{\cal A}'}_{UL,vis}}\left( {{d_{UL,0}}} \right)} \right|} \right) \notag\\
  	%	& \;\;\;\;\;\; \;  \times \left[ {1 - \exp \left( { - {\lambda _g}\left( {\left| {{{\cal A}_{UL,vis}}} \right| - \left| {{{{\cal A}'}_{UL,vis}}\left( \varphi ({{d_{UL,0}}}) \right)} \right|} \right)} \right)} \right] \notag\\
 & = \frac{{1 - \exp \left( { - {\lambda _g}\left| {{{{\cal A}'}_{UL,vis}}\left( {\varphi ({d_{UL,0}})} \right)} \right|} \right)}}{{1 - \exp \left( { - {\lambda _{\rm{g}}}\left| {{{\cal A}_{UL,vis}}} \right|} \right)}} \notag \\
 & \mathop  = \limits^{\left( b \right)}  \frac{{1 - \exp \left( { - \pi {\lambda _g}{R_E}\frac{{d_{UL,0}^2 + 2{R_E}{R_S} - R_E^2 - R_S^2}}{{{R_S}}}} \right)}}{{1 - \exp \left( { - \pi {\lambda _g}{R_E}  {\frac{{{{\left( {D_{UL,0}^{\max }} \right)}^{\ {2}}} + 2{R_E}{R_S} - R_E^2 - R_S^2}}{{{R_S}}}}  } \right)}},
  \end{align}
where (b) follows from (\ref{areaofUL}) with $\cos\varphi _{\max }= {\frac{{R_E^2 + R_S^2 - (D_{UL,0 }^{\max})^2}}{{2{R_S}{R_E}}}}$, and $ \left| {{{{\cal A}'}_{UL,vis}}\left( {{d_{UL,0}}} \right)} \right|=2 \pi R_E H_{UL}(\varphi (d_{UL,0}))$ with $\varphi  = \arccos \left({\frac{{R_E^2 + R_S^2 - d_{UL,0 }^{2} }}{{2{R_S}{R_E}}}}\right)$.

By taking derivative with respect to the variables in $F_{D_{DL,0}} (d_{DL,0})$
 and $F_{D_{UL,0}} (d_{UL,0})$, the conditional PDF of ${D_{DL,0}}$ and ${D_{UL,0}} $ are computed as
 \begin{align}
  {f_{{D_{DL,0}}}}\left( {{d_{DL,0}}} \right) %&= \frac{{\partial {F_{{D_{DL,0}}}}\left( {{d_{DL,0}}} \right)}}{{\partial {d_{DL,0}}}} \notag \\
  &= {\xi _{DL}}{d_{DL,0}}\exp \left( { - \pi {\lambda _s}\frac{{{R_S}}}{{{R_E}}}d_{DL,0}^2} \right),
 \end{align}
and
\begin{align}
 		{f_{{D_{UL,0}}}}\left( {{d_{UL,0}}} \right) %&= \frac{{\partial {F_{{D_{UL,0}}}}\left( {{d_{UL,0}}} \right)}}{{\partial {d_{UL,0}}}} \notag\\
	&	=  {{\xi _{UL}}{d_{UL,0}}\exp \left( { - \pi {\lambda _g}\frac{{{R_E}}}{{{R_S}}}d_{UL,0}^2} \right)},
\end{align}
  respectively, where $ d_{DL,0} \in \left[ {{D_{DL,0}^{min}},{D_{DL,0}^{max}}} \right]$,  $ d_{UL,0} \in \left[ {{D_{UL,0}^{min}},{D_{UL,0}^{max}}} \right]$,  ${\xi _{DL}}$ and ${\xi _{UL}}$ are given by
   \begin{align}
     {\xi _{DL}} = \frac{{\frac{{2\pi {\lambda _s}{R_s}}}{{{R_E}}}\exp \left( { - \pi{\lambda _s} {R_S}\frac{{2{R_E}{R_S} - R_E^2 - R_S^2}}{{{R_E}}}} \right)}}{{1 - \exp \left( { - \pi {\lambda _s}{R_S}\frac{{{{\left( {D_{DL,0}^{\max }} \right)}^{\rm{2}}} + 2{R_E}{R_S} - R_E^2 - R_S^2}}{{{R_E}}}} \right)}},
     \end{align}
 and
    \begin{align}
     	{\xi _{UL}} = \frac{\frac{{2\pi {\lambda _g}{R_E}}}{{{R_S}}}{\exp \left( { - \pi {\lambda _g}{R_E}\frac{{2{R_E}{R_S} - R_E^2 - R_S^2}}{{{R_S}}}} \right)}}{{1 - \exp \left( { - \pi {\lambda _g}{R_E}\frac{{{{\left( {D_{UL,0}^{\max }} \right)}^{ {2}}} + 2{R_E}{R_S} - R_E^2 - R_S^2}}{{{R_S}}}} \right)}}.
       \end{align}
  
  This completes the proof.

   \section{Proof of Lemma 3}
  \label{appendixD}
 
   Following the definition of the conditional aggregated interference at $G_{DL,0}$ presented as ${{\bar I}_G}$, ${\Phi _S^{'}}= {\Phi _S} \cap {{\cal A}_{DL,vis}}\backslash {\cal A}_{DL,vis}^{'}\left( {{d_{DL,0}}} \right) $ denotes the area on the spherical cap outside ${\cal A}{^{'}_{DL,vis}}\left( {{d_{DL,0}}} \right)$. Thus, the Laplace transform of ${{\bar I}_G}$ is computed by 
  \begin{align} 
  		&{{\cal L}_{{{\bar I}_G}}}\left( s \right) = {\mathbb{E}_{{{\bar I}_G}}}\left[ {\left. {{e^{ - s{{\bar I}_G}}}} \right|{D_{DL,0}=d_{DL,0}},{\Phi _s}\left( {{\cal A}{_{DL,vis}}} \right) > 0} \right] \notag\\
  %	&	= \mathbb{E}{_{{\Phi _S}\left( {{\cal A}{_{DL,vis}}} \right),{{\left| {{h_{DL,i}}} \right|}^2}}}\left[ {\exp \left( { - \sum\limits_{i \in {{{\Phi }_S^{'}}} } {s{{\bar g}_{DL}}{{\left| {{h_{DL,i}}} \right|}^2}d_{DL,i}^{ - 2}} } \right)} \right] \notag\\
 % 	&	= {\mathbb{E}_{{\Phi _S}\left( {{\cal A}{_{DL,vis}}} \right),{{\left| {{h_{DL,i}}} \right|}^2}}}\left[ {\prod\limits_{i \in {{{\Phi }_S^{'}}} } {\exp \left( { - s{{\bar g}_{DL}}{{\left| {{h_{DL,i}}} \right|}^2}d_{DL,i}^{ - 2}} \right)} } \right] \notag\\
  	&	\mathop  = \limits^{\left( a \right)} {\mathbb{E}_{{\Phi _S}\left( {{\cal A}{_{DL,vis}}} \right)}}\left[ {\prod\limits_{i \in {{\Phi} _S^{'}}} {{\mathbb{E}_{{{\left| h \right|}^2}}}\left[ {\exp \left( { - s{{\bar g}_{DL}}{{\left| h \right|}^2}d_{DL,i}^{ - 2}} \right)} \right]} } \right] \notag\\
  	&	\mathop  = \limits^{\left( b \right)} \exp \left( { - {\lambda _s}\int_{d \in \bar {\cal A}{_{DL,vis}}} {\rm{d}}\left( {\left|  \bar {\cal A}{_{DL,vis}} \right|} \right)}  \right) \notag\\
  	&	- \exp \left( { - {\lambda _s}\int_{d \in  \bar {\cal A}{_{DL,vis}}} {{\mathbb{E}_{{{\left| h \right|}^2}}}\left( {e ^{ { - s{{\bar g}_{DL}}{{\left| h \right|}^2}d_{}^{ - 2}} }} \right) {\rm{d}} \left( {\left| \bar {\cal A}{_{DL,vis}}\right|} \right)} } \right) \notag\\
  	&	\mathop  = \limits^{\left( c \right)} \exp \left( { - 2\pi {\lambda _s}\frac{{{R_S}}}{{{R_E}}}\int_{{d_{DL,0}}}^{D_{DL,0}^{\max }} {\left( {1 - {\mathbb{E}_{{{\left| h \right|}^2}}}\left( {e^ { { - s{{\bar g}_{DL}}{{\left| h \right|}^2}d_{}^{ - 2}} }} \right)} \right)  d{\rm{d}}d} } \right) \notag\\
  	&	\mathop  = \limits^{\left( d \right)} \exp \left( { - 2\pi {\lambda _s}\frac{{{R_S}}}{{{R_E}}}\int_{{d_{DL,0}}}^{D_{DL,0}^{\max }} {d\left( {1 - \frac{1}{{{{\left( {s{{\bar g}_{DL}}d_{}^{ - 2}\beta  + 1} \right)}^\alpha }}}} \right){\rm{d}}d} } \right) \notag\\
  	&	\mathop  = \limits^{\left( e \right)} \exp \left( { - \pi {\lambda _s}\frac{{{R_S}}}{{{R_E}}}\left( {{{\left( {D_{DL,0}^{\max }} \right)}^2} - d_{DL,0}^2} \right)} \right) \notag\\
  	&	\times \exp \left( {\pi {\lambda _s}\frac{{{R_S}}}{{{R_E}}}\int_{\frac{{d_{DL,0}^2}}{{s{{\bar g}_{DL}}\beta }}}^{\frac{{{{\left( {D_{DL,0}^{\max }} \right)}^2}}}{{s{{\bar g}_{DL}}\beta }}} {\left( {\frac{1}{{{{\left( {{t^{ - 1}} + 1} \right)}^\alpha }}}} \right){\rm{d}}\left( {s{{\bar g}_{DL}}\beta t} \right)} } \right)  	 \notag\\
%  &		= \exp \left( { - \pi {\lambda _s}\frac{{{R_S}}}{{{R_E}}}\left( {{{\left( {D_{DL,0}^{\max }} \right)}^2} - d_{DL,0}^2} \right)} \right)  \notag\\
  %	&	 \times \exp \left( {\pi {\lambda _s}s{{\bar g}_{DL}}\beta \frac{{{R_S}}}{{{R_E}}}\int_0^{\frac{{{{\left( {D_{DL,0}^{\max }} \right)}^2}}}{{s{{\bar g}_{DL}}\beta }}} {\frac{{{t^\alpha }}}{{{{\left( {t + 1} \right)}^\alpha }}}{\rm{d}}t} } \right) \notag\\
  	%&\div \exp \left( {\pi {\lambda _s}s{{\bar g}_{DL}}\beta \frac{{{R_S}}}{{{R_E}}}\int_0^{\frac{{d_{DL,0}^2}}{{s{{\bar g}_{DL}}\beta }}} {\frac{{{t^\alpha }}}{{{{\left( {t + 1} \right)}^\alpha }}}{\rm{d}}t} } \right) \notag \\
  	&	\mathop  = \limits^{\left( f \right)} \exp \left( { - \pi {\lambda _s}\frac{{{R_S}}}{{{R_E}}}\left( {{{\left( {D_{DL,0}^{\max }} \right)}^2} - d_{DL,0}^2} \right)} \right) \notag\\
  	&	\times \exp \left( {\pi {\lambda _s}\frac{{{R_S}}}{{{R_E}}}\frac{{ {{\left( {D_{DL,0}^{\max }} \right)}^{2\left( {\alpha  + 1} \right)}}}}{{{\left( {s{{\bar g}_{DL}}\beta } \right)}^{ \alpha }(\alpha  + 1)}}{ \times  {J\left( {s,{{\bar g}_{DL}},D_{DL,0}^{\max }} \right)}} } \right) \notag\\
  	&	\div \exp \left( {\pi {\lambda _s}\frac{{{R_S}}}{{{R_E}}}\frac{{d_{DL,0}^{2\left( {\alpha  + 1} \right)}}}{{{\left( {s{{\bar g}_{DL}}\beta } \right)}^{  \alpha }(\alpha  + 1)}}{  {J\left( {s,{{\bar g}_{DL}},d_{DL,0} } \right)}} } \right)
  \end{align}
  where (a) follows from the i.i.d. distribution of  $ {\left| {{h_{DL,i}}} \right|^2}$, (b) follows from the probability generating functional (PGFL) of the SPPP and  $ {  \bar {\cal A}{_{DL,vis}}  } = {{\cal A}{_{DL,vis}}\backslash {\cal A}{^{'}_{DL,vis}}\left( {{d_{DL,0}}} \right)} $, (c) comes from $ \frac{{\partial \left| {\bar {\cal A}{_{DL,vis}}} \right|}}{{\partial d_{DL,0}}} = 2\pi \frac{{{R_S}}}{{{R_E}}}{d_{DL,0}}$, (d) comes from the Laplace transform of ${{\left| h \right|}^2}$ that is  $ {\mathbb{E}_{{{\left| h \right|}^2}}}\left( {{e^{ - s{{\bar g}_{DL}}{{\left| h \right|}^2}d_{}^{ - 2}}}} \right) = {{\cal L}_{{{\left| h \right|}^2}}}\left( {s{{\bar g}_{DL}}d_{}^{ - 2}} \right) = \frac{1}{{{{\left( {s{{\bar g}_{DL}}d_{}^{ - 2}\beta  + 1} \right)}^\alpha }}}$, (e) is the change of variable ${t=\left( {s{{\bar g}_{DL}}\beta } \right)^{ - 1}}d_{}^2$, and (f) comes from $ \int_0^u {\frac{{{x^{\mu  - 1}}dx}}{{{{(1 + \beta x)}^\nu }}}}  = \frac{{{u^\mu }}}{\mu }{ \times _2}{F_1}\left( {\nu ,\mu ;1 + \mu ; - \beta u} \right)$, and the Gaussian hypergeometric function $ _2{F_1}\left( {.,. ;. ; .} \right)$ is expressed by $ J\left( {\varpi ,\tau ,\upsilon } \right){ = _2}{F_1}\left( {\alpha ,\alpha  + 1;2 + \alpha ; - \frac{{{\upsilon ^2}}}{{\varpi \tau \beta }}} \right)$.
  
  Similarly, according to the definition of  the conditional aggregated interference ${{\bar I}_S}$ at $S_{UL,0}$, ${\Phi _G^{'}}={\Phi _G} \cap {{\cal A}_{UL,vis}}\backslash {\cal A}_{UL,vis}^{'}\left( {{d_{UL,0}}} \right) $ denotes the area on the spherical cap outside ${\cal A}{^{'}_{UL,vis}}\left( {{d_{UL,0}}} \right)$. Thus, the Laplace transform of ${{\bar I}_S}$ is computed by 
   \begin{align} 
  	&{{\cal L}_{{{\bar I}_S}}}\left( s \right) = {\mathbb{E}_{{{\bar I}_S}}}\left[ {\left. {{e^{ - s{{\bar I}_S}}}} \right|{D_{UL,0}=d_{UL,0}},{\Phi _G}\left( {{\cal A}{_{UL,vis}}} \right) > 0} \right] \notag\\
  	&	= \mathbb{E}{_{{\Phi _G}\left( {{\cal A}{_{UL,vis}}} \right),{{\left| {{h_{UL,j}}} \right|}^2}}}\left[ {\exp \left( { - \sum\limits_{j \in {{{\Phi }_G^{'}}} } {s{{\bar g}_{UL}}{{\left| {{h_{UL,j}}} \right|}^2}d_{DL,j}^{ - 2}} } \right)} \right] \notag\\
   	& = \exp \left( { - \pi {\lambda _g}\frac{{{R_E}}}{{{R_S}}}\left( {{{\left( {D_{UL,0}^{\max }} \right)}^2} - d_{UL,0}^2} \right)} \right) \notag\\
  	&	\times \exp \left( {\pi {\lambda _g}\frac{{{R_E}}}{{{R_S}}}\frac{{ {{\left( {D_{UL,0}^{\max }} \right)}^{2\left( {\alpha  + 1} \right)}}}}{{{\left( {s{{\bar g}_{UL}}\beta } \right)}^{ \alpha }(\alpha  + 1)}}{ \times  {J\left( {s,{{\bar g}_{UL}},D_{UL,0}^{\max }} \right)}} } \right) \notag\\
  	&	\div \exp \left( {\pi {\lambda _g}\frac{{{R_E}}}{{{R_S}}}\frac{{d_{UL,0}^{2\left( {\alpha  + 1} \right)}}}{{{\left( {s{{\bar g}_{UL}}\beta } \right)}^{  \alpha }(\alpha  + 1)}}{    {J\left( {s,{{\bar g}_{UL}},d_{UL,0} } \right)}} } \right)
  \end{align}
 % where (a) follows from the i.i.d. distribution of  $ {\left| {{h_{UL,j}}} \right|^2}$, (b) follows from the PGFL of the SPPP and  $ {   \bar {\cal A}{_{UL,vis}} } = {{\cal A}{_{UL,vis}}\backslash {\cal A}{^{'}_{UL,vis}}\left( {{d_{UL,0}}} \right)} $, (c) comes from $ \frac{{\partial \left| {\bar {\cal A}{_{UL,vis}}} \right|}}{{\partial d_{UL,0}}} = 2\pi \frac{{{R_E}}}{{{R_S}}}{d_{UL,0}}$, (d) comes from the Laplace transform of ${{\left| h \right|}^2}$, (e) is the change of variable ${t=\left( {s{{\bar g}_{UL}}\beta } \right)^{ - 1}}d_{}^2$, (f) comes from   $ J\left( {\varpi ,\tau ,\upsilon } \right){ = _2}{F_1}\left( {\alpha ,\alpha  + 1;2 + \alpha ; - \frac{{{\upsilon ^2}}}{{\varpi \tau \beta }}} \right)$.
  
    This completes the proof.
    
    \section{}
    \label{AppendixE}
   Based on the expression of (\ref{POUTDL}), the outage probability of $G_{DL,0}$ conditioned on $\Phi_S({\cal A}_{DL,vis})>0$ is computed by
    \begin{align} \label{ConditionPOUTDL}
    	&	P_{out}^{DL}\left( {\gamma \left| {\Phi_S \left( {{{\cal A}_{DL,vis}}} \right) > 0} \right.} \right) \notag\\
    %	&= \Pr \left( {\left. {\frac{{{{\left| {{h_{DL,0}}} \right|}^2}D_{DL,0}^{ - 2}}}{{\bar \sigma _G^2 + {{\bar I}_G}}} < \gamma } \right|\Phi_S \left( {{{\cal A}_{DL,vis}}} \right) > 0} \right)\notag\\
    	&	= {\mathbb{E}_{{D_{DL,0}}}}\left[ {{\mathbb{E}_{{{\bar I}_G}}}\left[ {\Pr \left[ {{{\left| {{h_{DL,0}}} \right|}^2} < \gamma \left( {\bar \sigma _G^2 + {{\bar I}_G}} \right)D_{DL,0}^2} \right.} \right.} \right.\notag\\
    	&	\times \left. {\left. {\left. {\left| {D_{DL,0}^{},\Phi_S \left( {{{\cal A}_{DL,vis}}} \right) > 0,{{\bar I}_G}} \right.} \right]} \right]} \right]\notag\\
    	&	\mathop  = \limits^{\left( a \right)} 1 - {\mathbb{E}_{{D_{DL,0}}}}\left[ {\exp \left( { - {\beta ^{ - 1}}\bar \sigma _G^2\gamma D_{DL,0}^2} \right)} \right.\notag\\
    	&	\times {\mathbb{E}_{{{\bar I}_G}}}\left[ {{e^{ - {\beta ^{ - 1}}\gamma D_{DL,0}^2{{\bar I}_G}}}} \right.\sum\limits_{m = 0}^{\alpha  - 1} {\frac{{{{\left( {{\beta ^{ - 1}}\gamma D_{DL,0}^2\left( {\bar \sigma _G^2 + {{\bar I}_G}} \right)} \right)}^m}}}{{m!}}} \notag\\
    	&	\times \left| {\left. {\left. {D_{DL,0}^{},\Phi _S\left( {{{\cal A}_{DL,vis}}} \right) > 0,{{\bar I}_G}} \right]} \right]} \right.\notag\\
    	&	\mathop  = \limits^{\left( b \right)} 1 - {\mathbb{E}_{{D_{DL,0}}}}\left[ {\exp \left( { - {\beta ^{ - 1}}\bar \sigma _G^2\gamma D_{DL,0}^2} \right)} \right]\notag\\
    	&	\times {\mathbb{E}_{{D_{DL,0}}}}\left[ {\sum\limits_{m = 0}^{\alpha  - 1} {{\mathbb{E}_{{{\bar I}_G}}}\left[ {{e^{ - {\beta ^{ - 1}}\gamma D_{DL,0}^2{{\bar I}_G}}} \times \frac{{{{\left( {sD_{DL,0}^2} \right)}^m}}}{{m!}}} \right.} } \right.\notag\\
    	&	\times \left. {\left. {\left. {\sum\limits_{k = 0}^m {C_m^k{{\left( {\bar \sigma _G^2} \right)}^{m - k}}{{\left( {{{\bar I}_G}} \right)}^k}} } \right|D_{DL,0}^{},\Phi_S \left( {{{\cal A}_{DL,vis}}} \right) > 0,{{\bar I}_G}} \right]} \right]\notag\\
    	&	\mathop  = \limits^{\left( c \right)} 1 - {\mathbb{E}_{{D_{DL,0}}}}\left[ {\exp \left( { - {\beta ^{ - 1}}\bar \sigma _G^2\gamma D_{DL,0}^2} \right)} \right]\notag\\
    	&	\times {\mathbb{E}_{{D_{DL,0}}}}\left[ {\sum\limits_{m = 0}^{\alpha  - 1} {\left[ {\frac{{{{\left( {sD_{DL,0}^2} \right)}^m}}}{{m!}}\sum\limits_{k = 0}^m {C_m^k{{\left( {\bar \sigma _G^2} \right)}^{m - k}} } } \right.} } \right.\notag\\
    	&	\times \left. {\left. {\left. {{{\left( { - 1} \right)}^k}\frac{{{{\rm{d}}^k}{{\cal L}_{{{\bar I}_G}}}\left( {sD_{DL,0}^2} \right)}}{{{\rm{d}}{{\left( {sD_{DL,0}^2} \right)}^k}}}} \right|D_{DL,0}^{},\Phi_S \left( {{{\cal A}_{DL,vis}}} \right) > 0,{{\bar I}_G}} \right]} \right]\notag\\
    %	&	= 1 - \int_{{d_{_{DL,}0}} > 0} {e^{ { - {\beta ^{ - 1}}\bar \sigma _G^2\gamma d_{DL,0}^2} }{f_{{D_{DL,0}}}}\left( {{d_{DL,0}}} \right){\rm{d}}{d_{DL,0}}}\notag \\
    %	&	\times \int_{{d_{_{DL,}0}} > 0} {\sum\limits_{m = 0}^{\alpha  - 1} {\left[ {\frac{{{{\left( {sD_{DL,0}^2} \right)}^m}}}{{m!}}\sum\limits_{k = 0}^m {C_m^k{{\left( {\bar \sigma _G^2} \right)}^{m - k}}} } \right.} } \notag\\
    %	&	\times \left. {{{\left( { - 1} \right)}^k}\frac{{{{\rm{d}}^k}{{\cal L}_{{{\bar I}_G}}}\left( {sd_{DL,0}^2} \right)}}{{{\rm{d}}{{\left( {sd_{DL,0}^2} \right)}^k}}}} \right] \times {f_{{D_{DL,0}}}}\left( {{d_{DL,0}}} \right){\rm{d}}{d_{DL,0}},
    \end{align}
    where (a) follows the CDF of ${\left| h \right|}^2$ that is given as $ {F_{{{\left| h \right|}^2}}}\left( x \right) = \Pr \left( {{{\left| h \right|}^2} \le x} \right) = 1 - \exp \left( { - \frac{x}{\beta }} \right)\sum\limits_{m = 0}^{\alpha  - 1} {\frac{{{{\left( {\frac{x}{\beta }} \right)}^m}}}{{m!}}} $, (b) is the change of variable $ s = {\beta ^{ - 1}}\gamma $ and ${\left( {a + b} \right)^m} = \sum\limits_{k = 0}^m {C_m^k{a^{m - k}}{b^k}}  $, (c) follows $ {\mathbb{E}_x}\left[ {{{\left( x \right)}^k}{e^{ - sx}}} \right] = {\left( { - 1} \right)^k}\frac{{{{\rm{d}}^k}{{\cal L}_x}\left( s \right)}}{{{\rm{d}}{s^k}}}$. Here, the PDF of $D_{DL,0}$ is given in Lemma 2, and  the Laplace transform of ${\bar I}_G$ is given in Lemma 3.

Similarly, based on the expression of (\ref{POUTUL}), the outage probability of $S_{UL,0}$ conditioned on $\Phi_G({\cal A}_{UL,vis})>0$ is computed by
   \begin{align}\label{ConditionPOUTUL}
	&	P_{out}^{UL}\left( {\gamma \left| {\Phi_G \left( {{{\cal A}_{UL,vis}}} \right) > 0} \right.} \right) \notag\\
	&	= 1 - \int_{{d_{_{UL,}0}} > 0} {e^{ { - {\beta ^{ - 1}}\bar \sigma _S^2\gamma d_{UL,0}^2} }{f_{{D_{UL,0}}}}\left( {{d_{UL,0}}} \right){\rm{d}}{d_{UL,0}}}\notag \\
	&	\times \int_{{d_{_{UL,}0}} > 0} {\sum\limits_{m = 0}^{\alpha  - 1} {\left[ {\frac{{{{\left( {sD_{UL,0}^2} \right)}^m}}}{{m!}}\sum\limits_{k = 0}^m {C_m^k{{\left( {\bar \sigma _S^2} \right)}^{m - k}}} } \right.} } \notag\\
	&	\times \left. {{{\left( { - 1} \right)}^k}\frac{{{{\rm{d}}^k}{{\cal L}_{{{\bar I}_S}}}\left( {sd_{UL,0}^2} \right)}}{{{\rm{d}}{{\left( {sd_{UL,0}^2} \right)}^k}}}} \right] \times {f_{{D_{UL,0}}}}\left( {{d_{UL,0}}} \right){\rm{d}}{d_{UL,0}},
\end{align}
where %(a)-(c) is same with (\ref{ConditionPOUTDL}), 
the PDF of $D_{UL,0}$ is given in Lemma 2, and 
  the Laplace transform of ${\bar I}_S$ is given in Lemma 3.

 By multiplying  $\Pr (\Phi_S({\cal A}_{DL,vis})>0)$ in Lemma 1 with (\ref{ConditionPOUTDL})  and   $\Pr (\Phi_G({\cal A}_{UL,vis})>0)$ in Lemma 1 with (\ref{ConditionPOUTUL}),   $P_{out}^{DL}$ and  $P_{out}^{UL}$ are obtained as   
\begin{align}
	P_{out}^{DL}\left( \gamma  \right) &= \Pr \left( {\gamma _{DL,0}^{} < \gamma \left| {{\Phi _S}\left( {{{\cal A}_{DL,vis}}} \right) > 0} \right.} \right) \notag\\
	&{\rm{  }} \times \Pr \left( {{\Phi _s}\left( {{{\cal A}_{DL,vis}}} \right) > 0} \right) \notag\\
	&= 1 - \eta \left( {\gamma ,\bar \sigma _G^2,d_{DL,0}^{\min },d_{DL,0}^{\max },{{\bar g}_{DL}},\frac{{{R_S}}}{{{R_E}}},{\lambda _s}} \right) \notag\\
	&\times \left( {1 - e^  { - \pi {R_S}{\lambda _s}\left( {\frac{{{{  {d_{DL,0}^{\max }}  }^{\rm{2}}} + 2{R_E}{R_S} - R_E^2 - R_S^2}}{{{R_E}}}} \right)} } \right),
\end{align} 
and
\begin{align} 
	P_{out}^{UL}\left( \gamma  \right) &= \Pr \left( {\gamma _{UL,0}^{} < \gamma \left| {{\Phi _G}\left( {{{\cal A}_{UL,vis}}} \right) > 0} \right.} \right) \notag\\
	&{\rm{  }} \times \Pr \left( {{\Phi _G}\left( {{{\cal A}_{UL,vis}}} \right) > 0} \right) \notag\\
	&	= 1 - \eta \left( {\gamma ,\bar \sigma _S^2,d_{UL,0}^{\min },d_{UL,0}^{\max },{{\bar g}_{UL}},\frac{{{R_E}}}{{{R_S}}},{\lambda _g}} \right) \notag\\
	&	{\rm{    }} \times \left( {1 - \exp \left( { - 2\pi {\lambda _g}R_E^2\left( {1 - \cos {\varphi _{\max }}} \right)} \right)} \right),
\end{align} 
respectively. 
The expression of  $\eta \left( {\gamma ,\delta ,{d_{\max }},{d_{\min }},g,\kappa ,\lambda ,} \right)$ and the corresponding sub-functions are given in (\ref{OP1})-(\ref{OP4}). 

    This completes the proof.

\bibliographystyle{IEEEtran}
\bibliography{references}

\newpage

\vfill

\end{document}